\documentclass{article}

\usepackage[preprint, nonatbib]{neurips_2022}

\usepackage[utf8]{inputenc} %
\usepackage[T1]{fontenc}    %
\usepackage{hyperref}       %
\usepackage{url}            %
\usepackage{booktabs}       %
\usepackage{amsfonts}       %
\usepackage{nicefrac}       %
\usepackage{microtype}      %
\usepackage{xcolor}         %
\usepackage{graphicx}
\usepackage{caption}
\usepackage{amsmath}
\usepackage{subcaption}
\usepackage{placeins}
\hypersetup{
    colorlinks=true
}

\title{Disentangled Uncertainty and Out of Distribution Detection in Medical Generative Models}

\author{%
    \parbox{\linewidth}{
        Kumud Lakara$^1$, Matias Valdenegro-Toro$^2$\\
    }\\
    ~\\
    \parbox{\linewidth}{
        $^1$ Department of CSE, Manipal Institute of Technology, 576104 Manipal, India.\\
        $^2$ Department of AI, University of Groningen, 9747AG Groningen, The Netherlands.\\
        ~\\
        \texttt{lakara.kumud@gmail.com}, \texttt{m.a.valdenegro.toro@rug.nl}
    }
}

\begin{document}

\maketitle

\begin{abstract}
	Trusting the predictions of deep learning models in safety critical settings such as the medical domain is still not a viable option. Distentangled uncertainty quantification in the field of medical imaging has received little attention. In this paper, we study disentangled uncertainties in image to image translation tasks in the medical domain. We compare multiple uncertainty quantification methods, namely Ensembles, Flipout, Dropout, and DropConnect, while using CycleGAN to convert T1-weighted brain MRI scans to T2-weighted brain MRI scans. We further evaluate uncertainty behavior in the presence of out of distribution data (Brain CT and RGB Face Images), showing that epistemic uncertainty can be used to detect out of distribution inputs, which should increase reliability of model outputs.
\end{abstract}

\section{Introduction}
Translating an image from one domain to another is a challenging task as there may not necessarily be a one-one mapping between the two domains. Generative models are often used to learn this one-to-one mapping in an unsupervised manner with an additional constraint on the image or feature space. This constraint imposes structure on the underlying joint distribution of the two images being translated from different domains \cite{park2020contrastive, liu2017unsupervised, zhu2017unpaired, tran2018dist}. In the medical domain image-to-image translation is very significant. It has a myriad of applications including denoising \cite{gu2021adain}, motion correction \cite{armanious2020unsupervised} and MRI scan translation. Generative adversarial networks have achieved state of the art performance on these tasks. However, they usually do not provide uncertainty estimates for their predictions. In addition to this, existing models usually learn a deterministic mapping between the two domains being translated. In the medical domain such mapping cannot quantify the model uncertainty to predict trustworthy outputs. Moreover, most state-of-the-art methods do not account for out of distribution (OOD) data and often fail under such conditions thus making them unreliable and incapable of real world deployment. Incorporation of uncertainty into image to image translation tasks is of prime importance, especially in the medical domain in order to make informed medical decisions.  

When considering uncertainty, aleatoric uncertainty is caused by inherent noise and randomness in the data itself and is hence irreducible \cite{der2009aleatory, kendall2017uncertainties}, while epistemic uncertainty on the other hand is due to the model's inability to learn effectively from the data and can be reduced by training with more data. These values are usually predicted as a single value called predictive uncertainty \cite{gal2016uncertainty}. Disentangling epistemic and aleatoric uncertainties is of great value when it comes to out of distribution detection \cite{hendrycks2016baseline}. This is of particular importance in the case of critical applications including healthcare and medical diagnosis \cite{esteva2017dermatologist, filos2019systematic}.  

This paper shows how a CycleGAN \cite{zhu2017unpaired} generative model trained on medical images for image to image translation (Brain MRI to Brain CT) can be combined with uncertainty quantification methods to produce disentangled estimates of aleatoric and epistemic uncertainty, while at the same time performing out of distribution detection. The contributions are a simple formulation for uncertainty disentanglement in CycleGAN, and a comparison of different uncertainty quantification methods (Flipout, Dropout, DropConnect, Ensembles) for both image translation with (disentangled) uncertainty and out of distribution detection.

\section{Related Work}

Past work on image-to-image translation in the medical domain has focussed on convolutional neural networks \cite{chen2017low, dou2016multilevel, havaei2017brain, kamnitsas2017efficient, litjens2017survey} and generative adversarial networks \cite{armanious2019unsupervised, armanious2020medgan, isola2017image, nie2018medical, upadhyay2019mixed, upadhyay2019robust, wolterink2017deep, yang2018low, zhang2019image}. Some machine learning techniques also rely on explicit feature representations \cite{huynh2015estimating, kustner2017mr, rueda2013single}. The existing methods usually work by enforcing additional constraints on the joint distribution of the images from different domains. \cite{kendall2017uncertainties, upadhyay2021uncertainty, wang2019aleatoric} explore the shortcomings of using conditional GAN architectures with deterministic outputs. %
Though state-of-the-art models are capable of synthesizing high quality output images, the image is often very different from the ground-truth. This could result in various problems including over-confidence and mis-interpretation which could in turn lead to wrong diagnoses, misinformed decisions, delayed treatment and other severe consequences. Hence, in the medical domain confidence and certainty in the model's predictions are of utmost value. Moreover, being able to evaluate uncertainty as two separate constituents further enhances the quality of confidence in the model's predictions. Previous work \cite{kendall2017uncertainties, depeweg2018decomposition, hullermeier2021aleatoric} discusses methods to calculate the total uncertainty as well as individual epistemic and aleatoric uncertainty. %
Epistemic uncertainty arises in the model parameters due to training with finite data and aleatoric uncertainty arises due to the noise or uncertainty that is inherently present in the data itself \cite{kabir2018neural, kendall2017uncertainties}. There have been numerous attempts to quantify aleatoric and epistemic uncertainty in medical imaging tasks including classification, segmentation and more \cite{nair2020exploring, tanno2017bayesian, wang2019aleatoric, wang2018automatic, wang2019automatic} however, uncertainty quantification for image-to-image translation tasks still remains largely unexplored. 

Image-to-image translation tasks in the medical domain have recently come to the fore-front for computer vision based research in the medical field. \cite{yang2020mri, armanious2020medgan, kaji2019overview, welander2018generative, han2019combining}. Uncertainty aware image-to-image translation models have also been proposed with special emphasis on CT and MRI images. Upadhyay et. al\cite{upadhyay2021uncertainty} propose a probabilistic method that models the per-pixel residual by generalized Gaussian distribution which helps increase the model's robustness to OOD data. \cite{upadhyay2021uncertainty2} propose an uncertainty guided progressive learning scheme for image-to-image translation. However, none of the current methods focus on the individual behavior of epistemic and aleatoric uncertainties for image-to-image translation tasks in the medical domain and that is what we address in this work.   

\section{CycleGAN and Uncertainty Disentanglement}

CycleGAN provides a method to tranlate an image from a source domain (say $P$) to a target domain (say $Q$) in the absence of paired examples. The basic idea is to learn a mapping $G: P \rightarrow Q$ in a manner such that the distribution of resultant images ($G(P)$) is identical to the distribution of $Q$. This is achieved by  enforcing an additional structure on the joint distribution. The initial $G: P \rightarrow Q$ mapping is coupled with an inverse $F: Q \rightarrow P$ mapping and a \emph{cycle consistency} loss is used.
\begin{equation}
    L_{\text{cyc}} = \mathbb{E}_{x \sim p_{\text{data}(x)}}[||F(G(x)) - x||_1] + \mathbb{E}_{y \sim p_{\text{data}(y)}}[||G(F(y)) - y||_1]
\end{equation}
To disentangle the uncertainty value into epistemic and aleatoric uncertainty we consider the mean ($\mu$) and variance ($\sigma^2$) output from the model. The variance head of the model is trained using the Gaussian negative log-likelihood loss:
\begin{equation}
	L_{NLL}(y_n, \mathbf{x}_n) = \frac{\log \sigma^2_i(\mathbf{x}_n)}{2} + \frac{(\mu_i(\mathbf{x}_n) - y_n)^2}{2 \sigma^2_i(\mathbf{x}_n)} .
	\label{eq:gaussian_nll}
\end{equation}
This replaces the L1 distance in the cycle consistency loss, which now becomes:
\begin{equation}
    L_{\text{cyc}} = \mathbb{E}_{x \sim p_{\text{data}(x)}}[L_{NLL}(F(G(x)), x)] + \mathbb{E}_{y \sim p_{\text{data}(y)}}[L_{NLL}(G(F(y)), y)]
\end{equation}
During inference, we sample weights from a distribution $\theta \sim p(\theta|x,y)$, which produce different predictions for $\mu$ and $\sigma^2$ which correspond to the approximate predictive posterior distribution of the model. These obtained samples are then combined to obtain the Gaussian mixture distribution:
\begin{equation}
	\mu_*(x) = M^{-1}\sum_i\mu_i(x) \qquad \sigma_*^2(x) = M^{-1}\sum_i(\sigma_i^2(x)+\mu_i^2(x)) - \mu_*^2(x)
\end{equation}
The predictive variance can further be decomposed as follows:
\begin{equation}
	\begin{aligned}
	\sigma^2_*(x) = M^{-1}\sum_i\mu^2_i(x) - \mu_*^2(x) &= \mathbb{E}_i[\sigma^2_i(x)] + \mathbb{E}_i[\mu^2_i(x)] - \mathbb{E}_i[\mu_i(x)]^2	\\
	& = \underbrace{\mathbb{E}_i[\sigma^2_i(x)]}_\text{Aleatoric Uncertainty} + \underbrace{\text{Var}_i[\mu_i(x)]}_\text{Epistemic Uncertainty}
	\end{aligned}
\end{equation}
From the above derivation, we posit that the mean of the variances corresponds to aleatoric uncertainty where as the variance of the means gives us the epistemic uncertainty. The common intuition here is that using the Gaussian NLL loss, each variance head estimates aleatoric uncertainty, so taking the mean of these variance heads, produces a combined aleatoric uncertainty, while taking the variance through  ensemble members or samples from a forward pass corresponds to disagreement between ensembles/samples, large disagreement indicates that the whole ensemble/samples are not sure of the prediction, while a low disagreement indicates a large confidence in the prediction, both of these concepts measure epistemic uncertainty.

\section{Experiments and Results}
We provide a comprehensive overview of different uncertainty quantification methods as described in section \ref{unc_methods} in their ability to disentangle epistemic uncertainty from aleatoric uncertainty. For the purpose of our experiments we take 3 datasets: IXI Brain MRI Scans \cite{ixi_data}, Computed Tomography Images for Intracranial Hemorrhage Detection and Segmentation version 1.3.1( referred to as Brain CT Scans dataset hereafter) \cite{hssayeni2020intracranial, ct_dataset}, UTKFace dataset \cite{zhifei2017cvpr}. The IXI Brain MRI Scans dataset is used for training and the CT scans and UTKFace datasets are used for out of distribution analyses of the model. The uncertainty values are normalized using min-max normalization for the purpose of analyses.

\begin{figure}[ht]
    \centering
	\begin{subfigure}[b]{0.49\textwidth}		
		\includegraphics[width=0.49\textwidth]{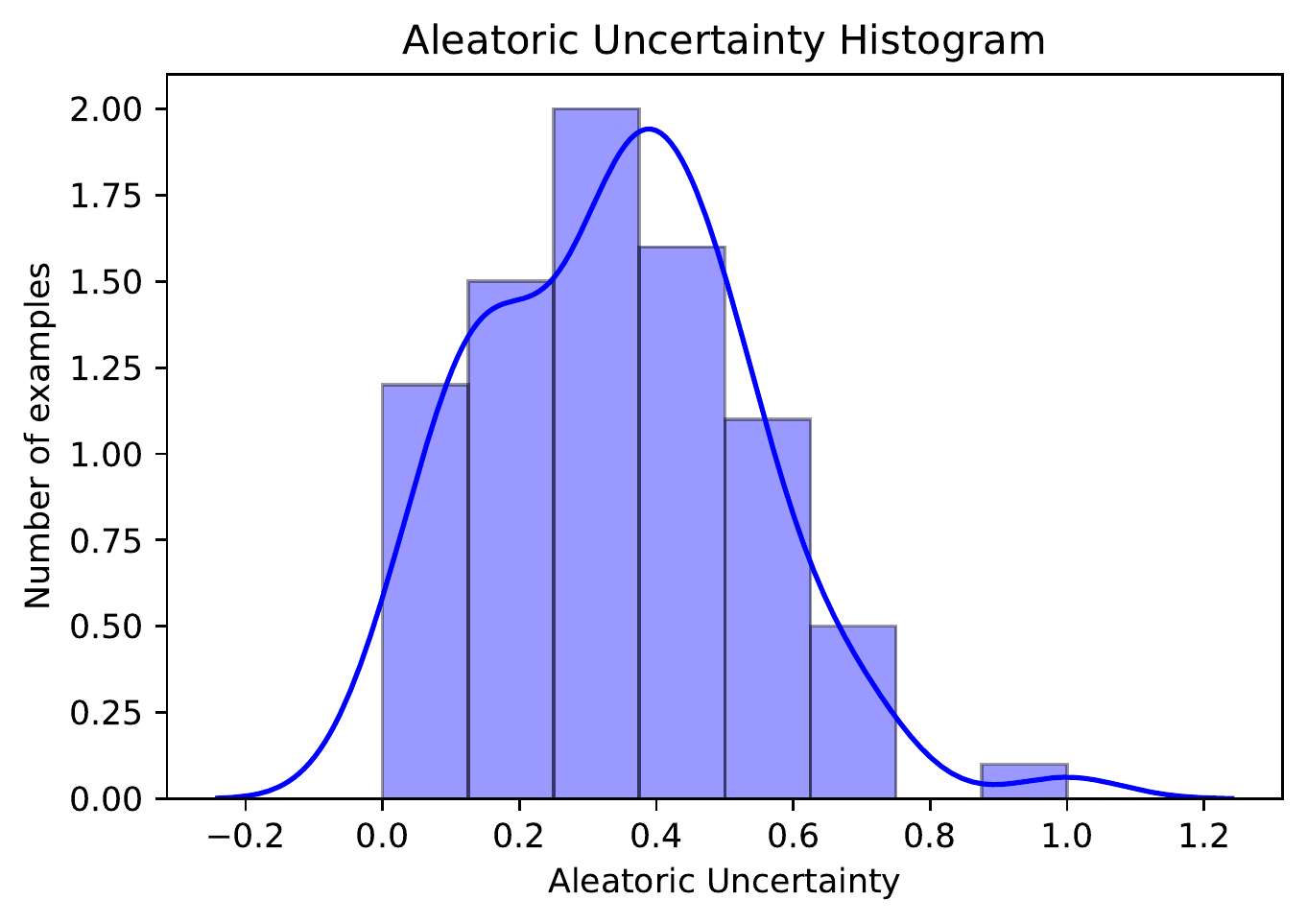}
        \includegraphics[width=0.49\textwidth]{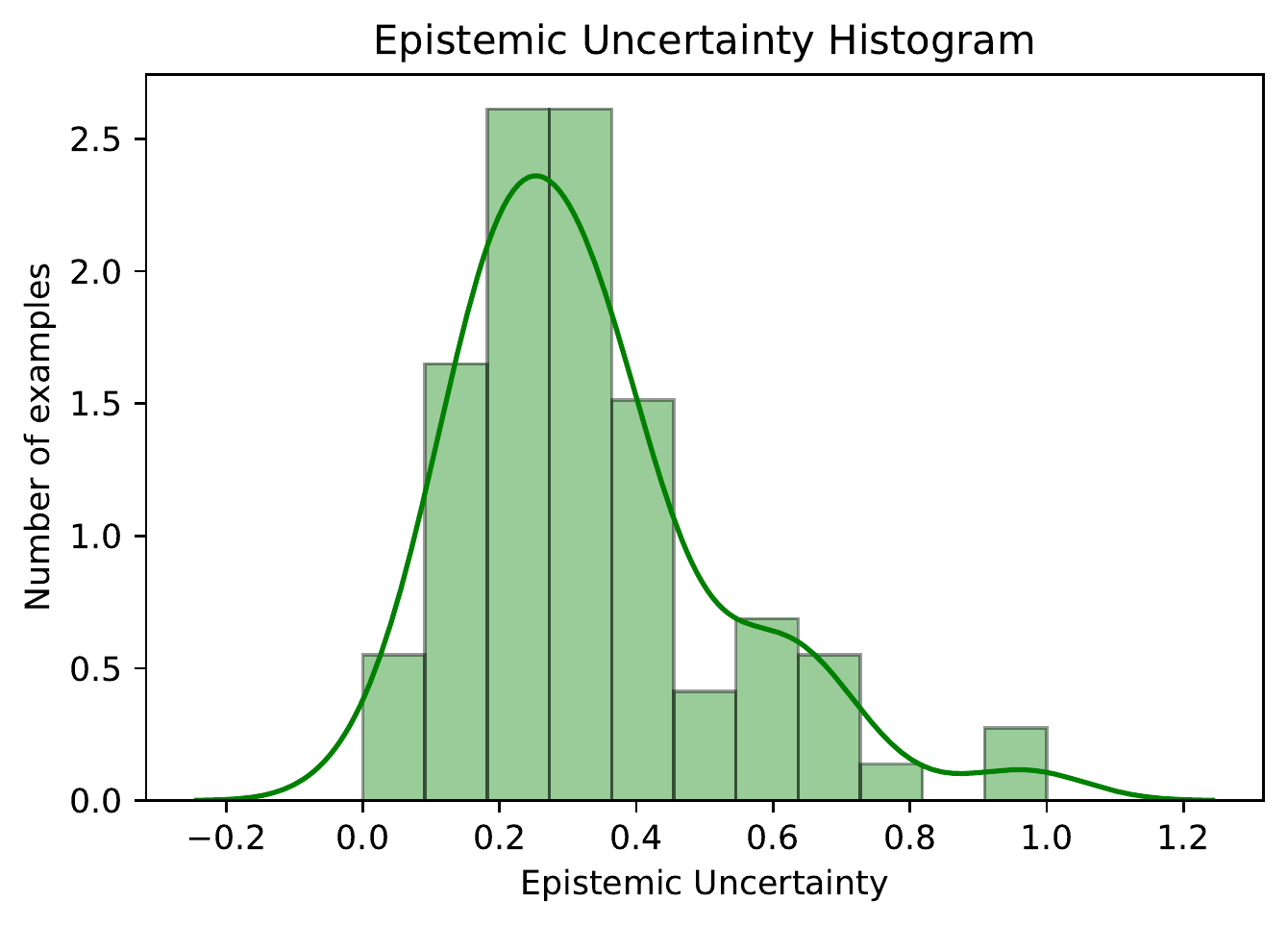}
        \caption{IXI Brain MRI Scans}
	\end{subfigure}
    \begin{subfigure}[b]{0.49\textwidth}		
        \includegraphics[width=0.49\textwidth]{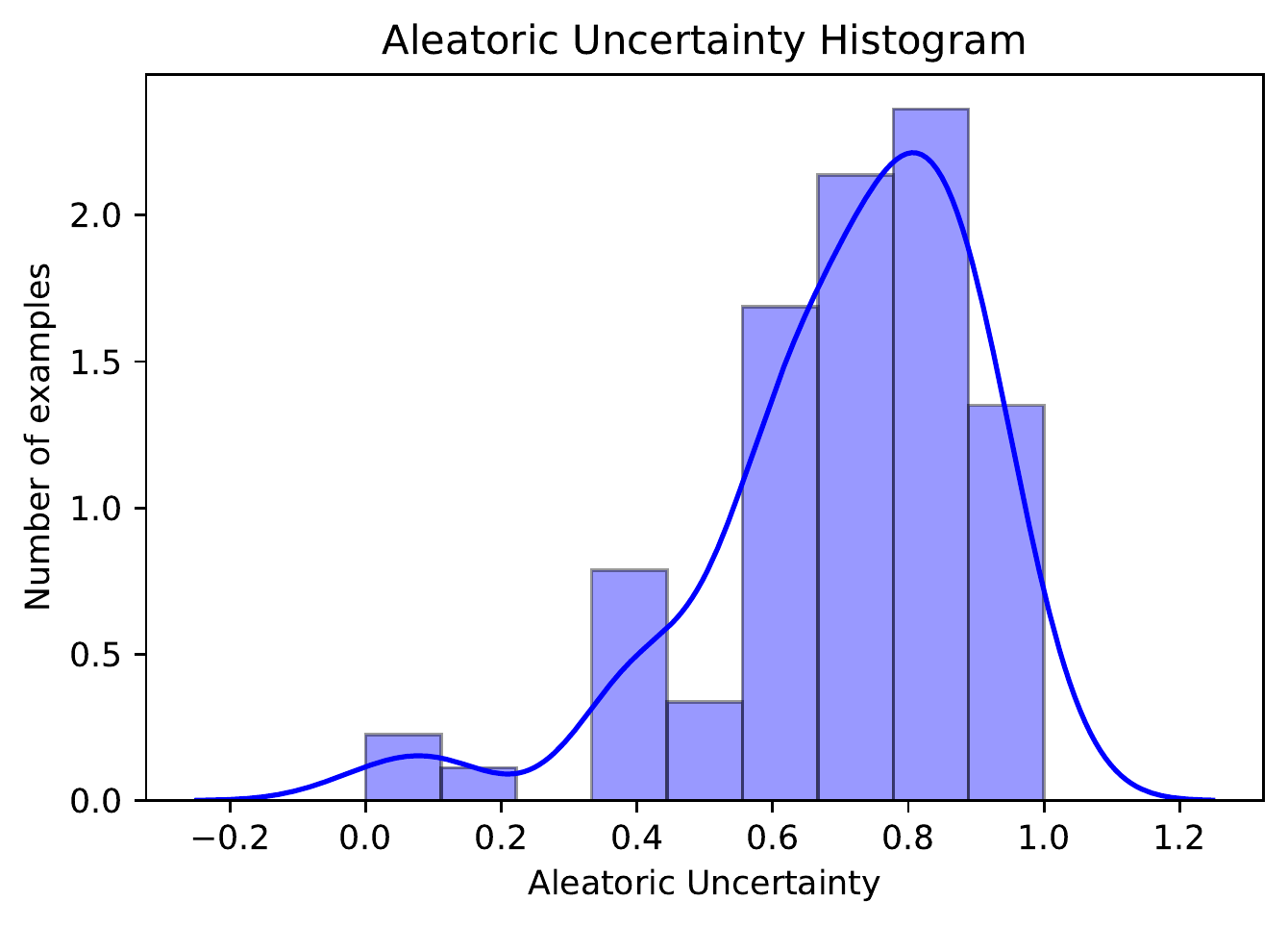}
        \includegraphics[width=0.49\textwidth]{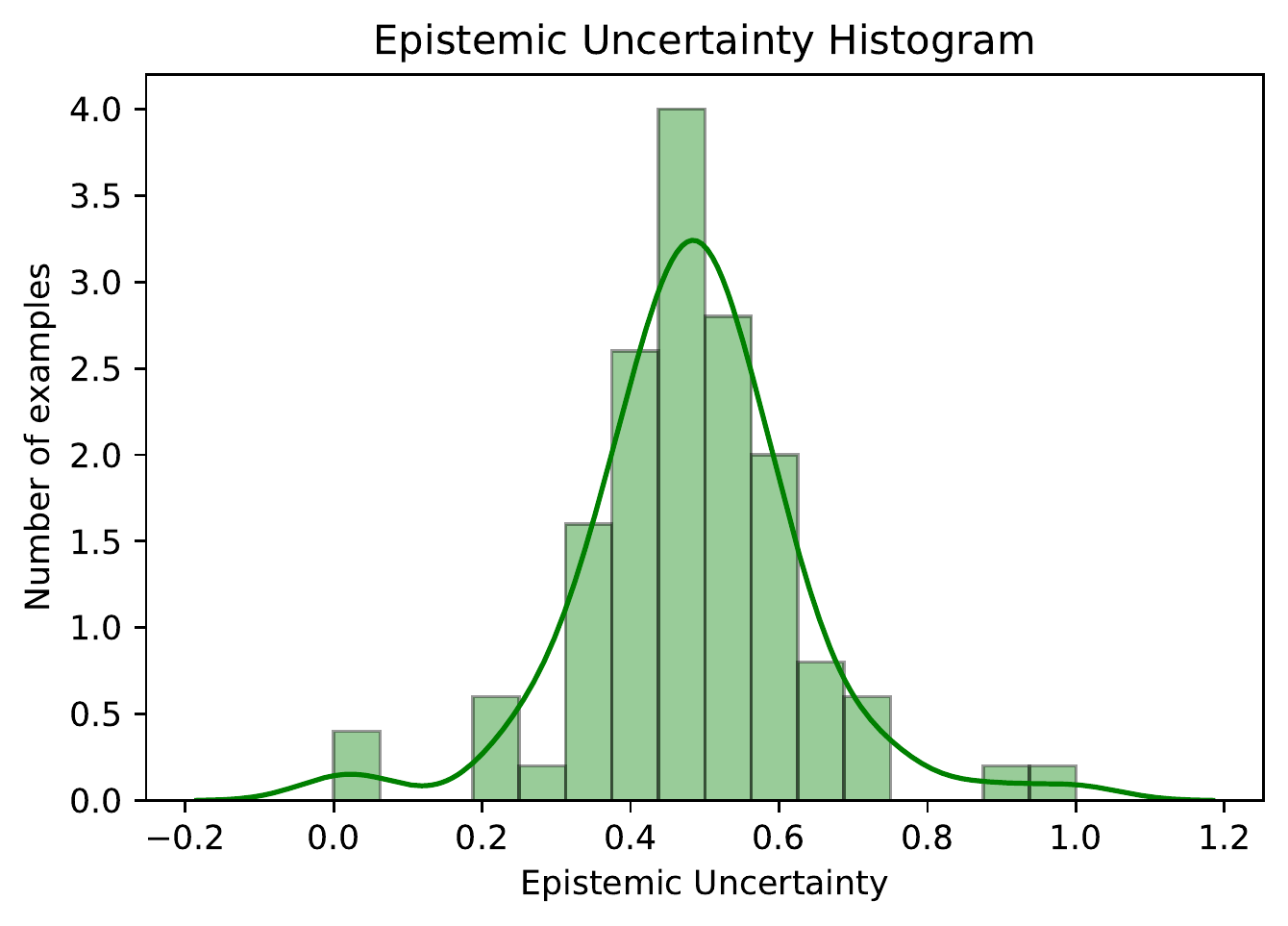}
        \caption{Brain CT Scans}
    \end{subfigure}

    \begin{subfigure}[b]{0.49\textwidth}		
        \includegraphics[width=0.49\textwidth]{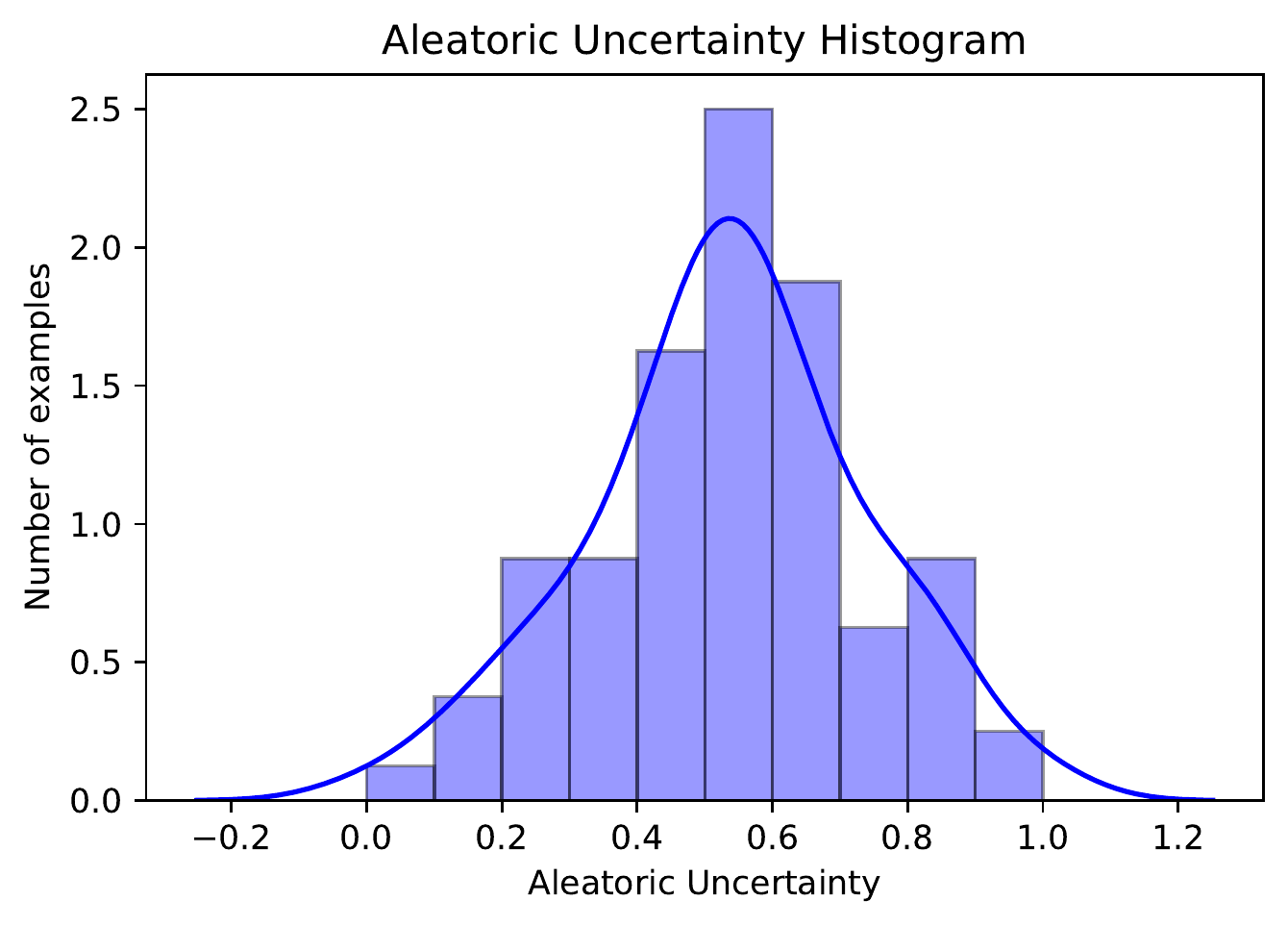}
        \includegraphics[width=0.49\textwidth]{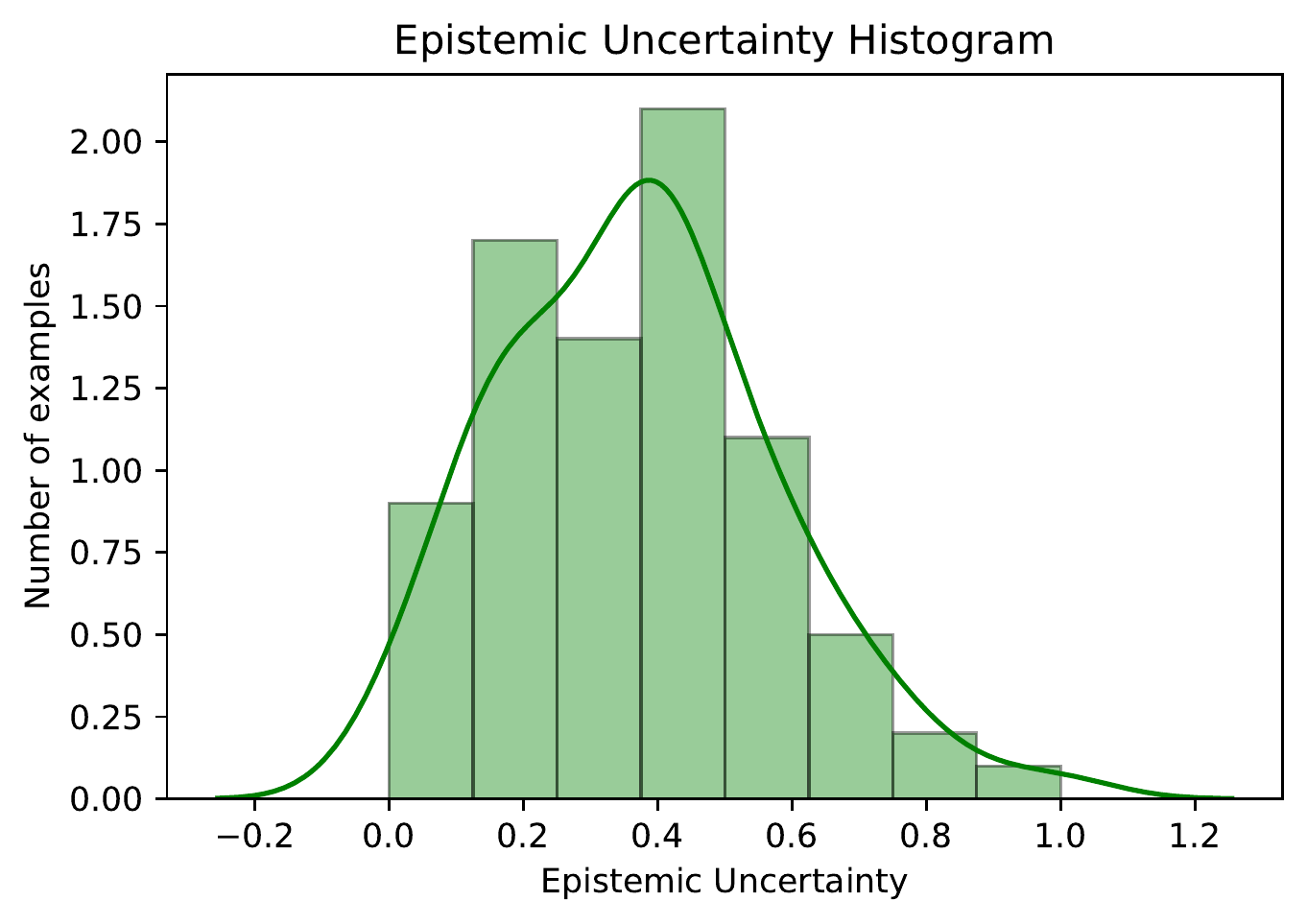}
        \caption{UTKFace}
    \end{subfigure}
	\caption{Aleatoric and epistemic uncertainty comparison of individual datasets with MC-Dropout}
	\label{mcdropout_results}
\end{figure}

\begin{figure}[ht]
    \centering
    \begin{subfigure}[b]{0.49\textwidth}		
        \includegraphics[width=0.49\textwidth]{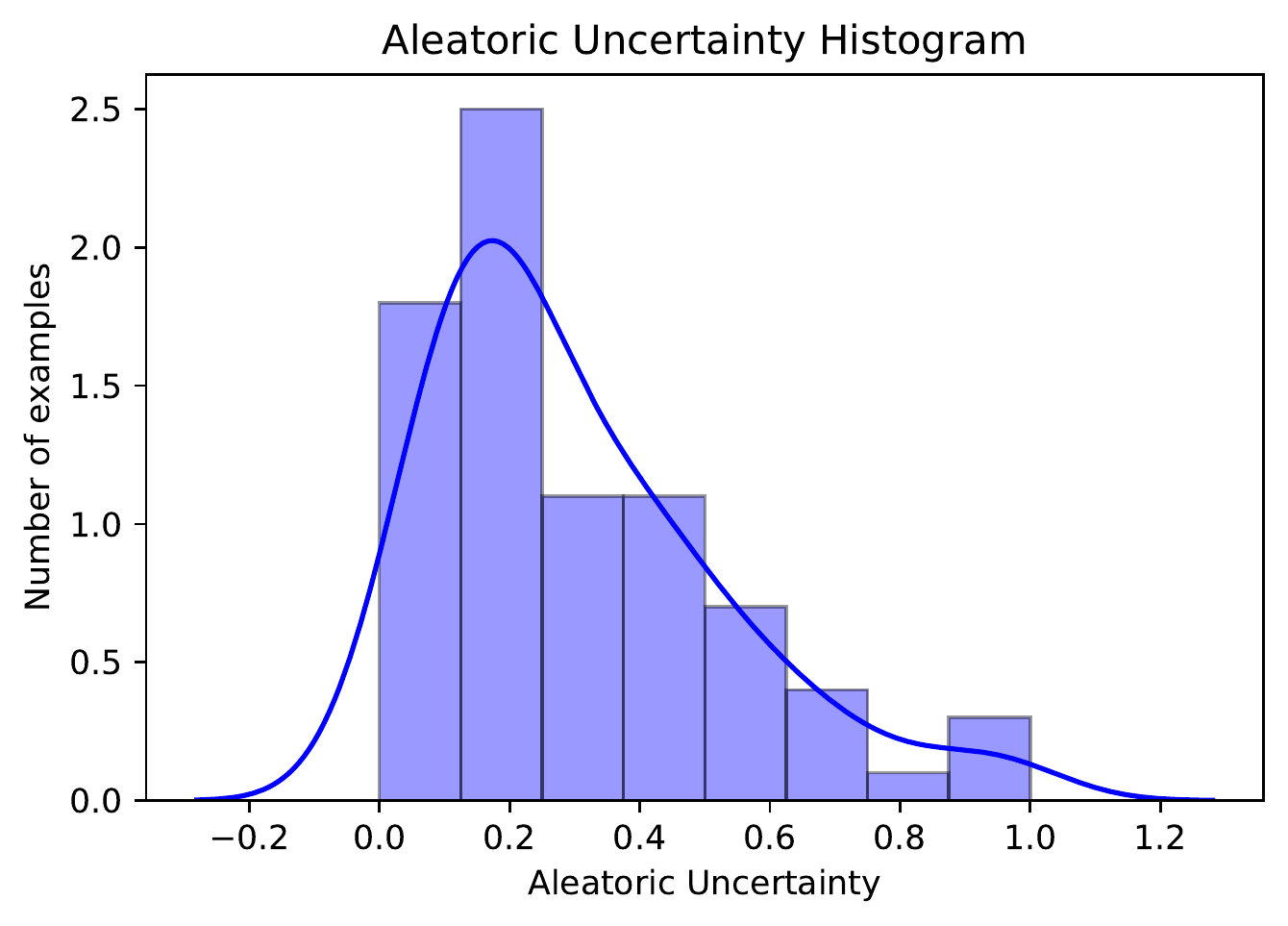}
        \includegraphics[width=0.49\textwidth]{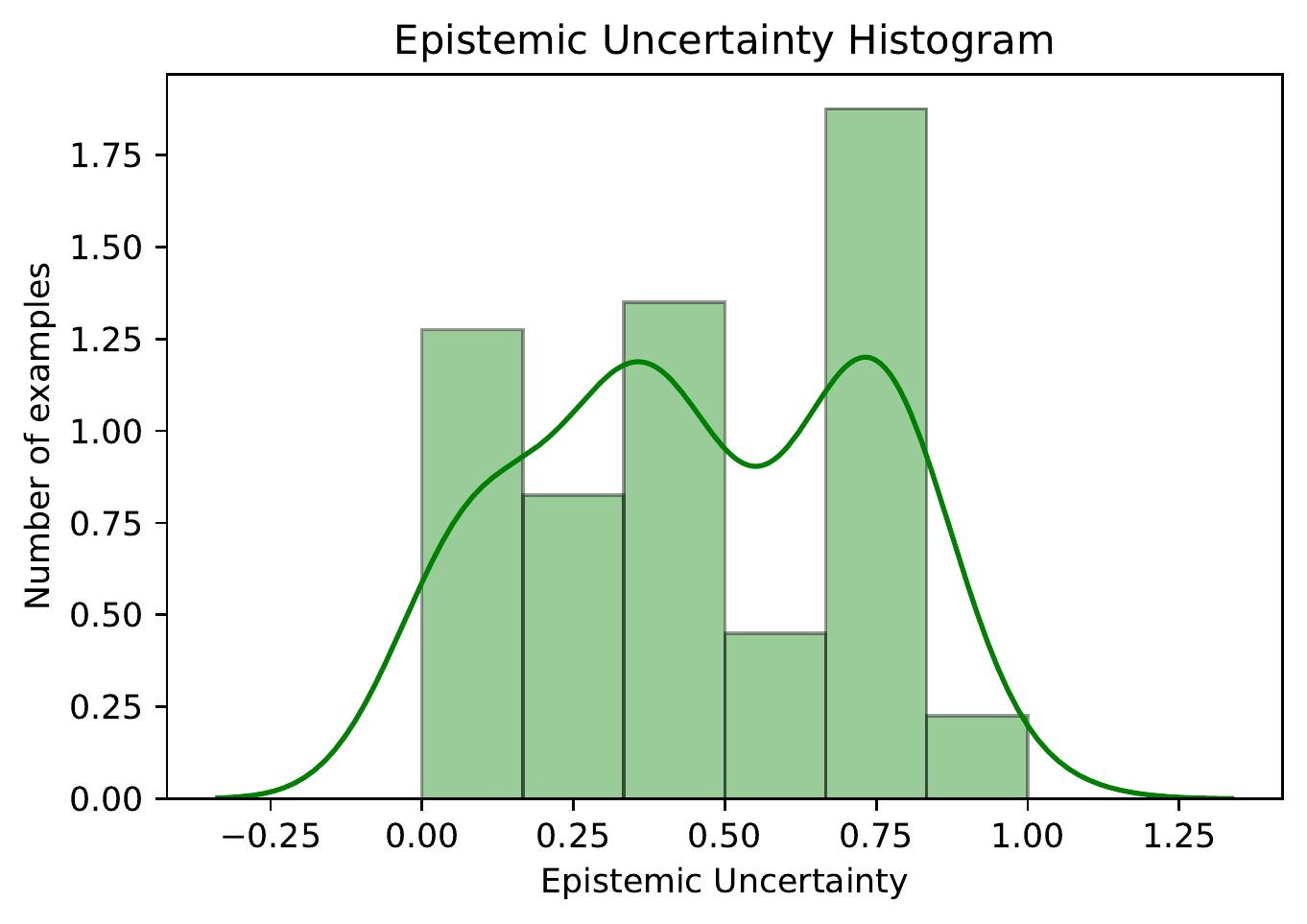}
        \caption{IXI Brain MRI Scans}
    \end{subfigure}
    \begin{subfigure}[b]{0.49\textwidth}		
        \includegraphics[width=0.49\textwidth]{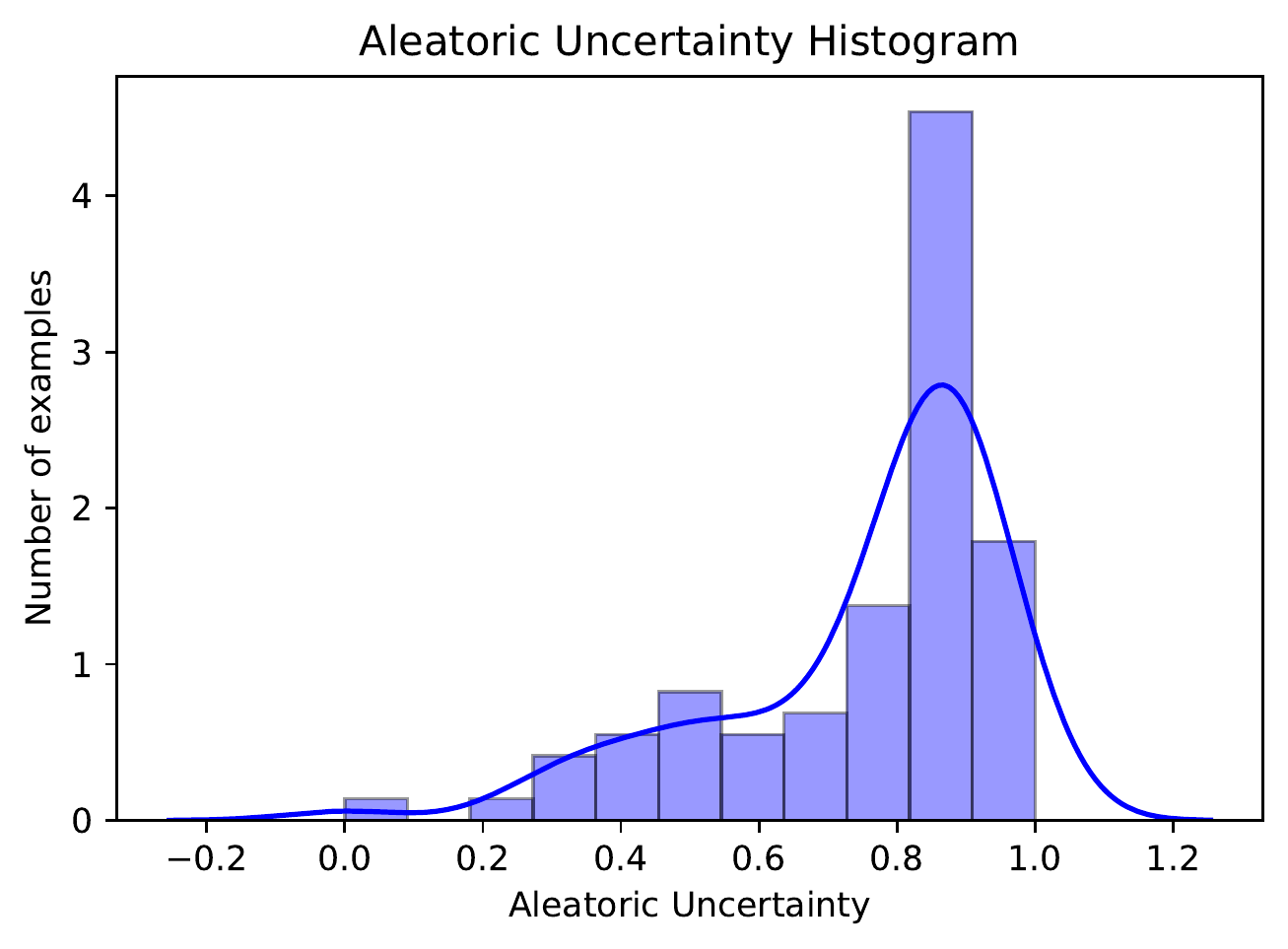}
        \includegraphics[width=0.49\textwidth]{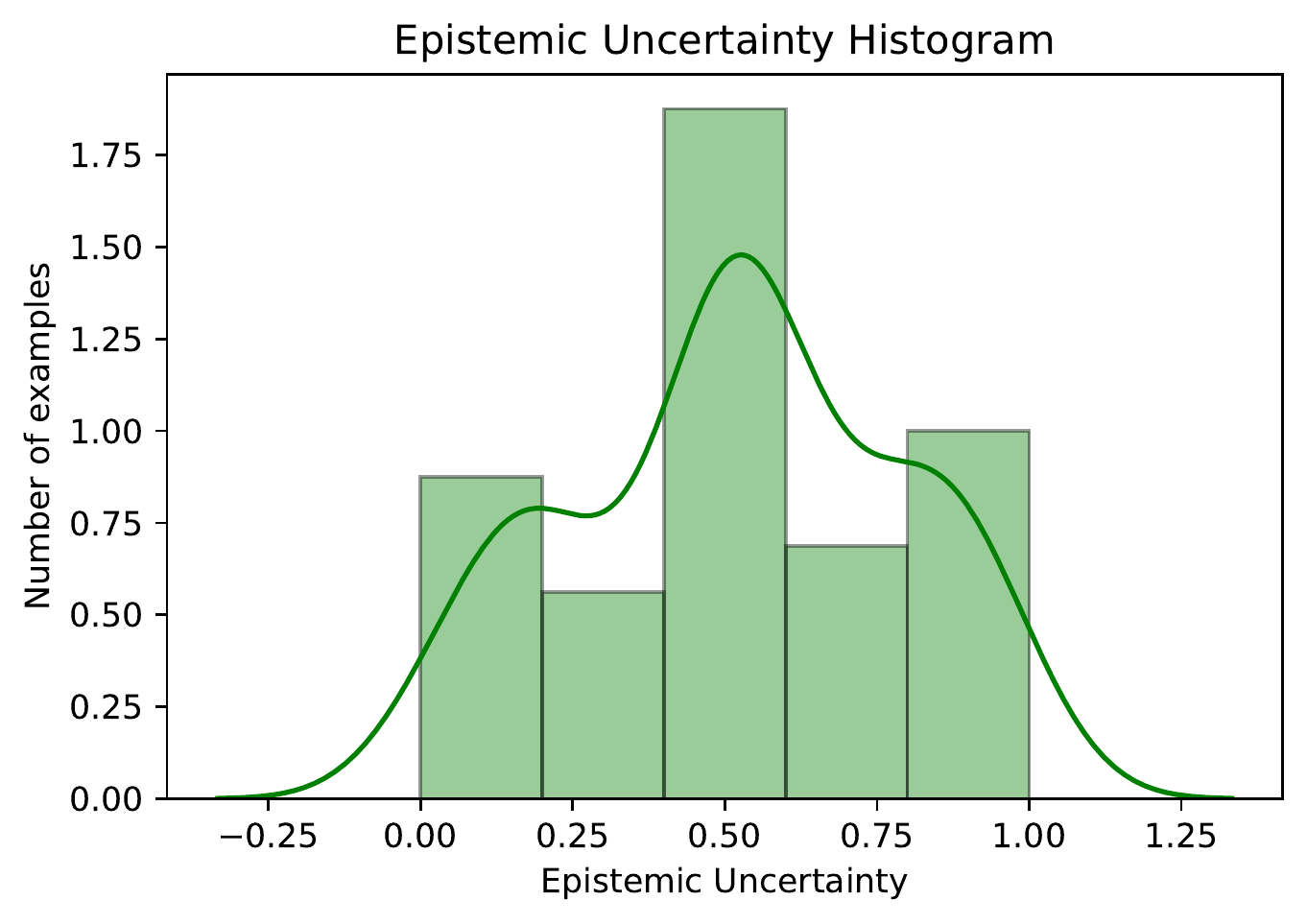}
        \caption{Brain CT Scans}
    \end{subfigure}
    
    \begin{subfigure}[b]{0.49\textwidth}		
        \includegraphics[width=0.49\textwidth]{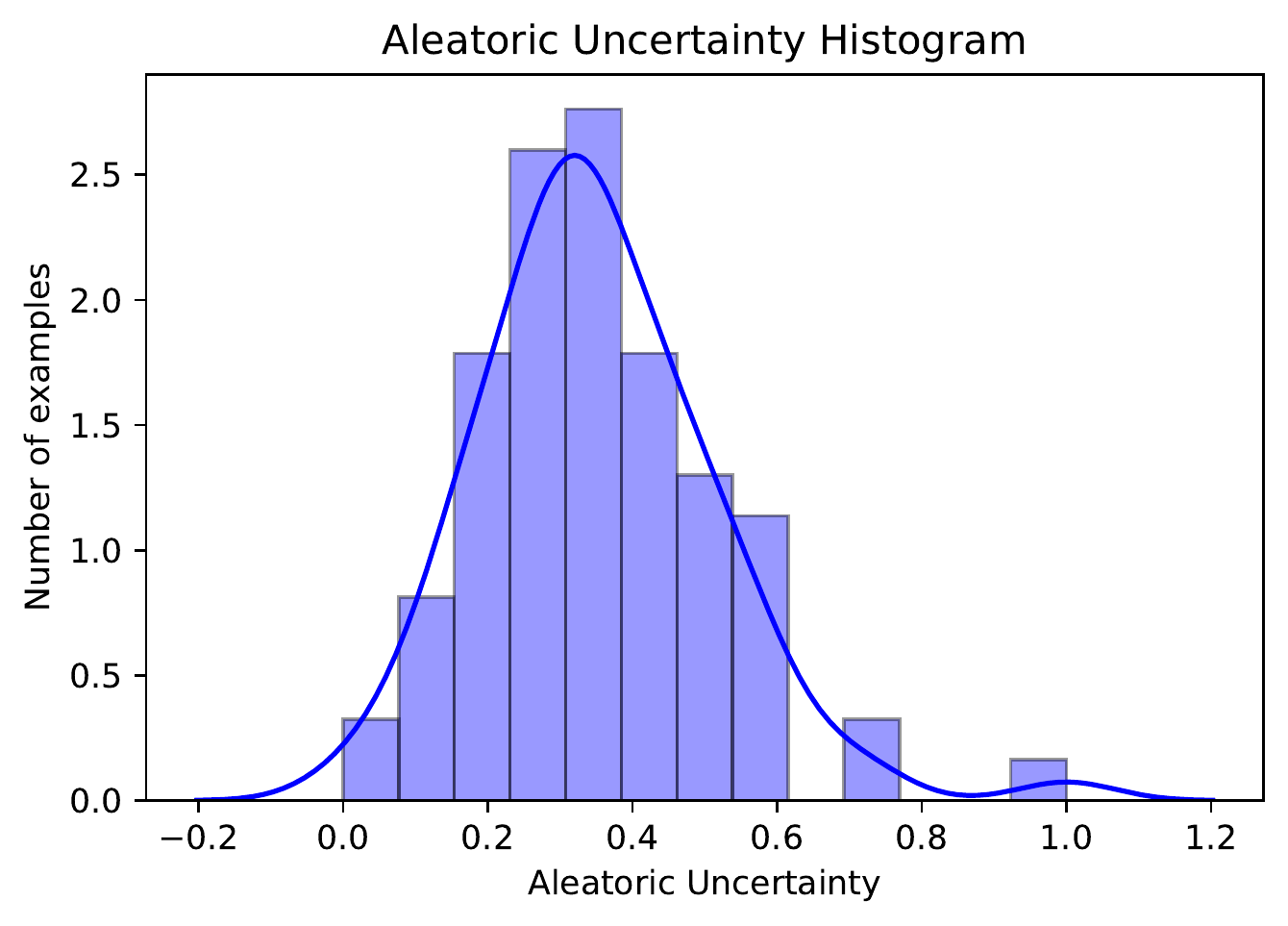}
        \includegraphics[width=0.49\textwidth]{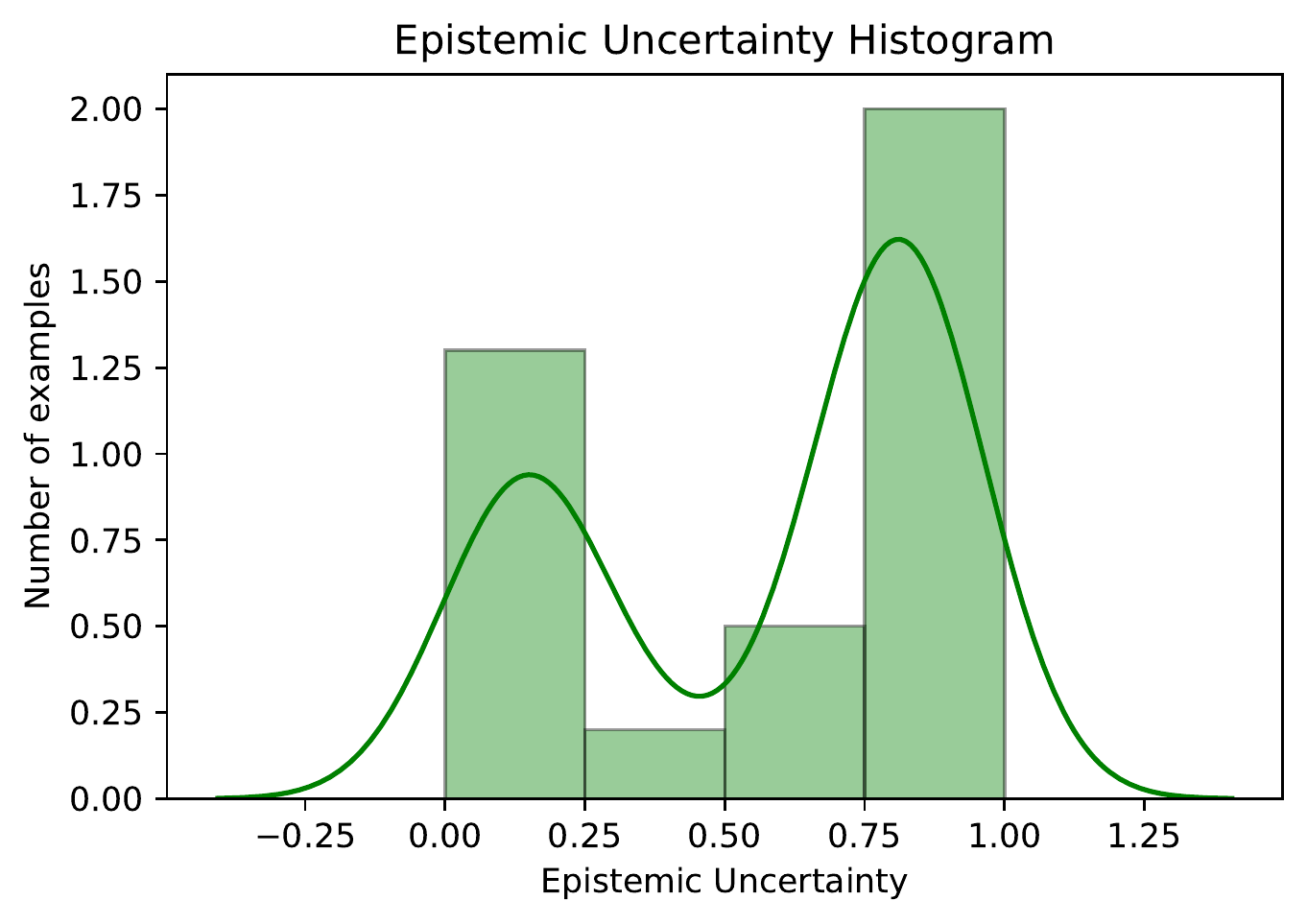}
        \caption{UTKFace}
    \end{subfigure}
    \caption{Aleatoric and epistemic uncertainty comparison of individual datasets with Flipout}
    \label{flipout_results}
\end{figure}

\begin{figure}[!h]
    \centering
    \begin{subfigure}[b]{0.49\textwidth}		
        \includegraphics[width=0.49\textwidth]{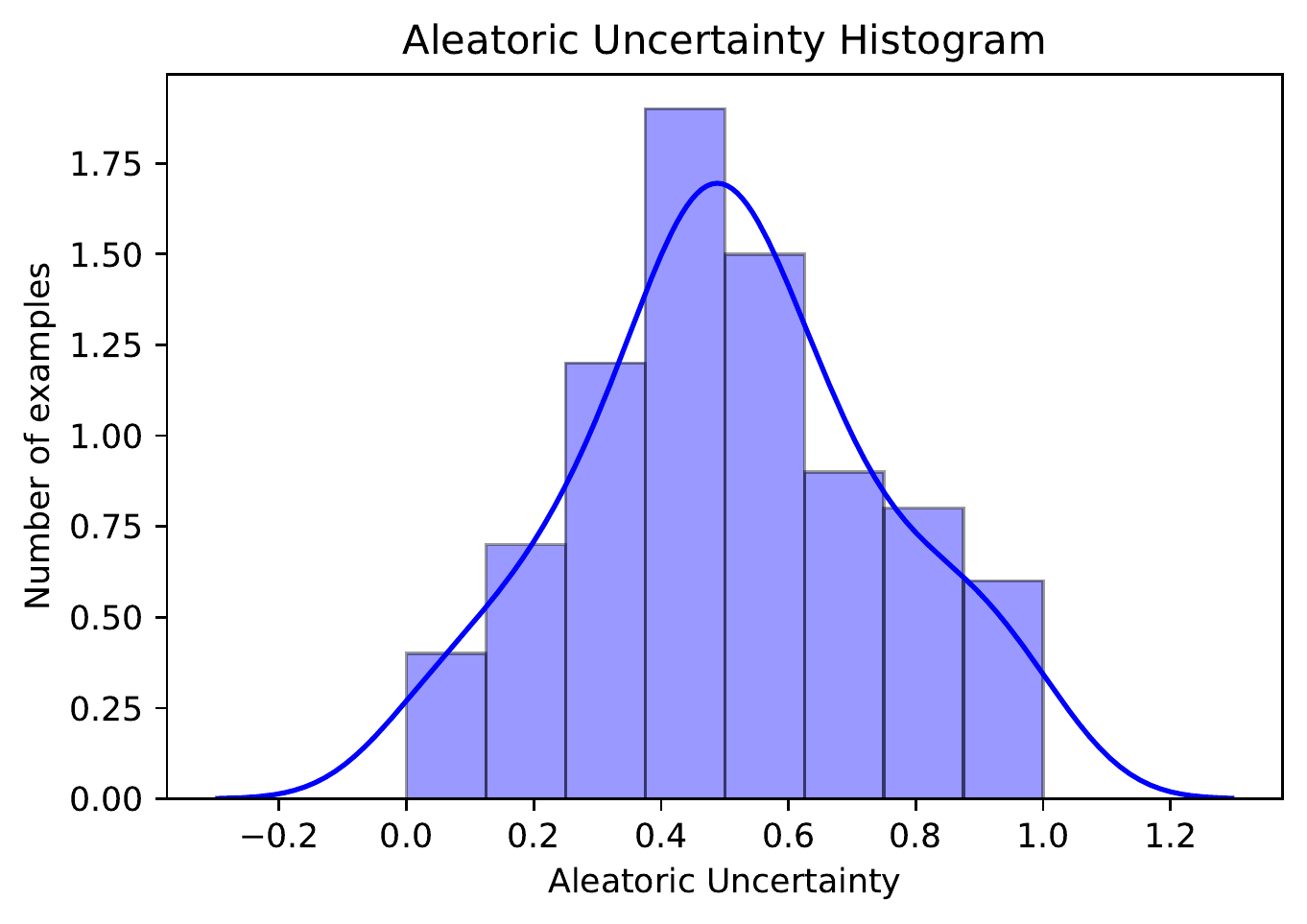}
        \includegraphics[width=0.49\textwidth]{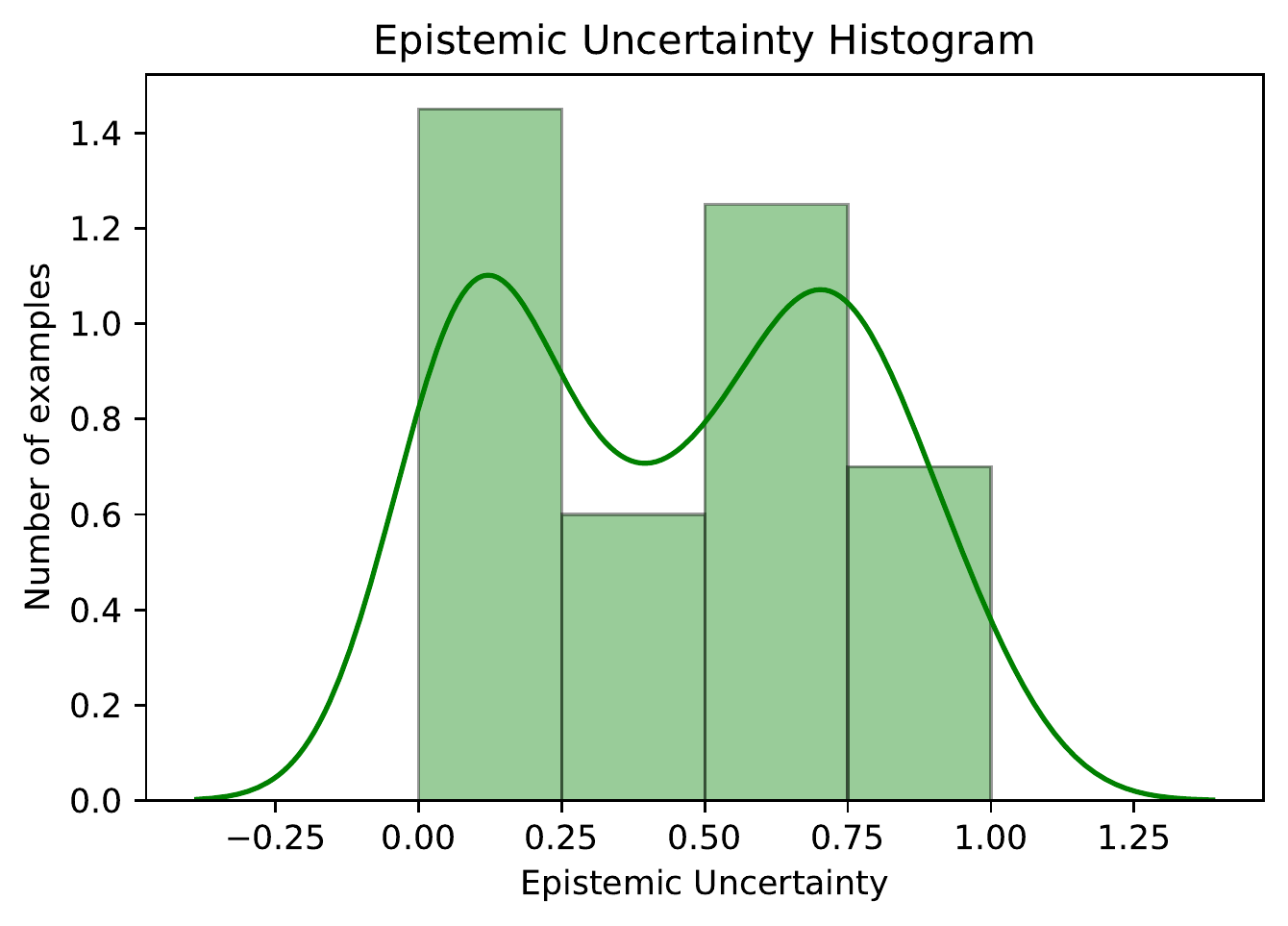}
        \caption{IXI Brain MRI Scans}
    \end{subfigure}
    \begin{subfigure}[b]{0.49\textwidth}		
        \includegraphics[width=0.49\textwidth]{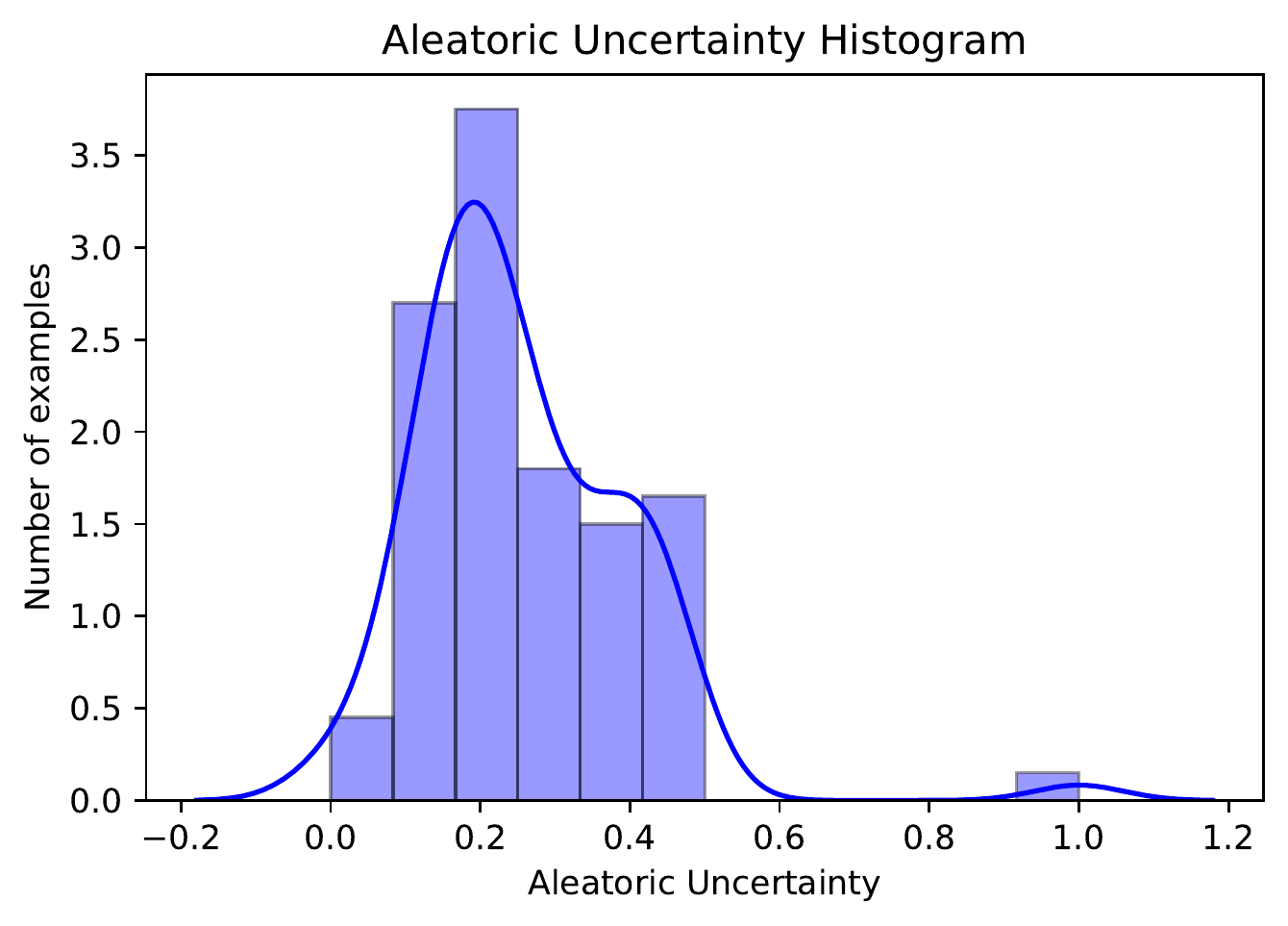}
        \includegraphics[width=0.49\textwidth]{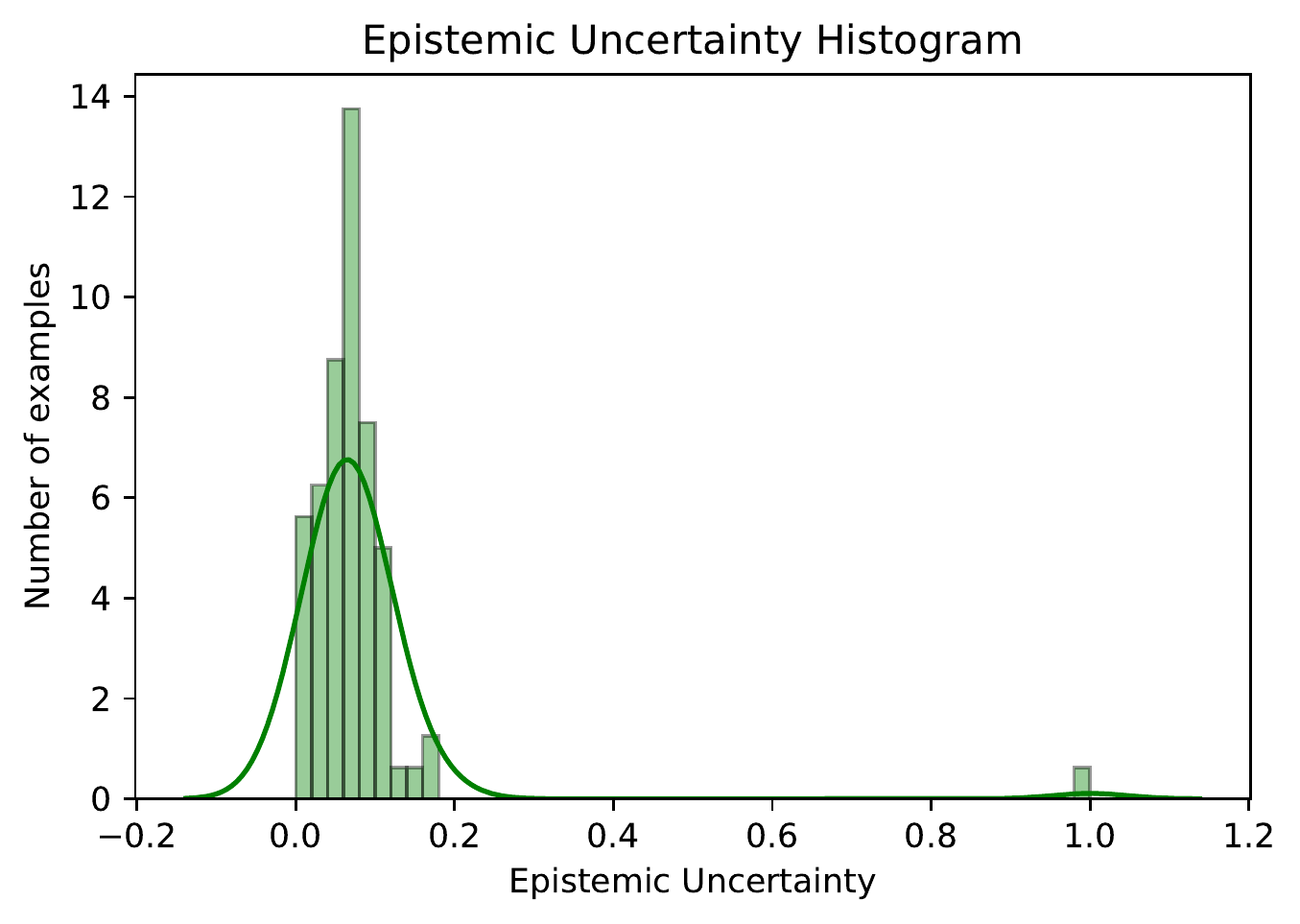}
        \caption{Brain CT Scans}
    \end{subfigure}
    
    \begin{subfigure}[b]{0.49\textwidth}		
        \includegraphics[width=0.49\textwidth]{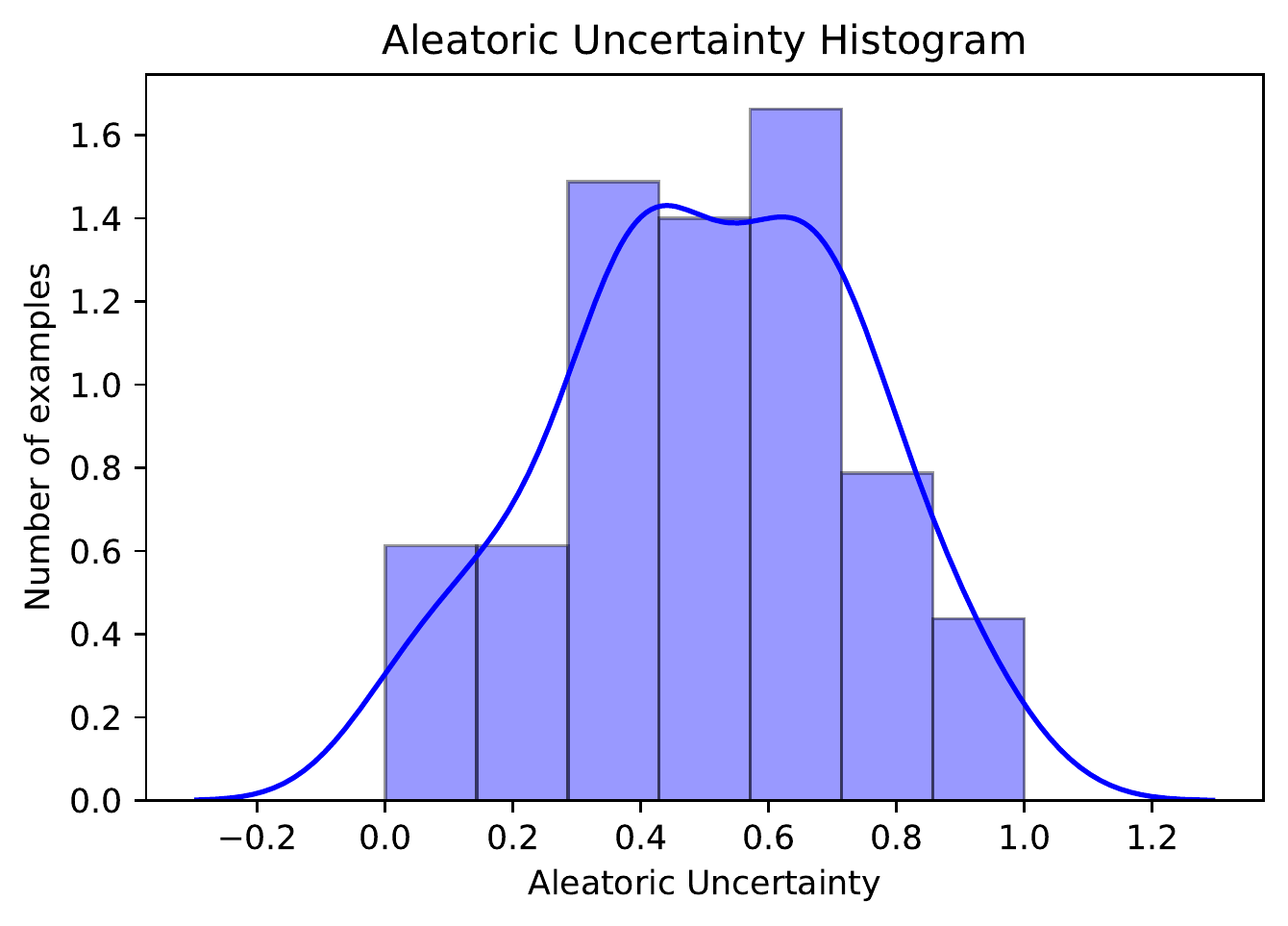}
        \includegraphics[width=0.49\textwidth]{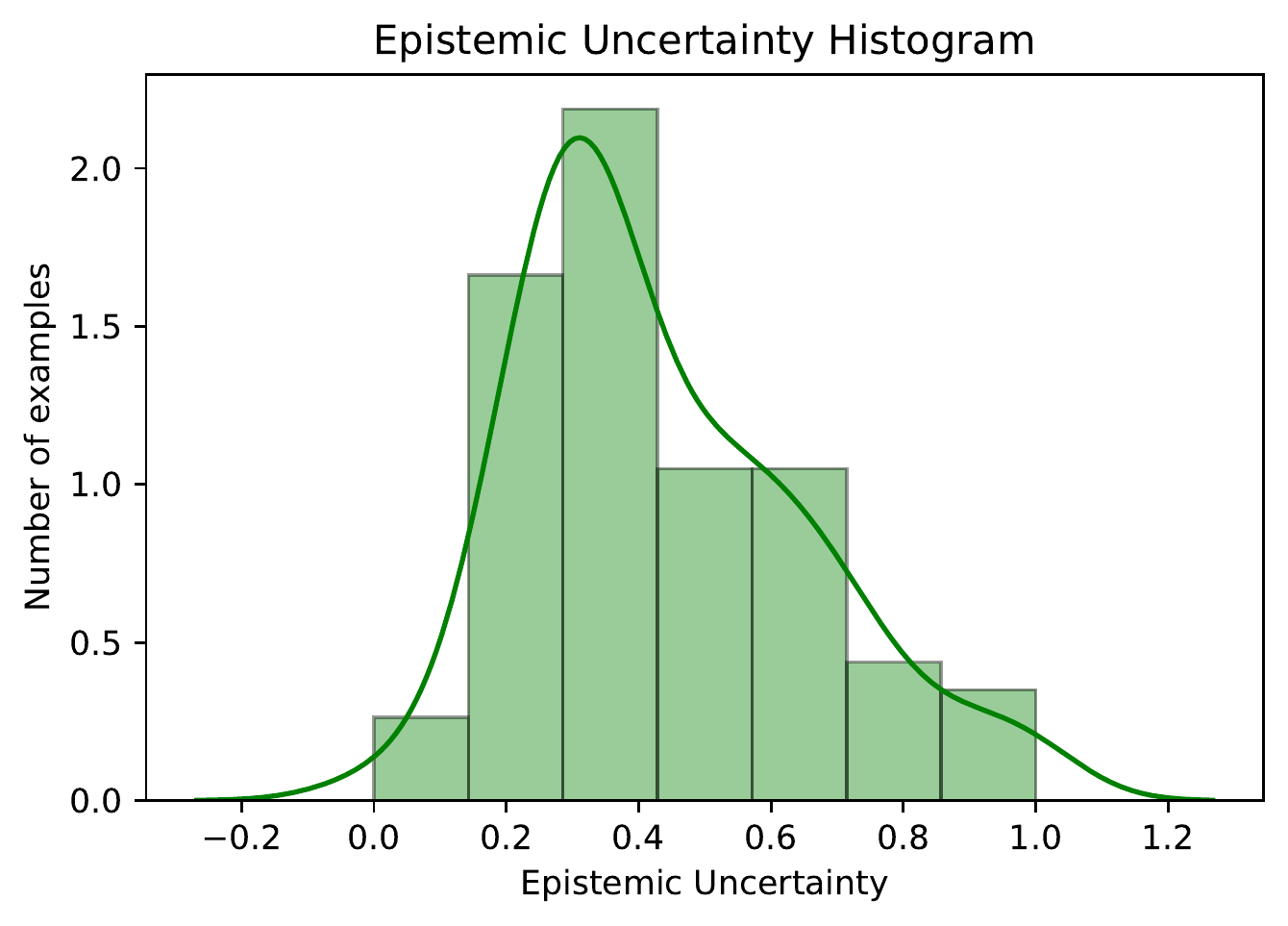}
        \caption{UTKFace}
    \end{subfigure}
    \caption{Aleatoric and epistemic uncertainty comparison of individual datasets with Ensembles}
    \label{ensemble_results}
\end{figure}

\begin{figure}[h]
    \centering
    \begin{subfigure}[b]{0.49\textwidth}		
        \includegraphics[width=0.49\textwidth]{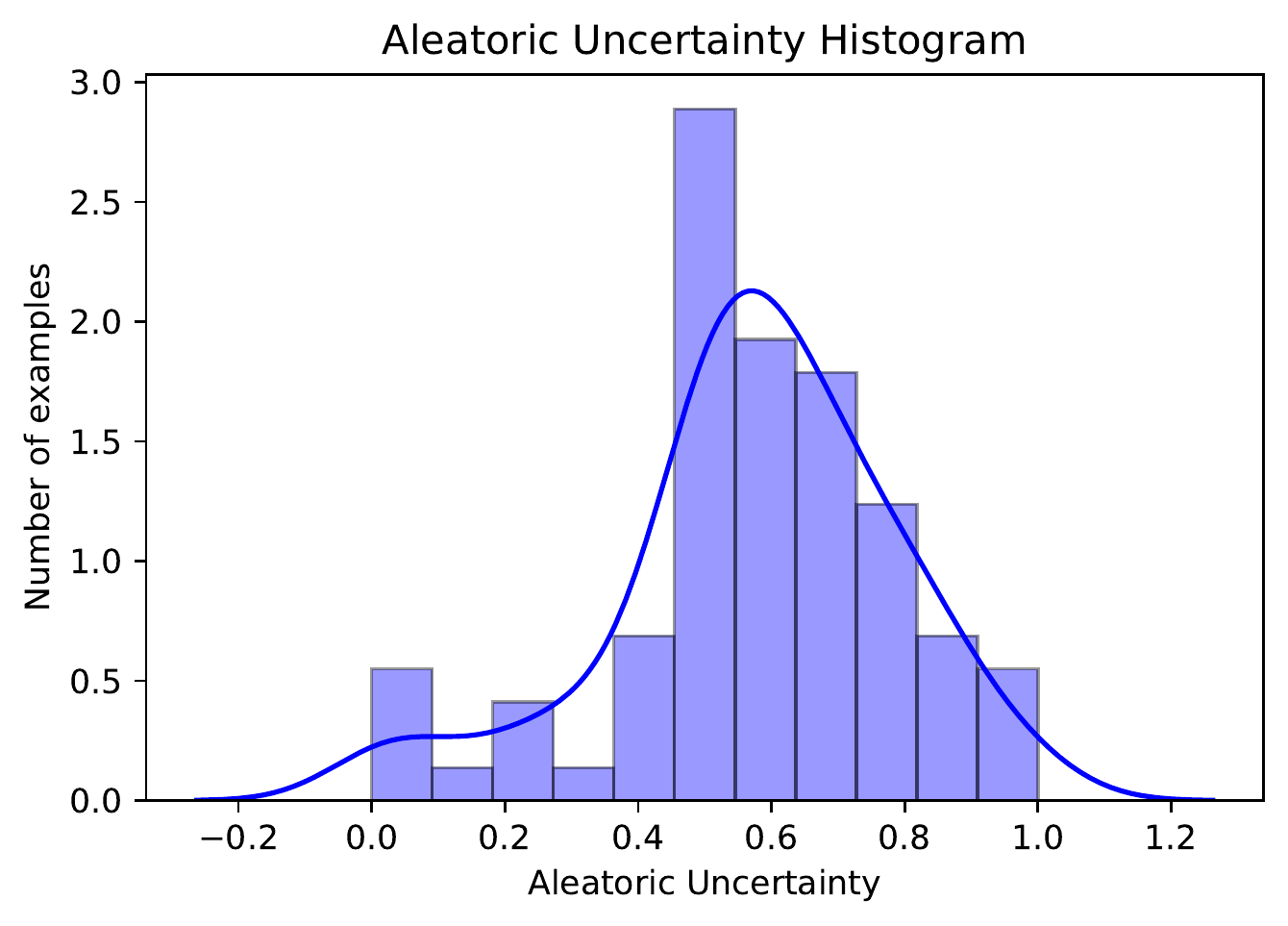}
        \includegraphics[width=0.49\textwidth]{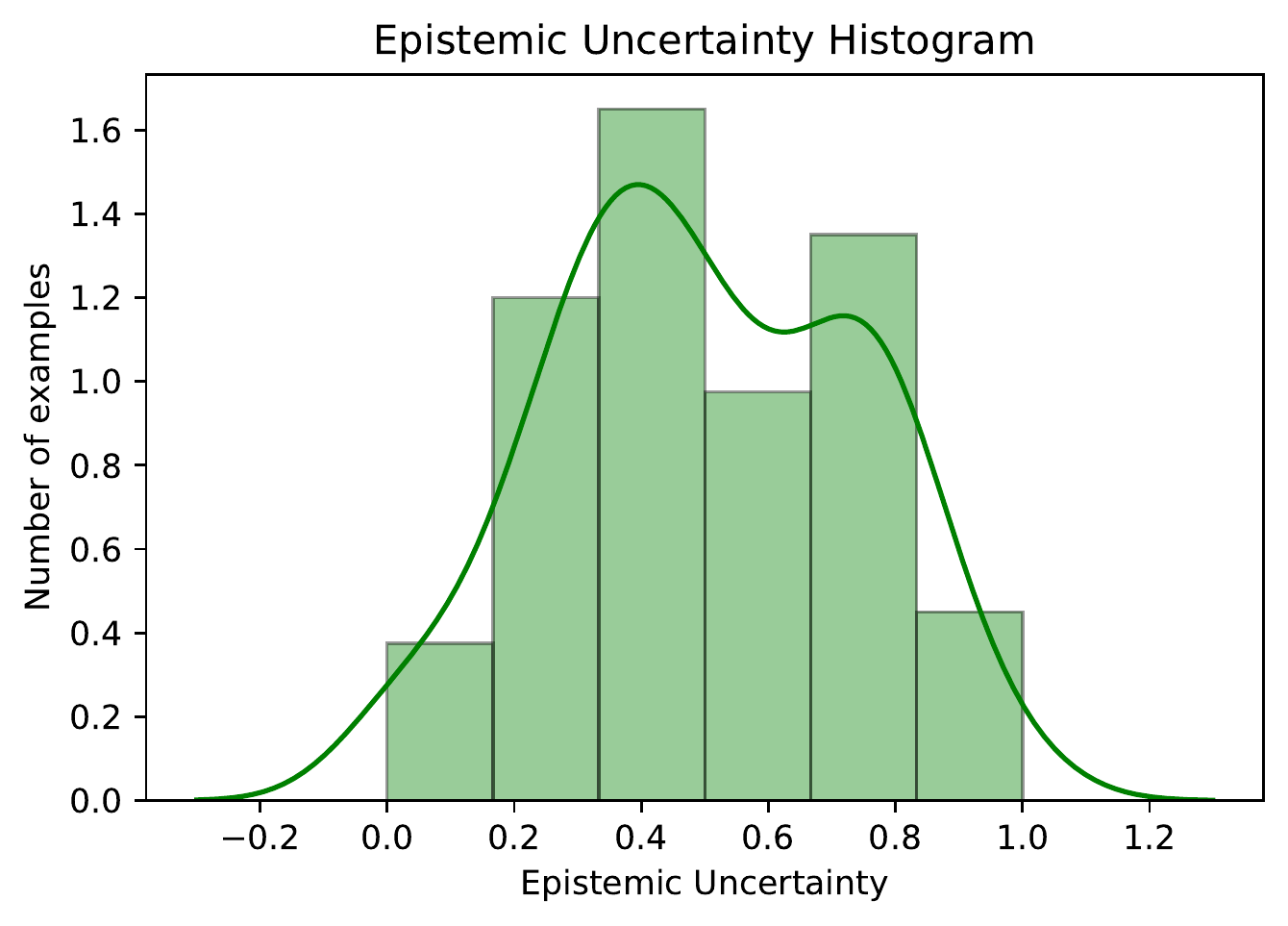}
        \caption{IXI Brain MRI Scans}
    \end{subfigure}
    \begin{subfigure}[b]{0.49\textwidth}		
        \includegraphics[width=0.49\textwidth]{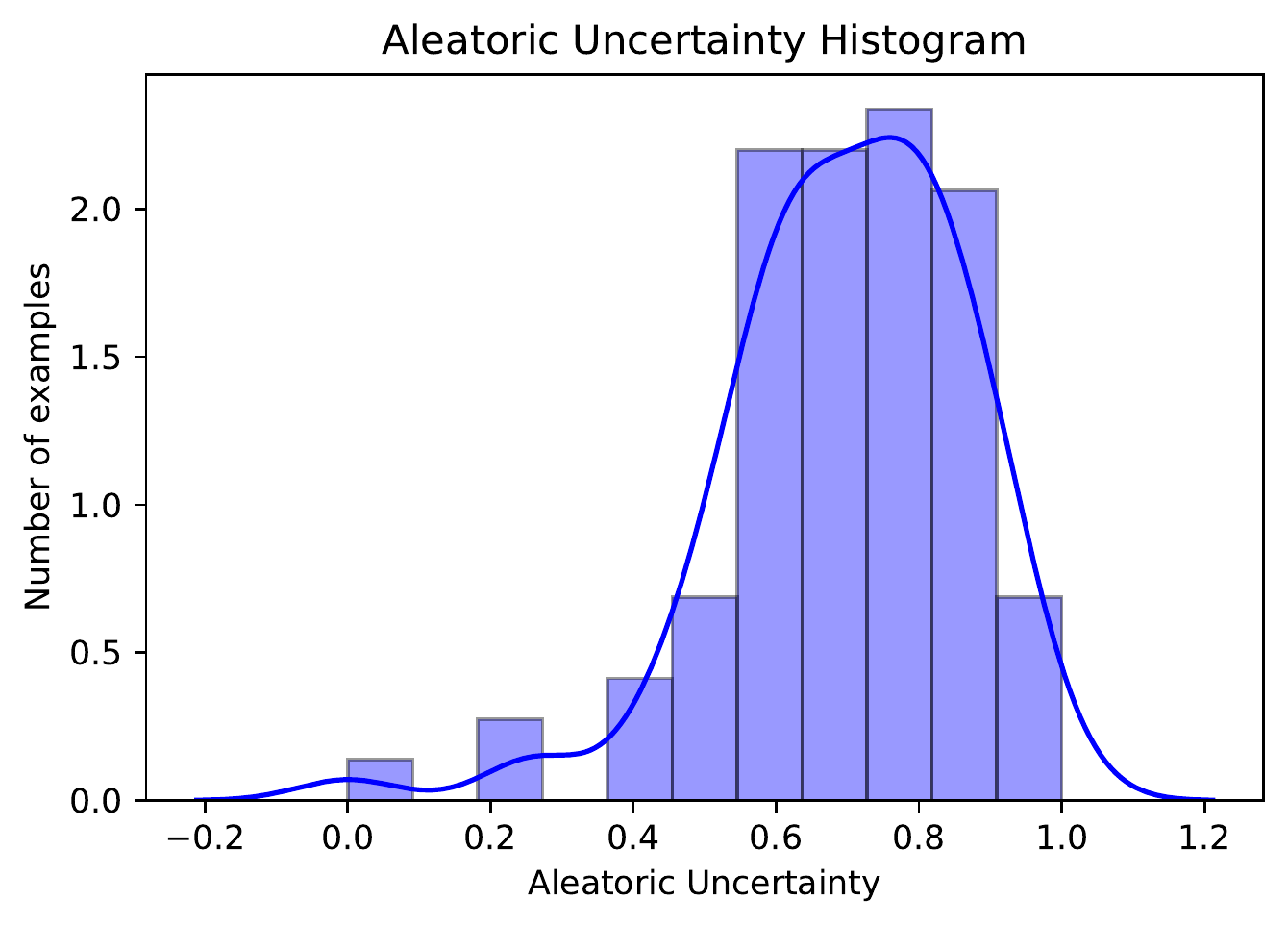}
        \includegraphics[width=0.49\textwidth]{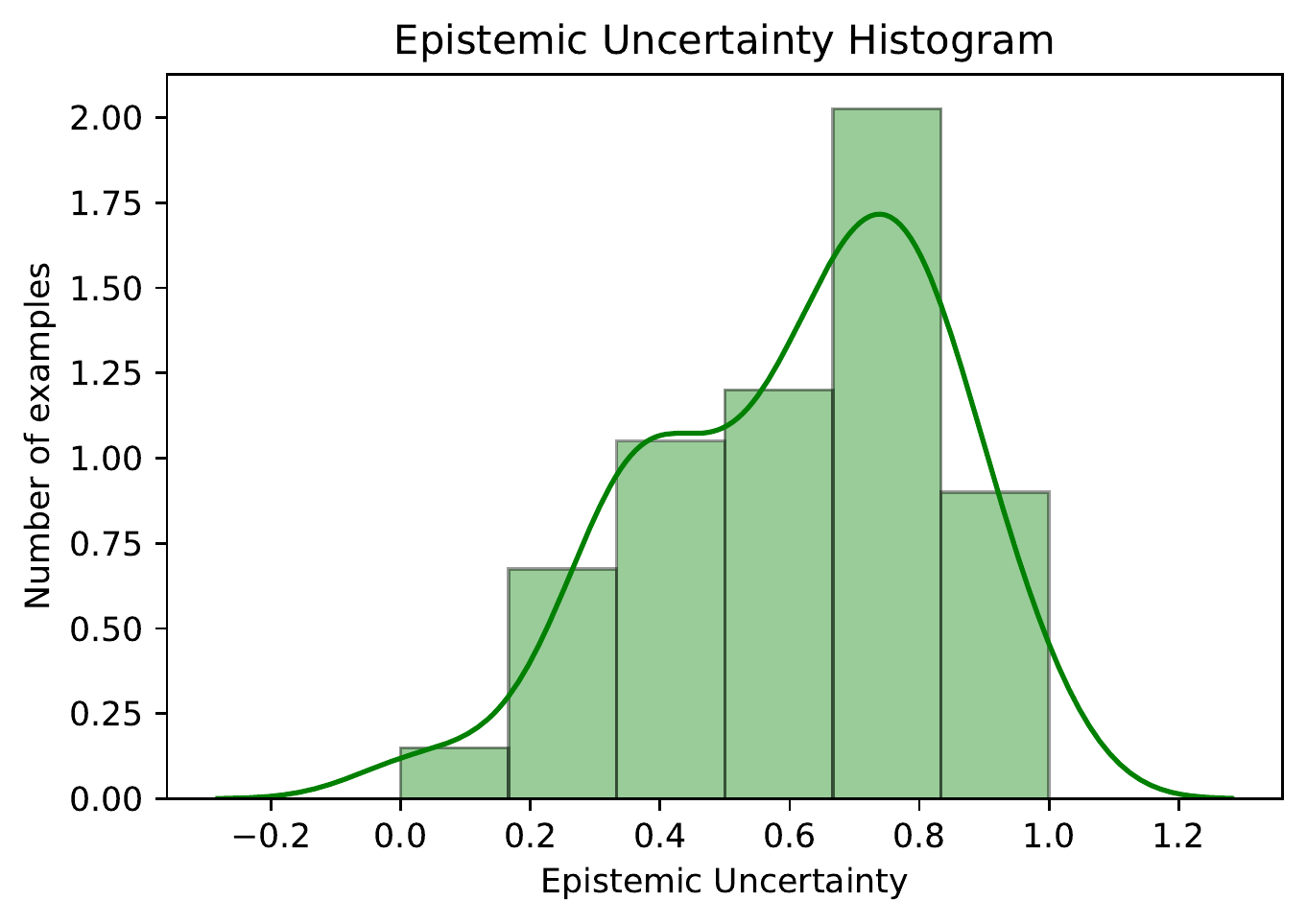}
        \caption{Brain CT Scans}
    \end{subfigure}
    
    \begin{subfigure}[b]{0.49\textwidth}		
        \includegraphics[width=0.49\textwidth]{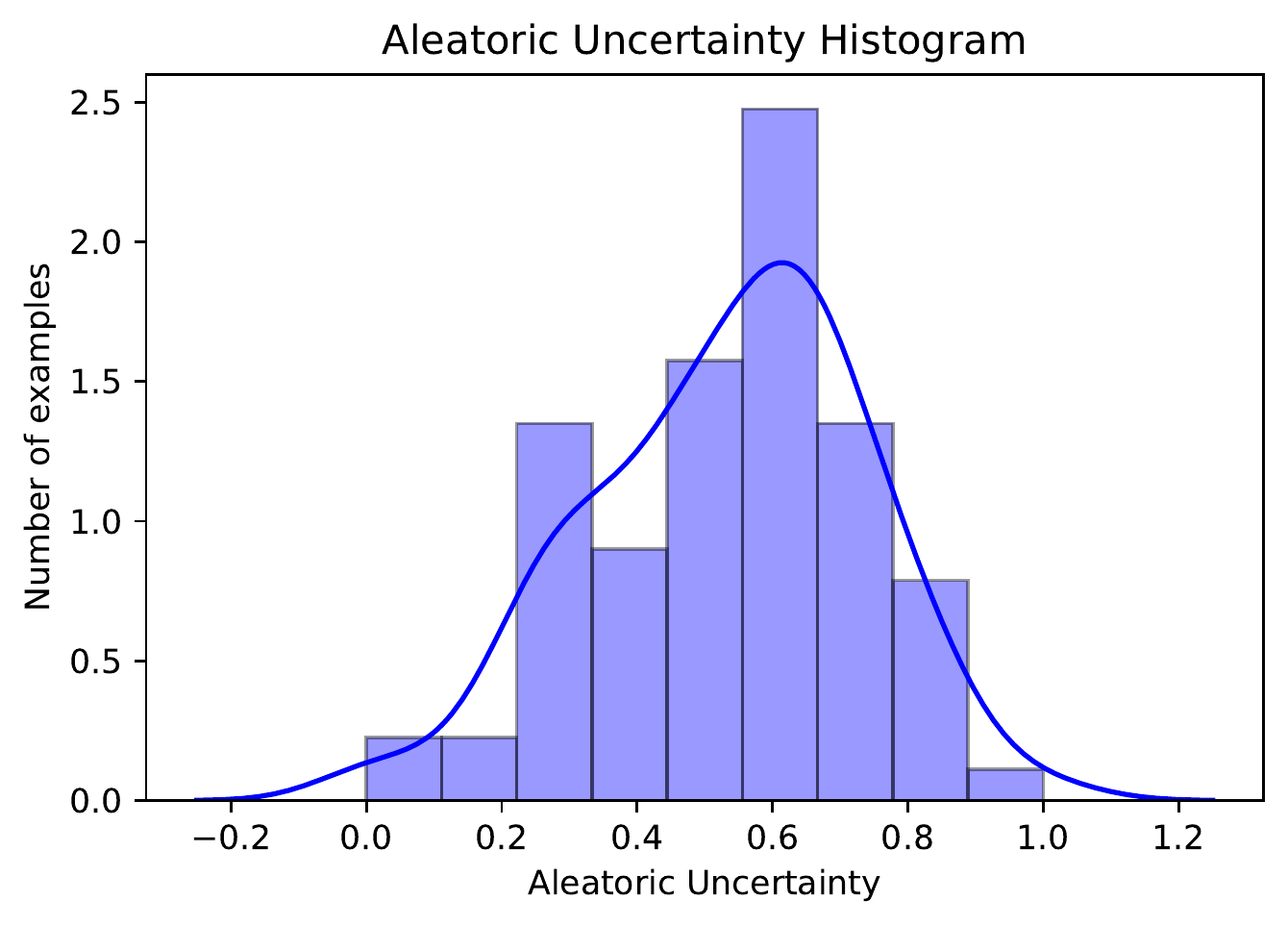}
        \includegraphics[width=0.49\textwidth]{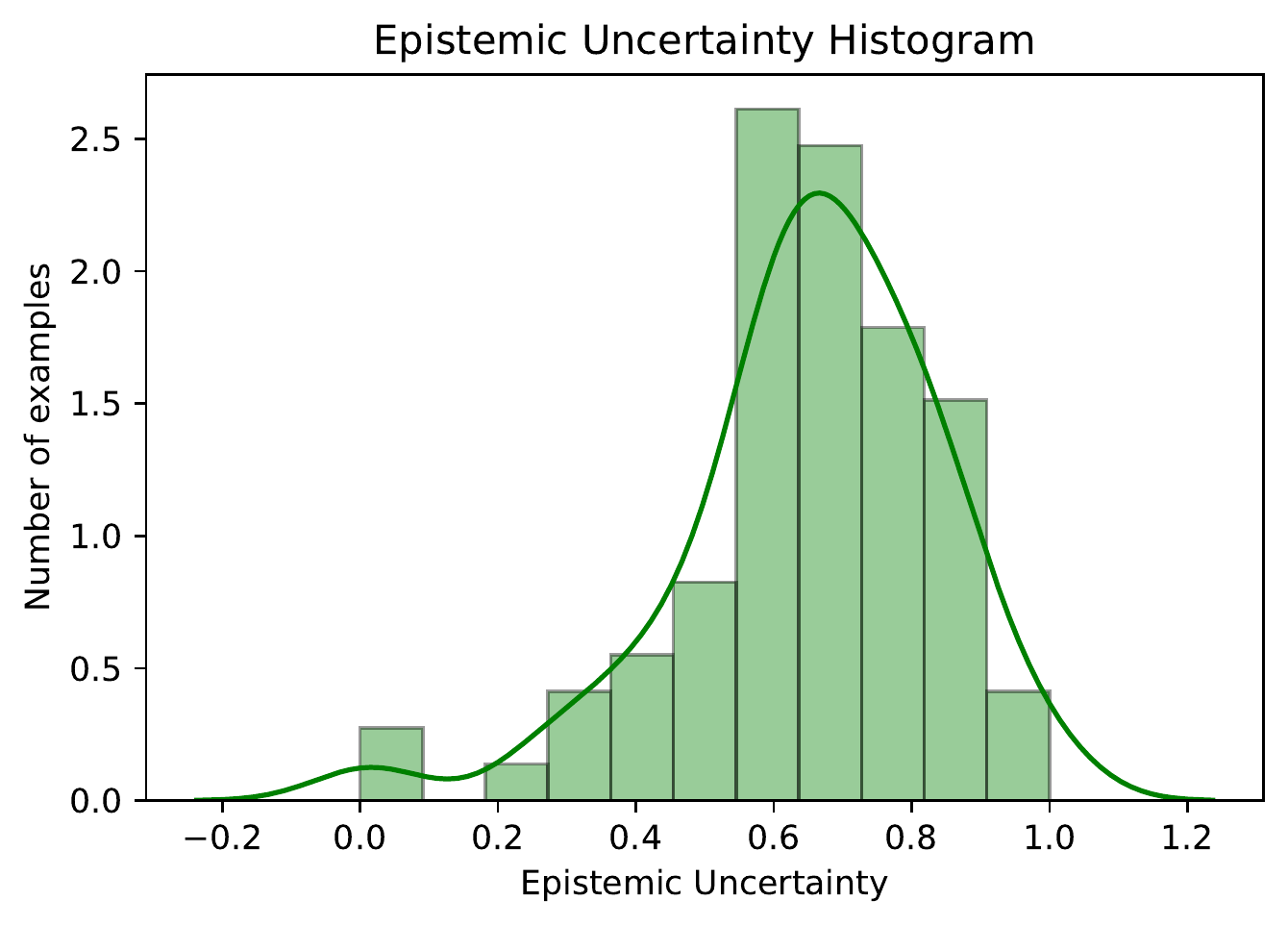}
        \caption{UTKFace}
    \end{subfigure}
    \caption{Aleatoric and epistemic uncertainty comparison of individual datasets with MC-DropConnect}
    \label{dropconnect_results}
\end{figure}

\begin{figure}[h]
    \centering
    \begin{subfigure}[b]{0.49\textwidth}		
        \includegraphics[width=0.49\textwidth]{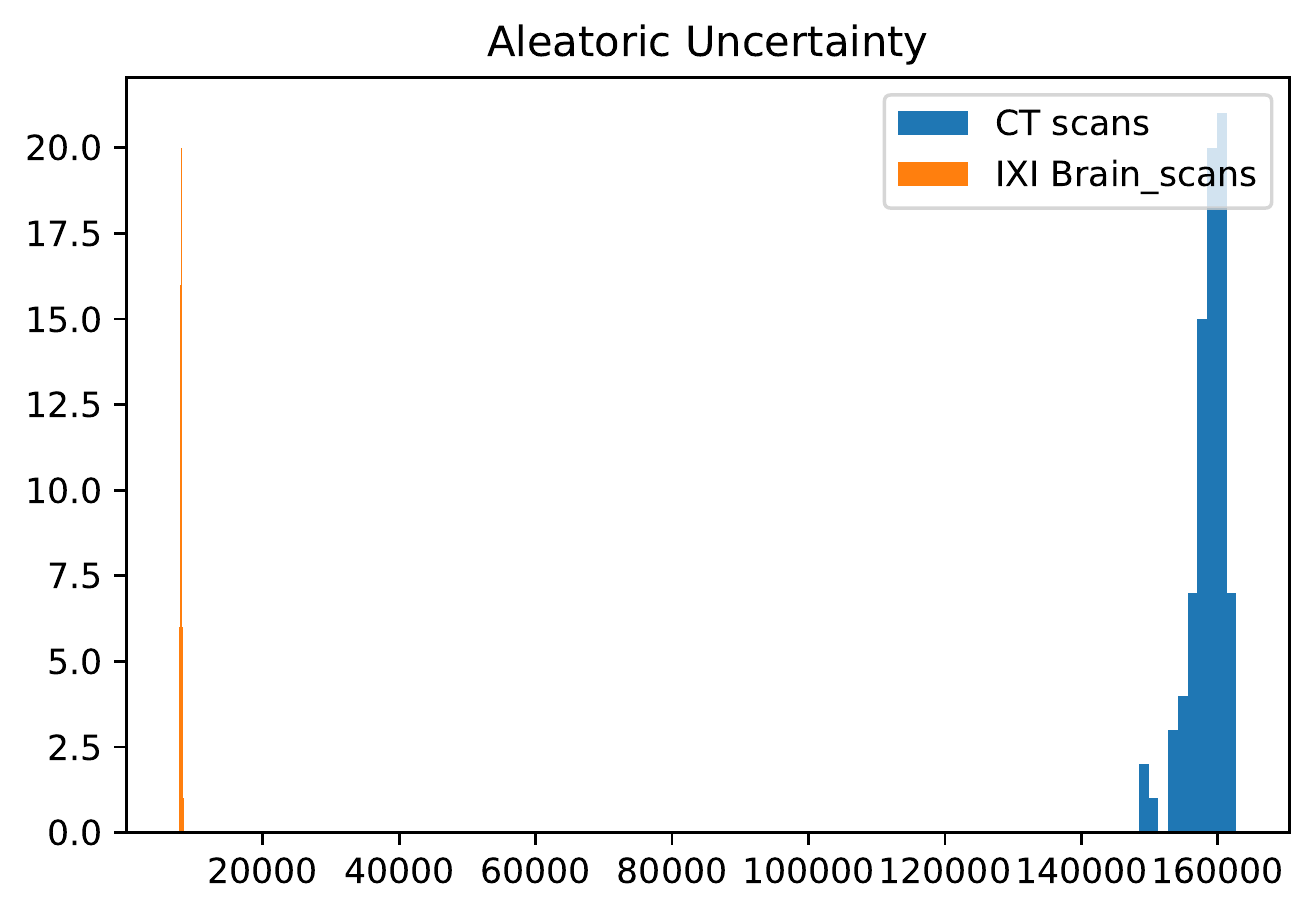}
        \includegraphics[width=0.49\textwidth]{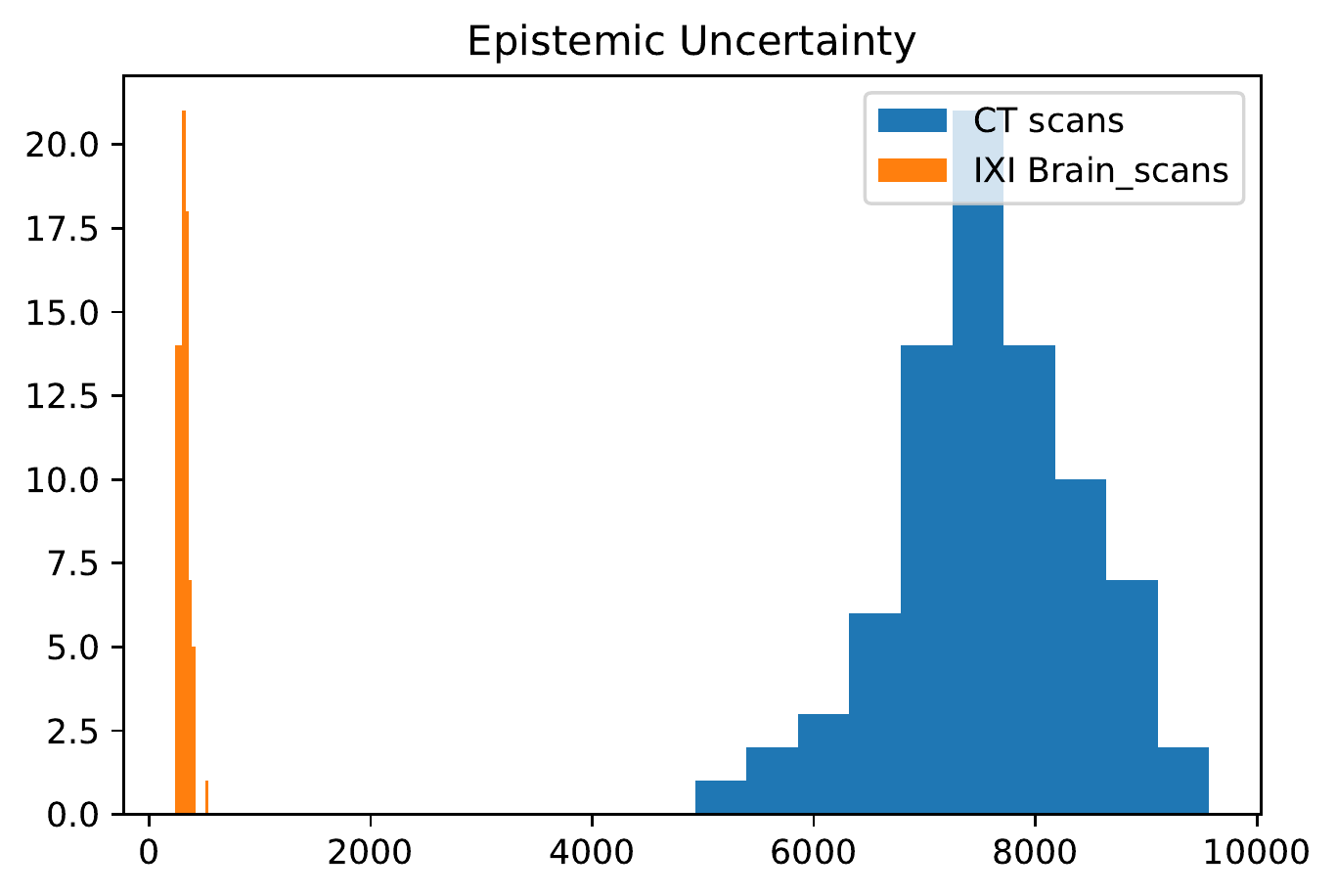}
        \caption{MC-Dropout}
    \end{subfigure}
    \begin{subfigure}[b]{0.49\textwidth}		
        \includegraphics[width=0.49\textwidth]{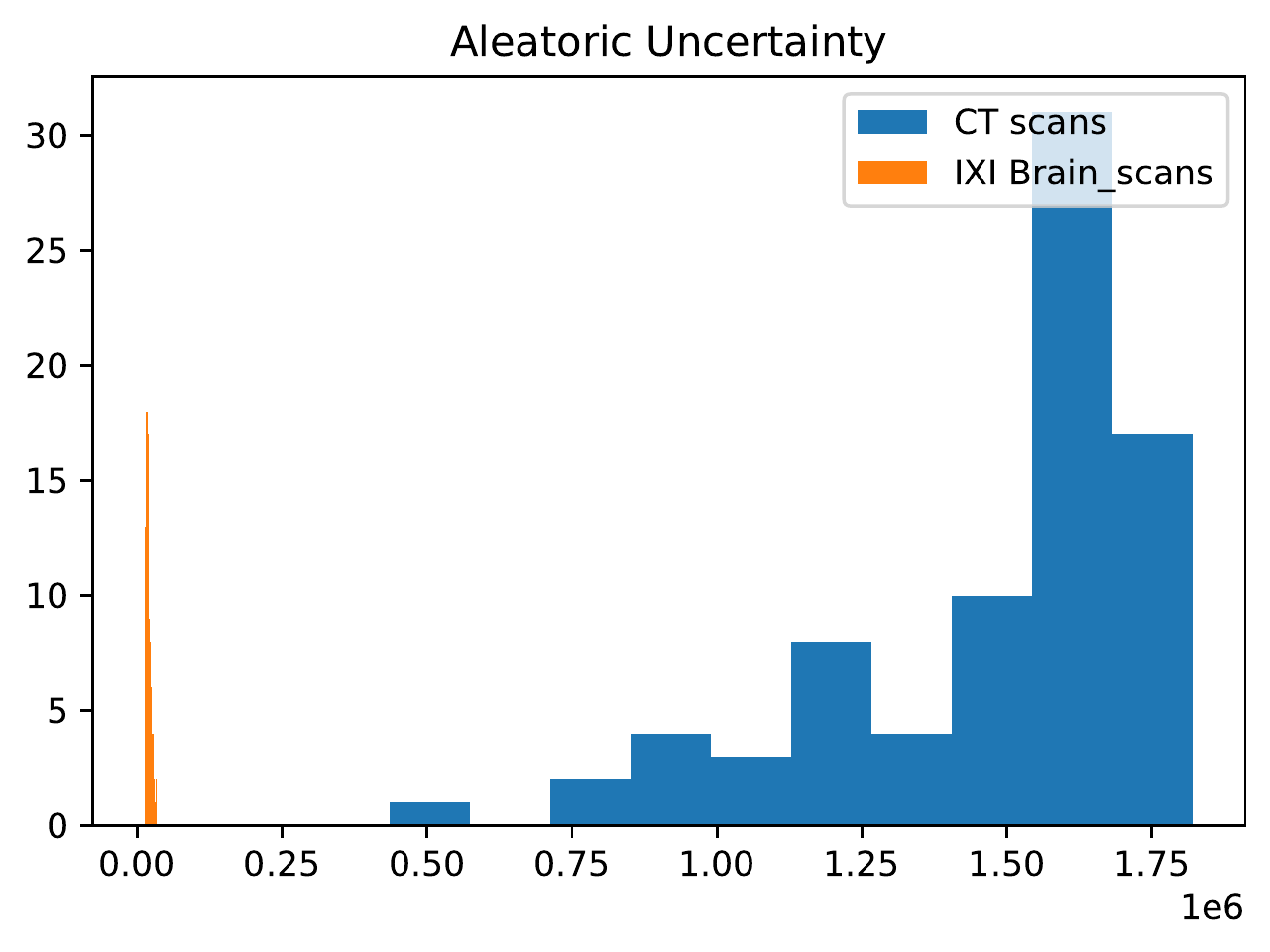}
        \includegraphics[width=0.49\textwidth]{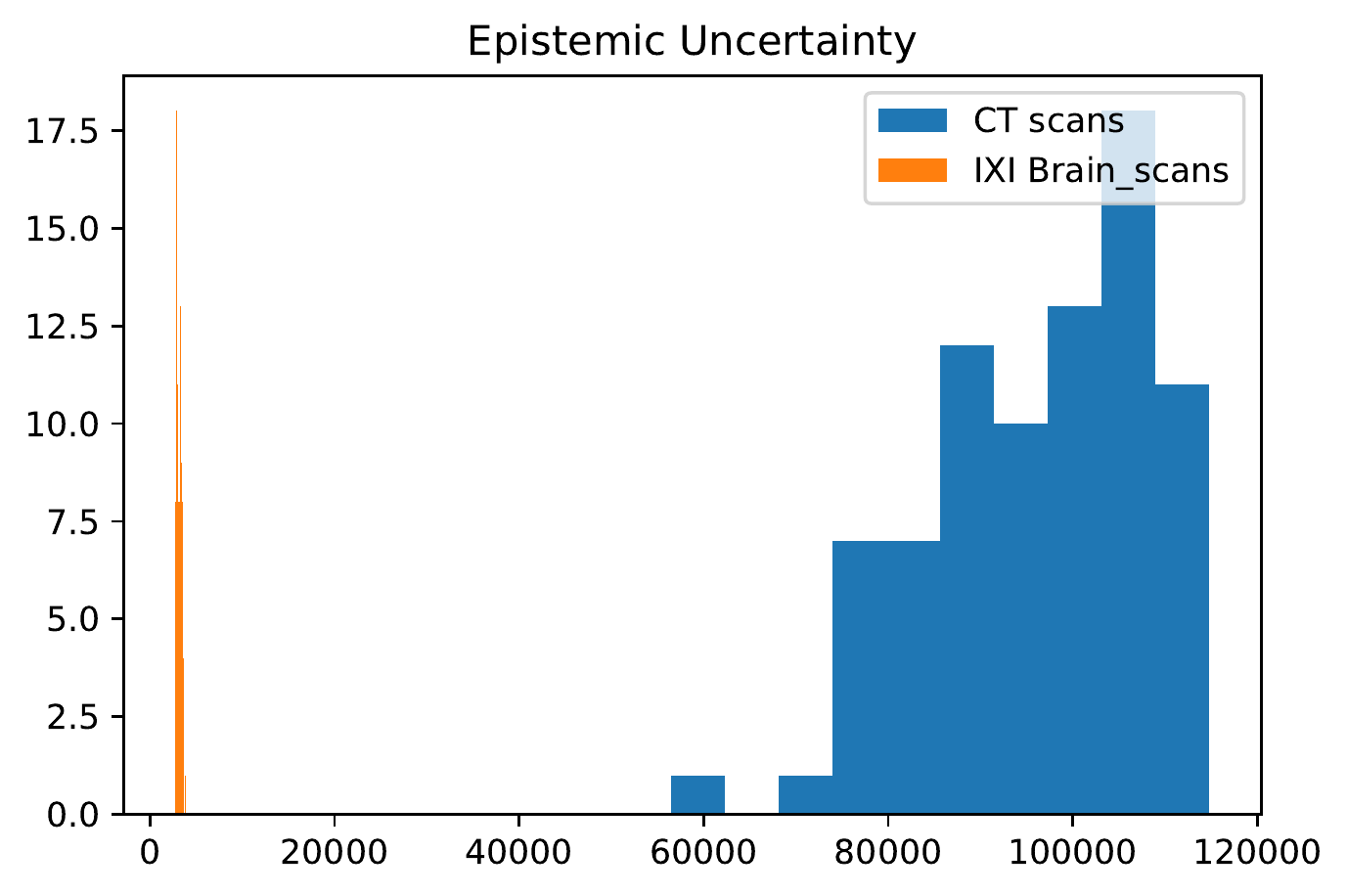}
        \caption{Flipout}
    \end{subfigure}

    \begin{subfigure}[b]{0.49\textwidth}		
        \includegraphics[width=0.49\textwidth]{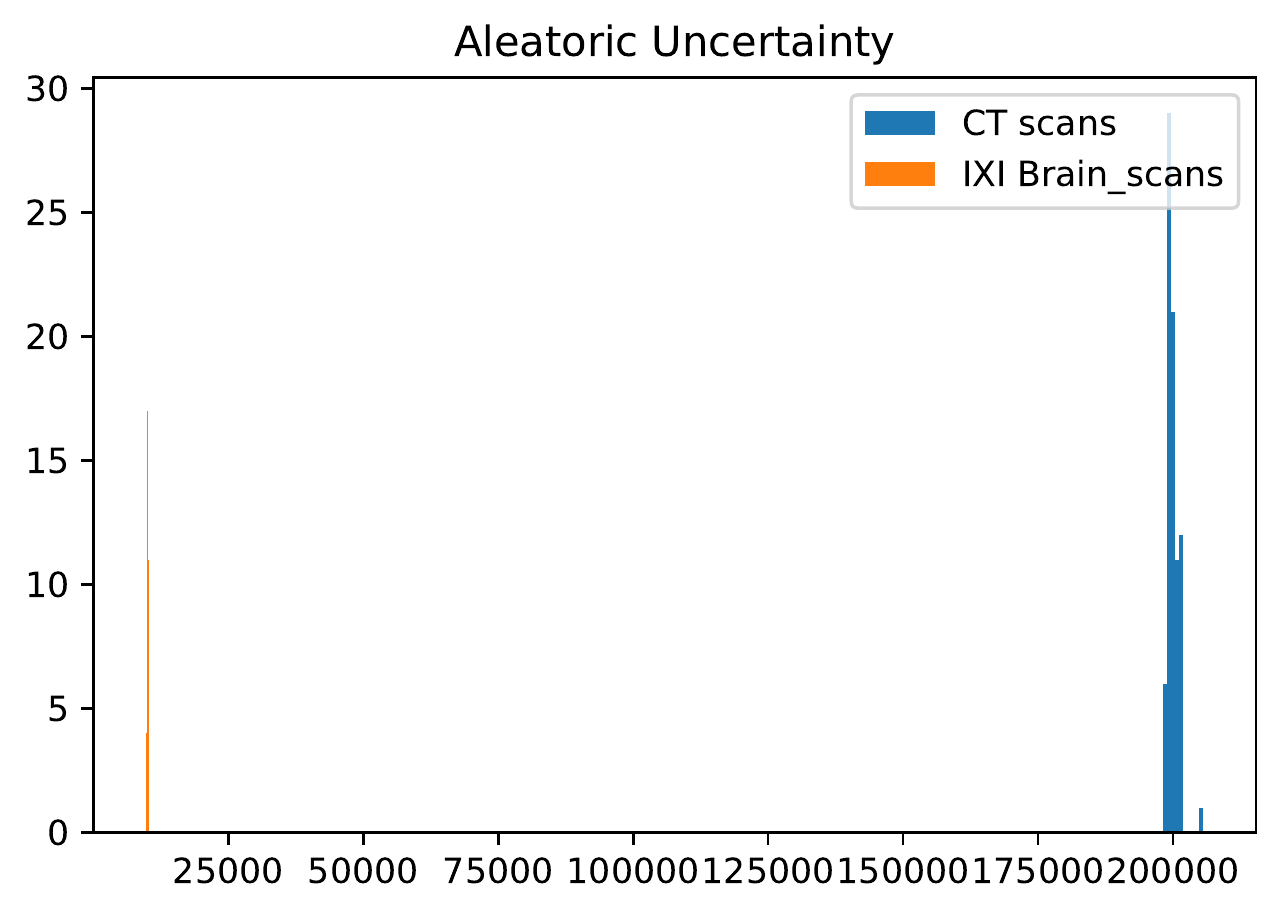}
        \includegraphics[width=0.49\textwidth]{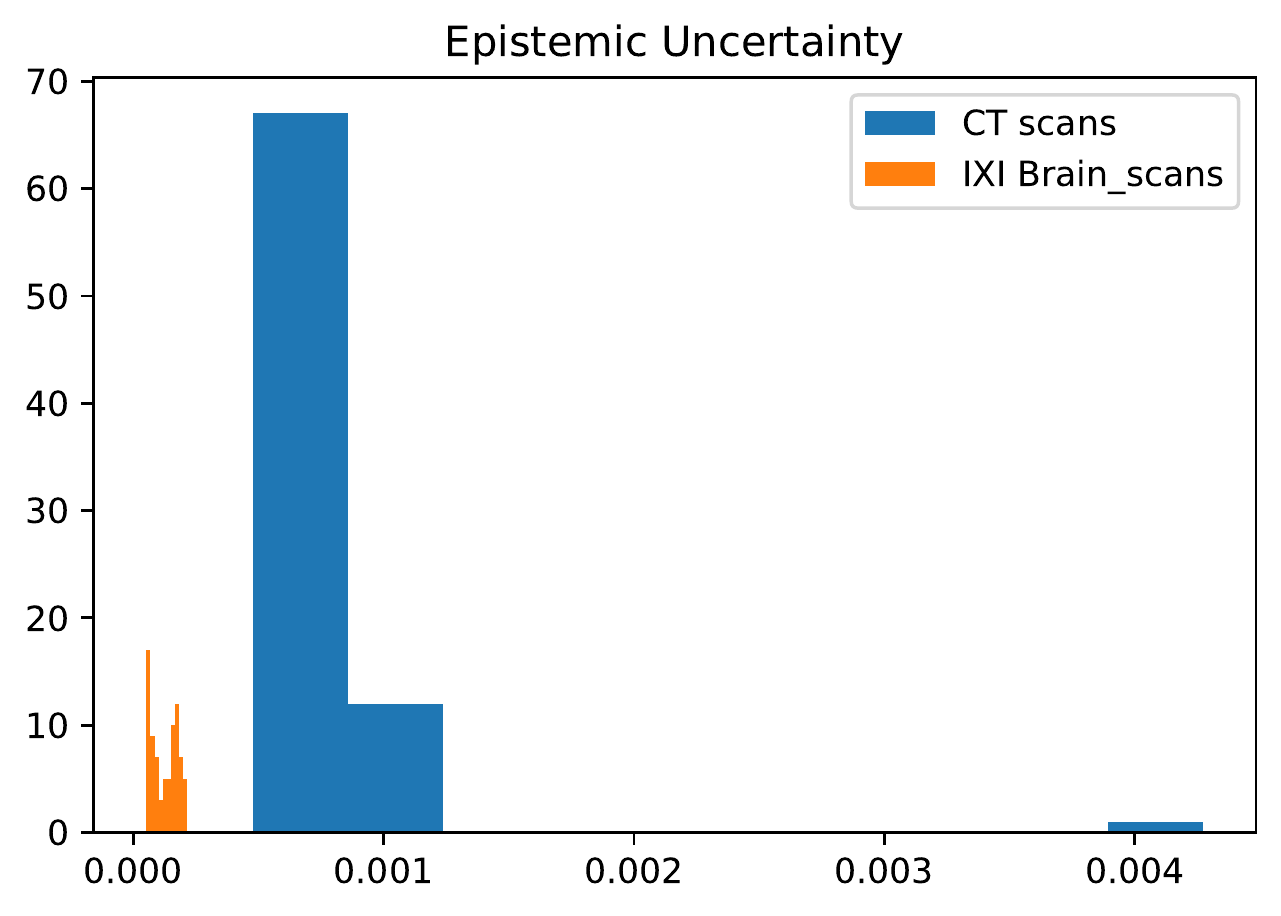}
        \caption{Ensembles}
    \end{subfigure}
    \begin{subfigure}[b]{0.49\textwidth}		
        \includegraphics[width=0.49\textwidth]{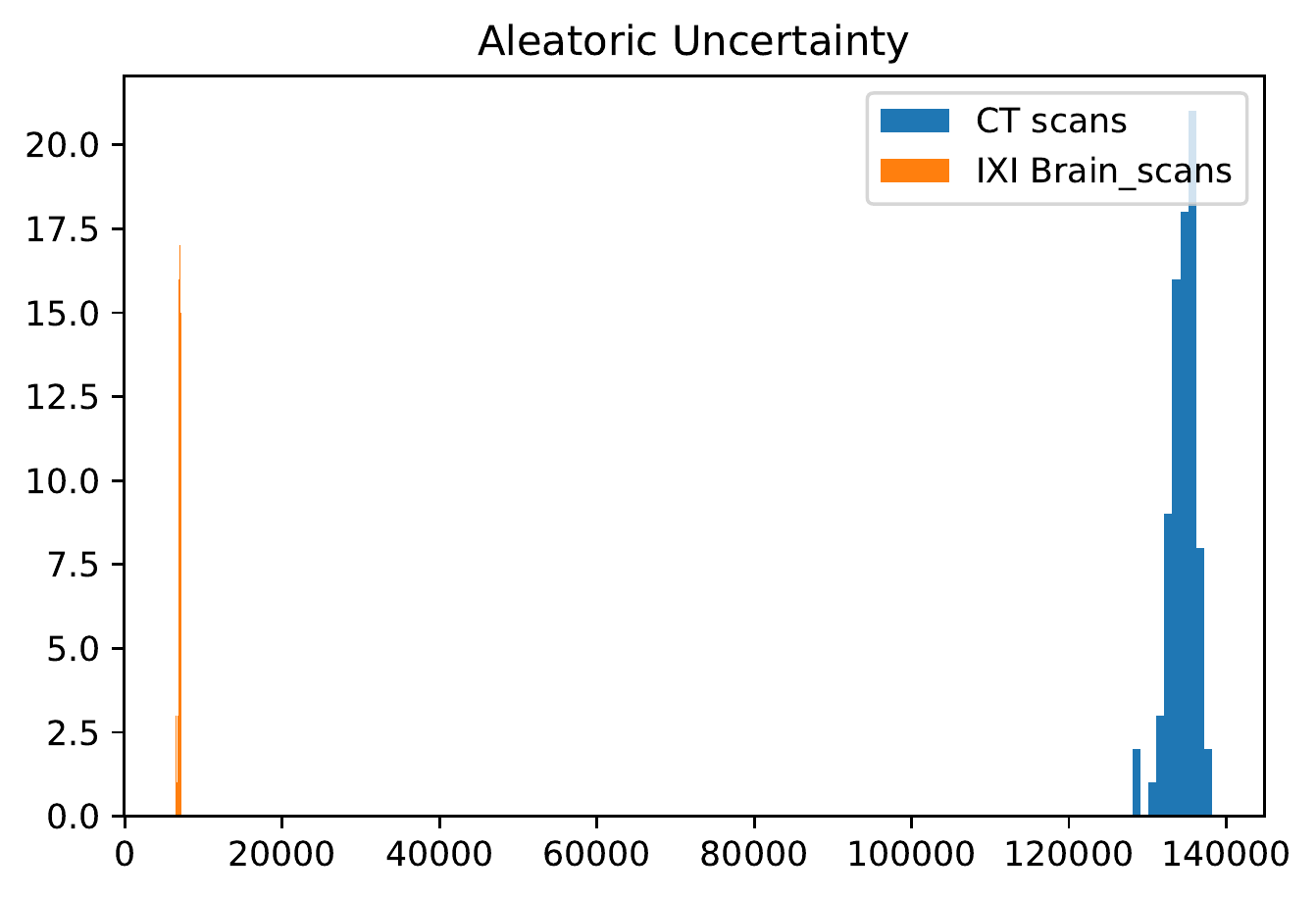}
        \includegraphics[width=0.49\textwidth]{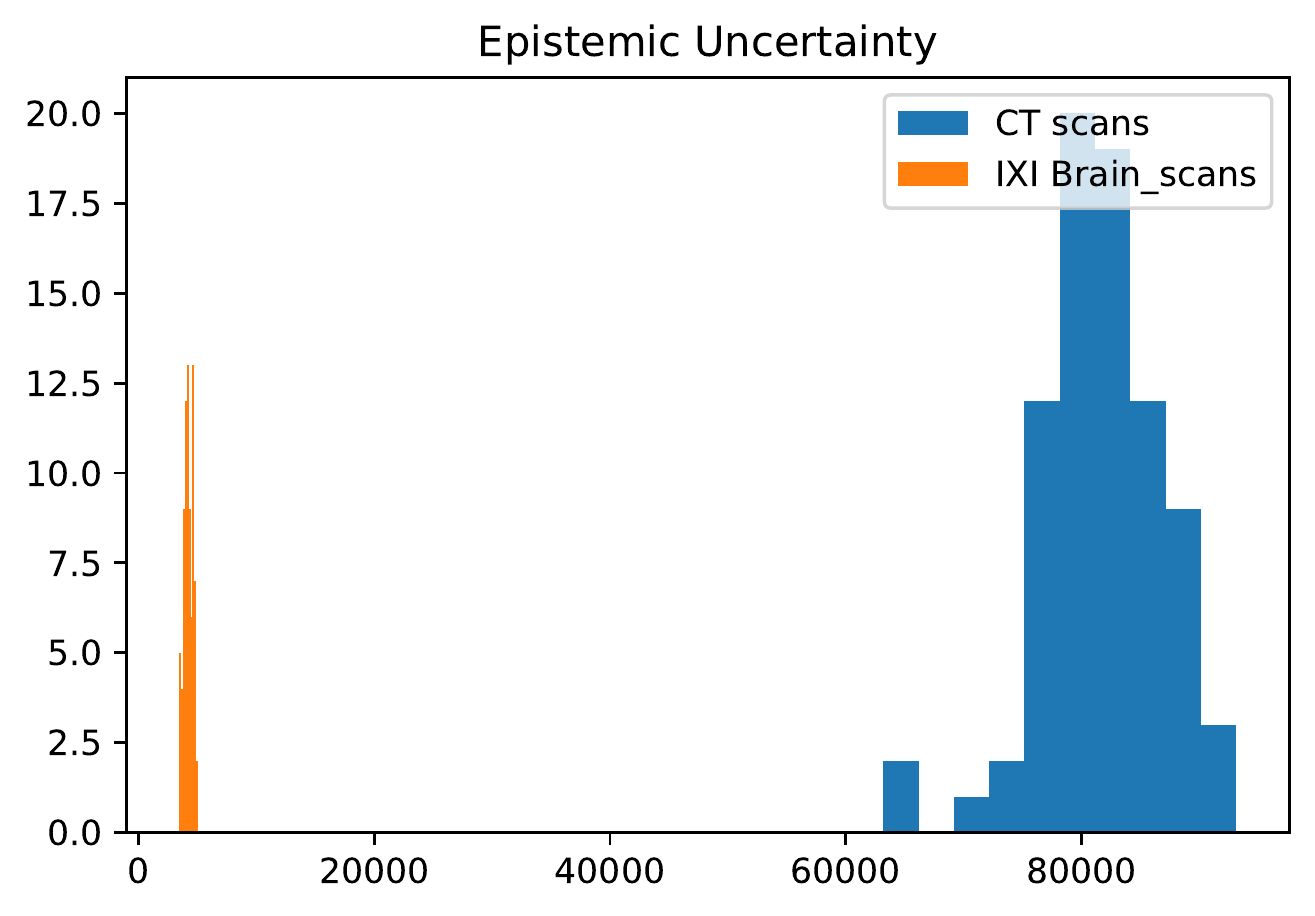}
        \caption{MC-DropConnect}
    \end{subfigure}
    \caption{Aleatoric and epistemic uncertainty comparison between IXI Brain Scans and Brain CT Scans}
    \label{ixi_ct}
\end{figure}

\begin{figure}[t]
    \centering
    \begin{subfigure}[b]{0.49\textwidth}		
        \includegraphics[width=0.49\textwidth]{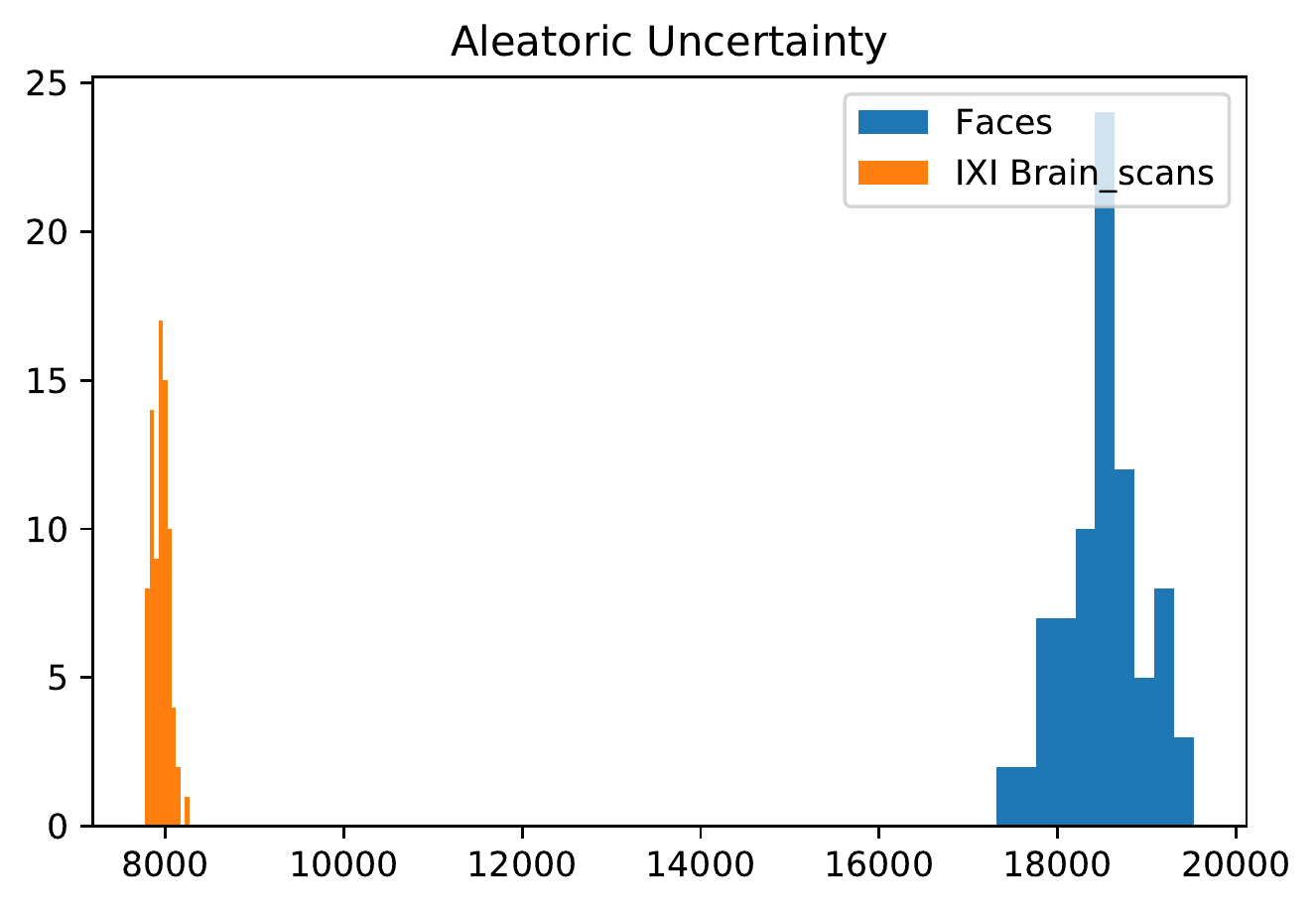}
        \includegraphics[width=0.49\textwidth]{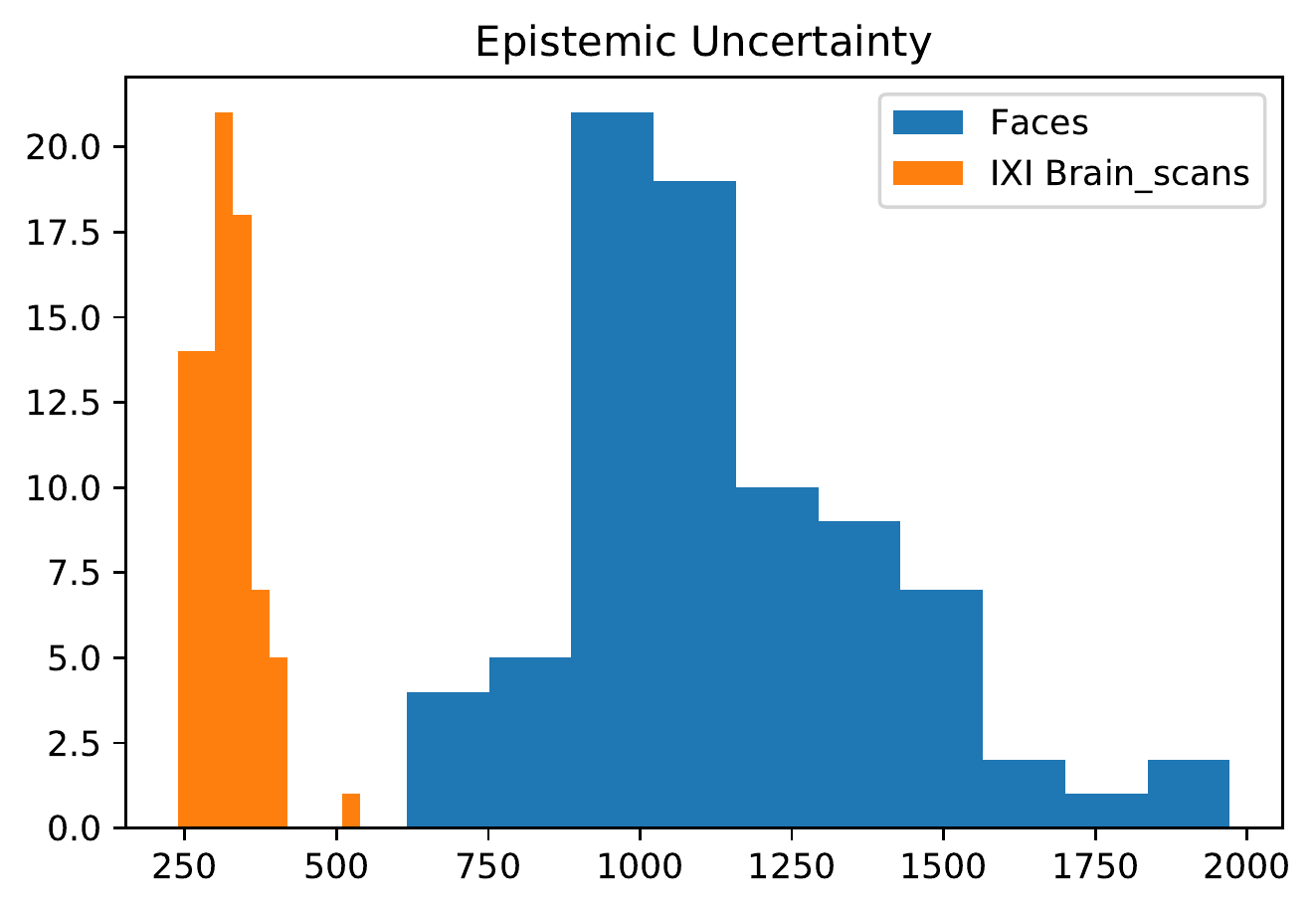}
        \caption{MC-Dropout}
    \end{subfigure}
    \begin{subfigure}[b]{0.49\textwidth}		
        \includegraphics[width=0.49\textwidth]{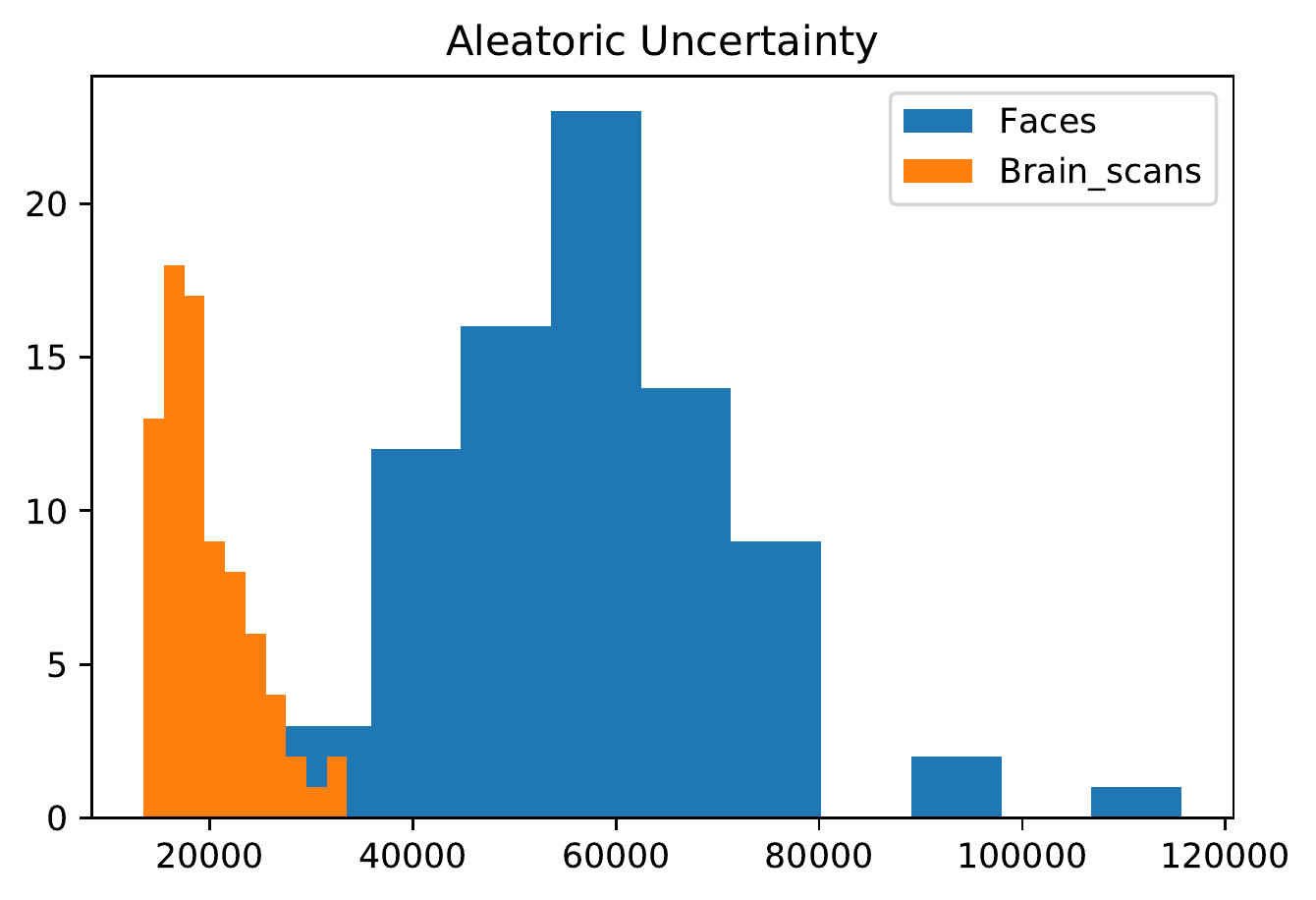}
        \includegraphics[width=0.49\textwidth]{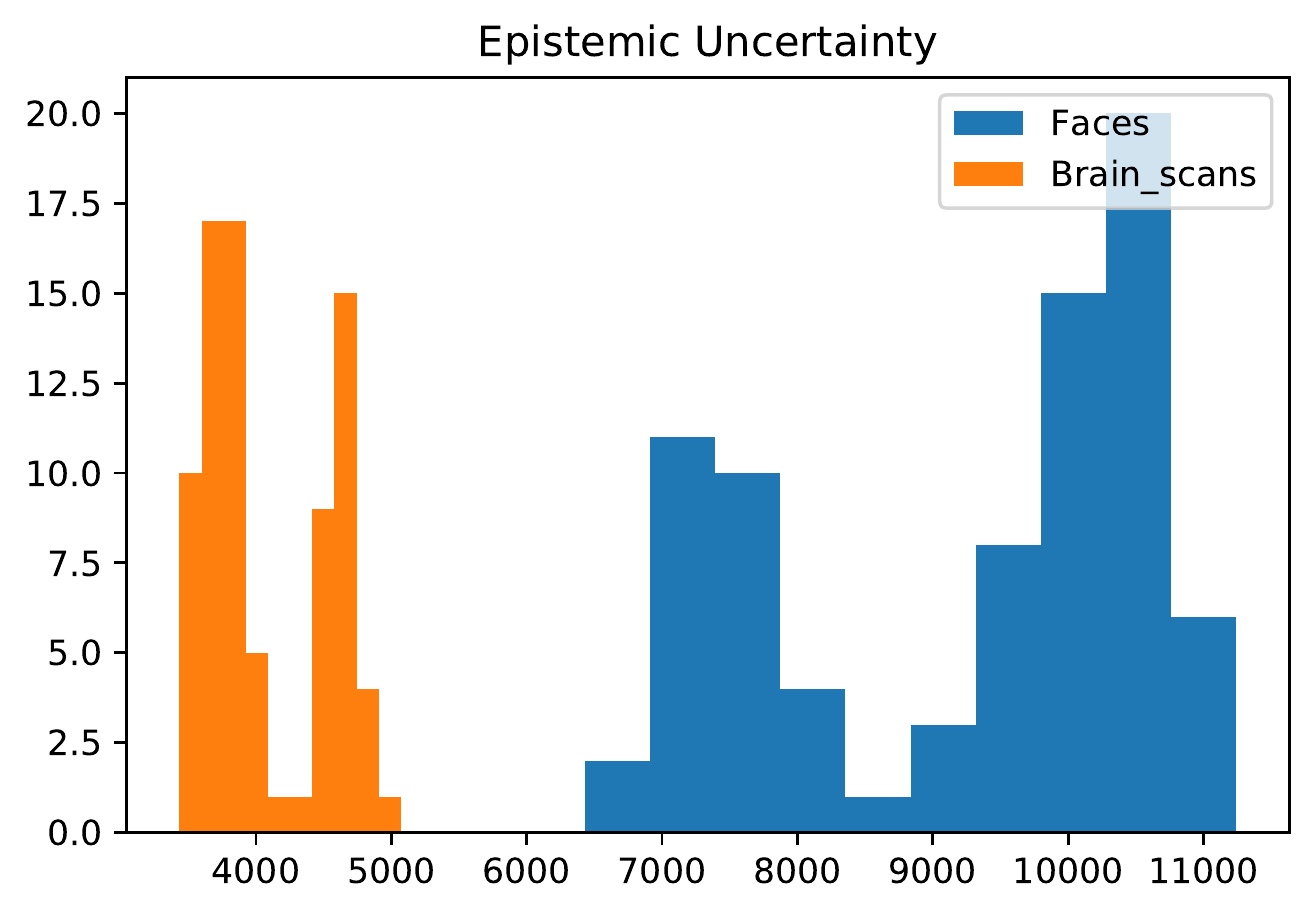}
        \caption{Flipout}
    \end{subfigure}
    
    \begin{subfigure}[b]{0.49\textwidth}		
        \includegraphics[width=0.49\textwidth]{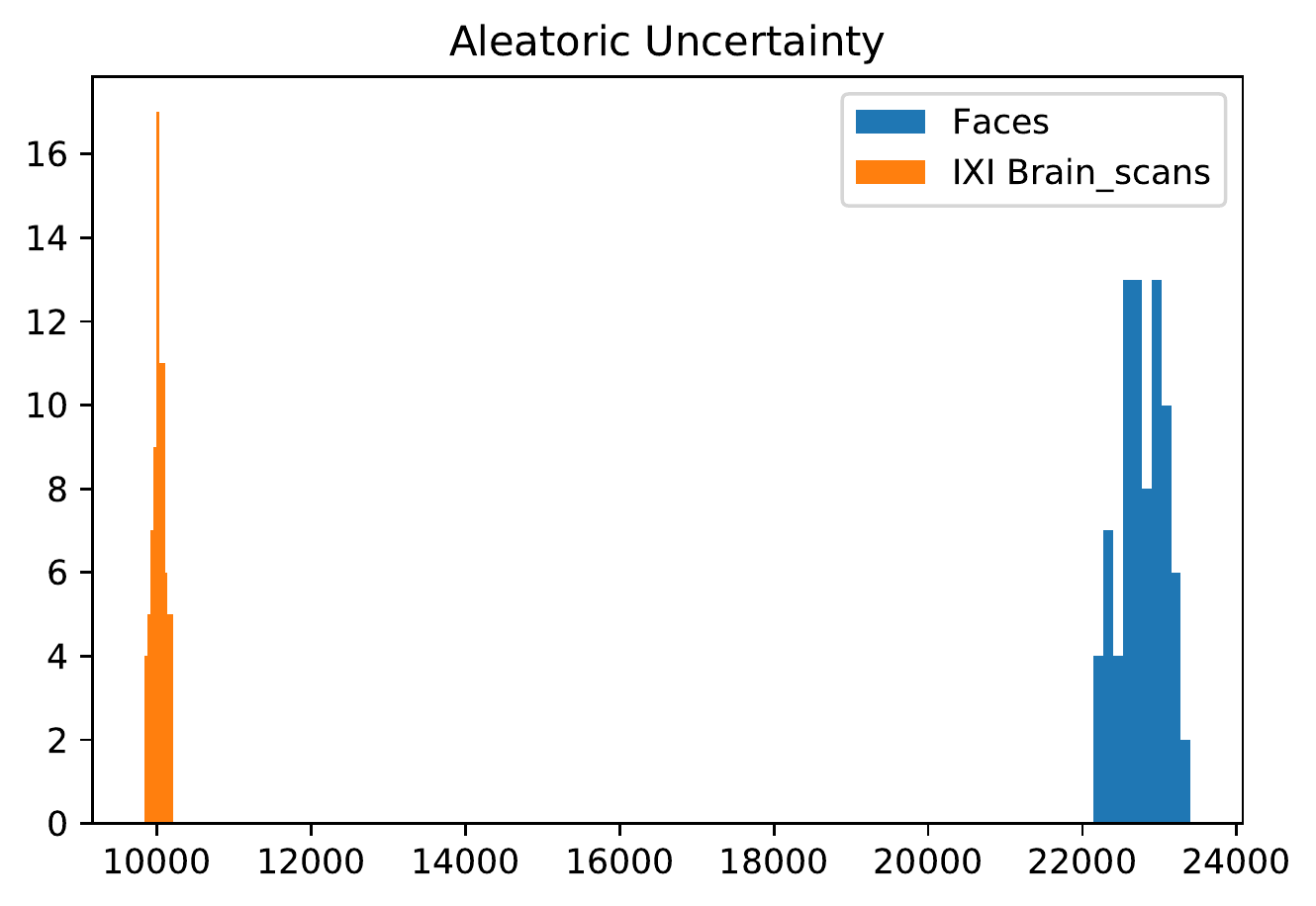}
        \includegraphics[width=0.49\textwidth]{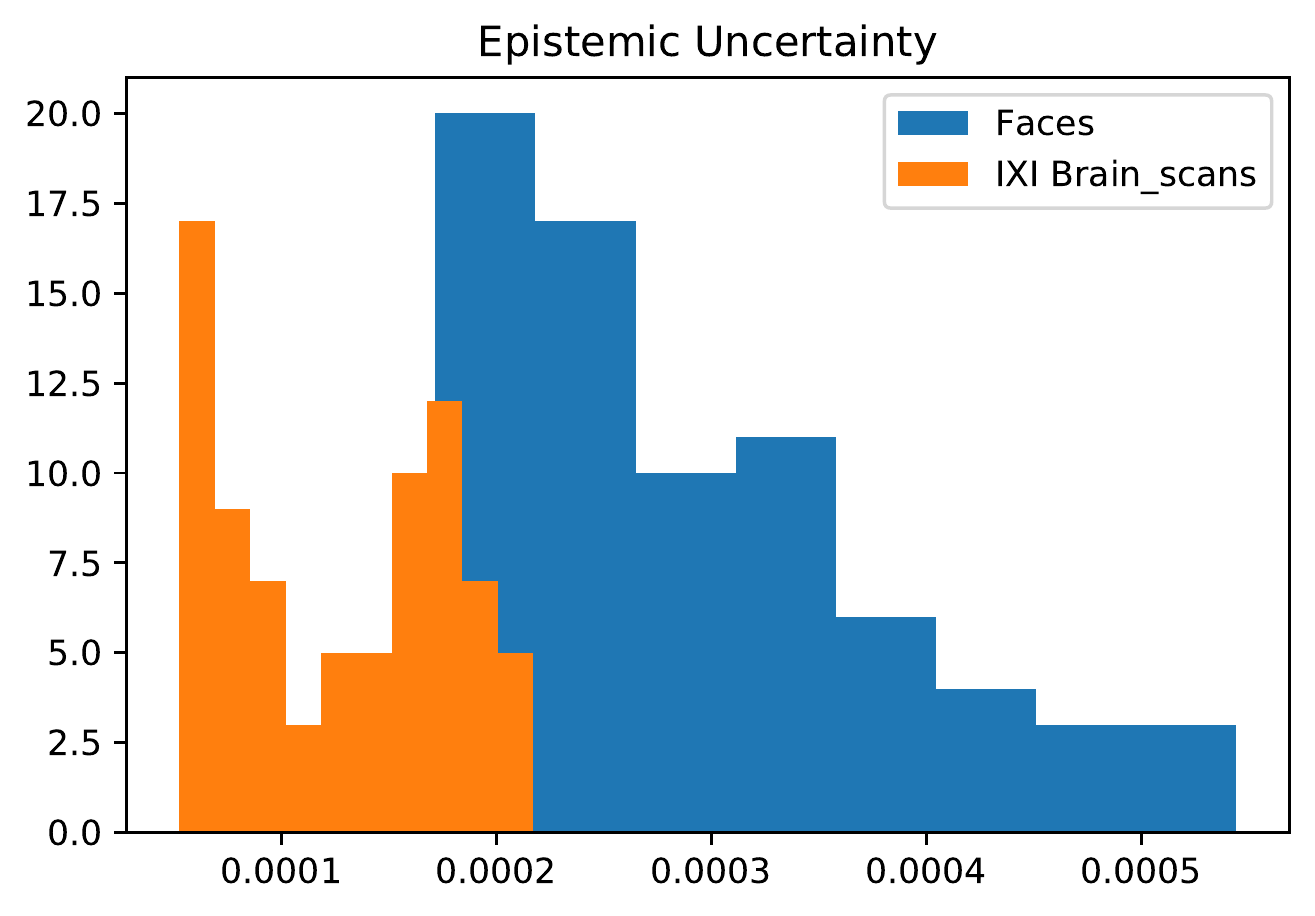}
        \caption{Ensembles}
    \end{subfigure}
    \begin{subfigure}[b]{0.49\textwidth}		
        \includegraphics[width=0.49\textwidth]{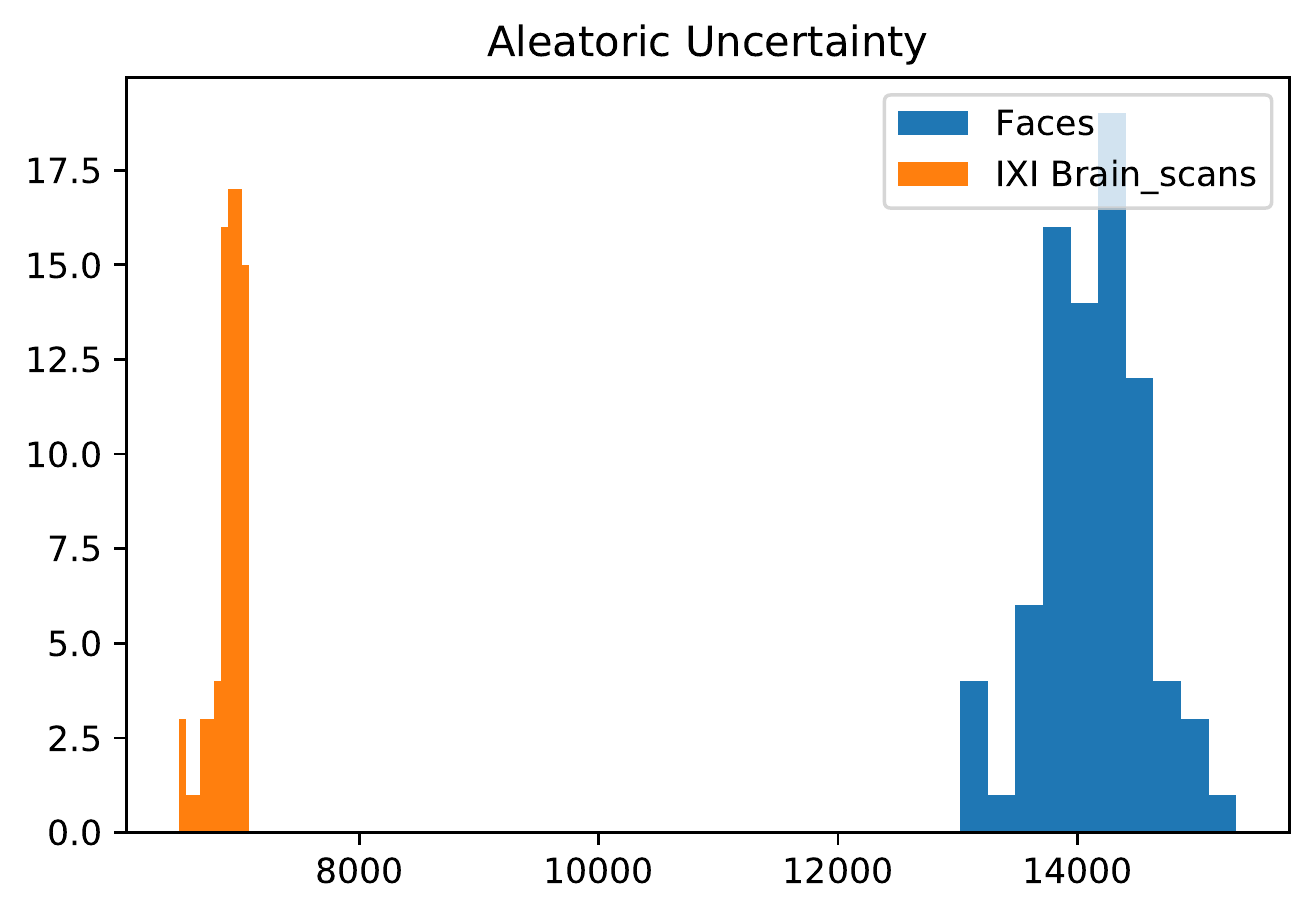}
        \includegraphics[width=0.49\textwidth]{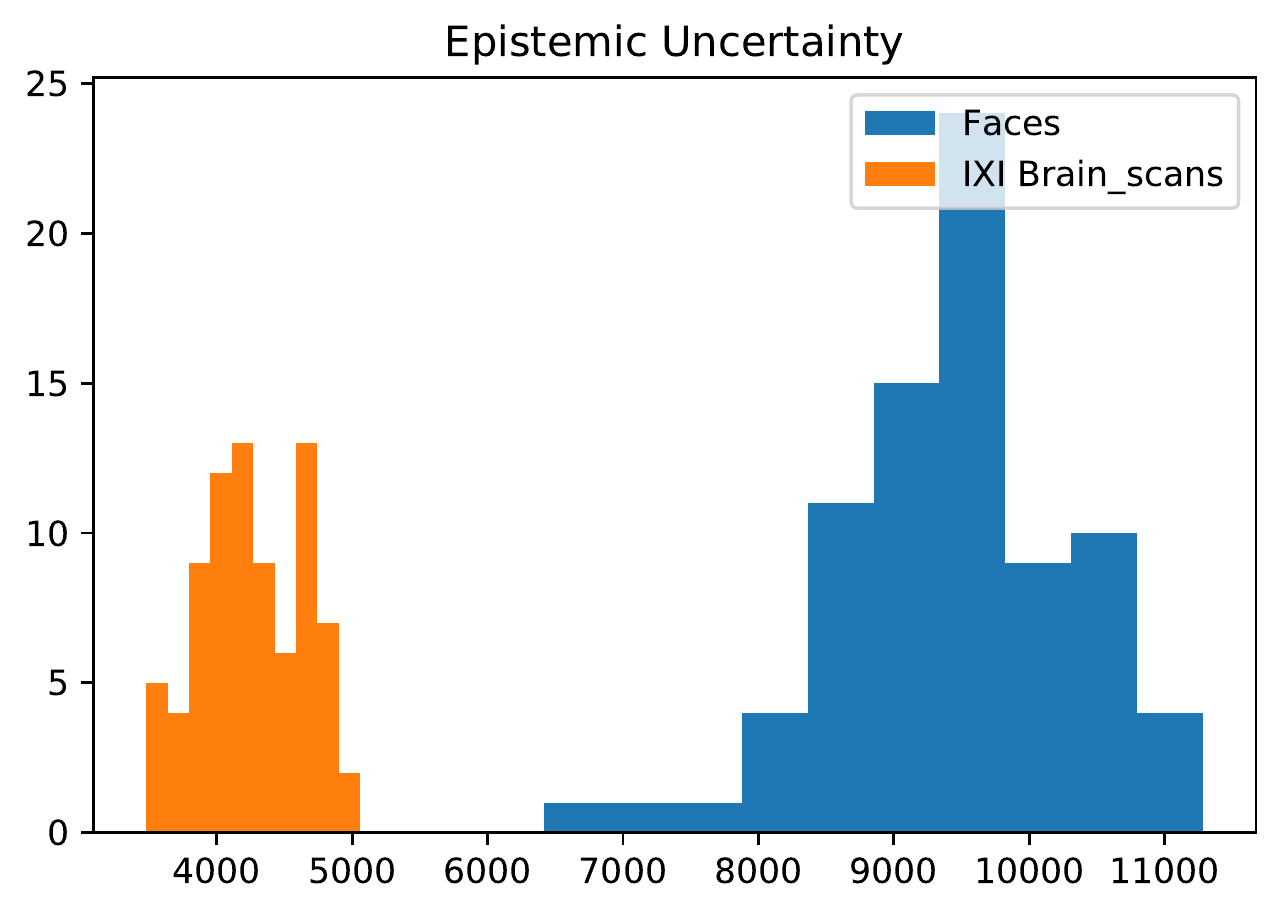}
        \caption{MC-DropConnect}
    \end{subfigure}
    \caption{Aleatoric and epistemic uncertainty comparison between IXI Brain Scans and UTKFace}
    \label{ixi_faces}
\end{figure}

Figures \ref{mcdropout_results} through \ref{dropconnect_results} show the respective histograms for the various method-dataset combinations. We notice that for most cases, when a model is provided out of distribution data, the epistemic uncertainty increases (sometimes very drastically as seen in figure \ref{ensemble_results}) where as the aleatoric uncertainty remains in a relatively similar range compared to the non-OOD data counterpart. This is also clearly visible in figures \ref{ixi_ct} through \ref{ixi_faces} where the uncertainty histograms for the in-distribution (ID) dataset (IXI brain scans) and the OOD datasets are plotted together. For most cases a clear distinction can be made between the plots for in-distribution and OOD data. Moreover, epistemic uncertainty can be seen to increase in the case of OOD data.  This can be used to detect outliers and further improves model robustness to OOD data.

Qualitative results are presented in tables \ref{ct_table} through \ref{faces_table}. Table \ref{roc_table} shows the ROC curves for aleatoric and epistemic uncertainty for various methods. Epistemic uncertainty is more reliable for OOD detection in most but not all methods, while aleatoric uncertainty fails to separate ID from OOD data in all but one case (Flipout).

We believe these results show that out of distribution inputs can be successfully detected using a image to image translation generative model such as CycleGAN, when combining it with a uncertainty quantification method, and that disentangled estimates of aleatoric and epistemic uncertainty allow for separate modeling of model errors and noise in the data, which we believe can improve interpretation of visual results, which is generally not possible with standard predictive uncertainty quantification.

\section{Conclusions and Future Work}
In this paper we presented a novel method for uncertainty disentanglement using CycleGAN for image-to-image translation in the medical domain. We analyzed various uncertainty quantification methods and their ability to disentangle predictive uncertainty into epistemic and aleatoric uncertainty estimates.

We drew insights related to uncertainty values/distributions and out of distribution data. Our results show the differences between different uncertainty quantification methods and how they behave when exposed to out of distribution data. In particular when combining disentangled uncertainties with a generative model, out of distribution inputs can be detected by looking at the epistemic uncertainty of the output image, while aleatoric uncertainty does not produce relevant information about out of distribution inputs, but does provide insights about noise in the input data.

We believe these results show the potential of disentangled uncertainties in image to image generative models for medical data. Without uncertainty quantification, the medical worker that has to interpret these images and obtain useful information for diagnosis, would not be aware if the model is unsure of its output, if out of distribution inputs were provided, or if the data is too noisy, which would propagate through the model (garbage in, garbage out).

We see multiple avenues of future research including fast prediction of uncertainty estimates, further evaluation of the quality of uncertainty, and evaluating these methods on more realistic medical imaging settings. We hope our work inspires further research in the field of uncertainty and out of distribution detection particularly in the field of medical imaging.

\bibliography{references}

\begin{thebibliography}{10}

\bibitem{armanious2019unsupervised}
Karim Armanious, Chenming Jiang, Sherif Abdulatif, Thomas K{\"u}stner, Sergios
  Gatidis, and Bin Yang.
\newblock Unsupervised medical image translation using cycle-medgan.
\newblock In {\em 2019 27th European Signal Processing Conference (EUSIPCO)},
  pages 1--5. IEEE, 2019.

\bibitem{armanious2020medgan}
Karim Armanious, Chenming Jiang, Marc Fischer, Thomas K{\"u}stner, Tobias Hepp,
  Konstantin Nikolaou, Sergios Gatidis, and Bin Yang.
\newblock Medgan: Medical image translation using gans.
\newblock {\em Computerized medical imaging and graphics}, 79:101684, 2020.

\bibitem{armanious2020unsupervised}
Karim Armanious, Aastha Tanwar, Sherif Abdulatif, Thomas K{\"u}stner, Sergios
  Gatidis, and Bin Yang.
\newblock Unsupervised adversarial correction of rigid mr motion artifacts.
\newblock In {\em 2020 IEEE 17th International Symposium on Biomedical Imaging
  (ISBI)}, pages 1494--1498. IEEE, 2020.

\bibitem{blundell2015weight}
Charles Blundell, Julien Cornebise, Koray Kavukcuoglu, and Daan Wierstra.
\newblock Weight uncertainty in neural network.
\newblock In {\em International conference on machine learning}, pages
  1613--1622. PMLR, 2015.

\bibitem{chen2017low}
Hu~Chen, Yi~Zhang, Mannudeep~K Kalra, Feng Lin, Yang Chen, Peixi Liao, Jiliu
  Zhou, and Ge~Wang.
\newblock Low-dose ct with a residual encoder-decoder convolutional neural
  network.
\newblock {\em IEEE transactions on medical imaging}, 36(12):2524--2535, 2017.

\bibitem{depeweg2018decomposition}
Stefan Depeweg, Jose-Miguel Hernandez-Lobato, Finale Doshi-Velez, and Steffen
  Udluft.
\newblock Decomposition of uncertainty in bayesian deep learning for efficient
  and risk-sensitive learning.
\newblock In {\em International Conference on Machine Learning}, pages
  1184--1193. PMLR, 2018.

\bibitem{der2009aleatory}
Armen Der~Kiureghian and Ove Ditlevsen.
\newblock Aleatory or epistemic? does it matter?
\newblock {\em Structural safety}, 31(2):105--112, 2009.

\bibitem{ixi_data}
Brain Development.
\newblock Ixi dataset.
\newblock Available at \url{http://brain-development.org/ixi-dataset}
  (2022/09/28).

\bibitem{dou2016multilevel}
Qi~Dou, Hao Chen, Lequan Yu, Jing Qin, and Pheng-Ann Heng.
\newblock Multilevel contextual 3-d cnns for false positive reduction in
  pulmonary nodule detection.
\newblock {\em IEEE Transactions on Biomedical Engineering}, 64(7):1558--1567,
  2016.

\bibitem{esteva2017dermatologist}
Andre Esteva, Brett Kuprel, Roberto~A Novoa, Justin Ko, Susan~M Swetter,
  Helen~M Blau, and Sebastian Thrun.
\newblock Dermatologist-level classification of skin cancer with deep neural
  networks.
\newblock {\em nature}, 542(7639):115--118, 2017.

\bibitem{filos2019systematic}
Angelos Filos, Sebastian Farquhar, Aidan~N Gomez, Tim~GJ Rudner, Zachary
  Kenton, Lewis Smith, Milad Alizadeh, Arnoud De~Kroon, and Yarin Gal.
\newblock A systematic comparison of bayesian deep learning robustness in
  diabetic retinopathy tasks.
\newblock {\em arXiv preprint arXiv:1912.10481}, 2019.

\bibitem{gal2016uncertainty}
Yarin Gal et~al.
\newblock Uncertainty in deep learning.
\newblock 2016.

\bibitem{gal2016dropout}
Yarin Gal and Zoubin Ghahramani.
\newblock Dropout as a bayesian approximation: Representing model uncertainty
  in deep learning.
\newblock In {\em international conference on machine learning}, pages
  1050--1059. PMLR, 2016.

\bibitem{gu2021adain}
Jawook Gu and Jong~Chul Ye.
\newblock Adain-based tunable cyclegan for efficient unsupervised low-dose ct
  denoising.
\newblock {\em IEEE Transactions on Computational Imaging}, 7:73--85, 2021.

\bibitem{han2019combining}
Changhee Han, Leonardo Rundo, Ryosuke Araki, Yudai Nagano, Yujiro Furukawa,
  Giancarlo Mauri, Hideki Nakayama, and Hideaki Hayashi.
\newblock Combining noise-to-image and image-to-image gans: Brain mr image
  augmentation for tumor detection.
\newblock {\em Ieee Access}, 7:156966--156977, 2019.

\bibitem{havaei2017brain}
Mohammad Havaei, Axel Davy, David Warde-Farley, Antoine Biard, Aaron Courville,
  Yoshua Bengio, Chris Pal, Pierre-Marc Jodoin, and Hugo Larochelle.
\newblock Brain tumor segmentation with deep neural networks.
\newblock {\em Medical image analysis}, 35:18--31, 2017.

\bibitem{hendrycks2016baseline}
Dan Hendrycks and Kevin Gimpel.
\newblock A baseline for detecting misclassified and out-of-distribution
  examples in neural networks.
\newblock {\em arXiv preprint arXiv:1610.02136}, 2016.

\bibitem{ct_dataset}
Murtadha~D Hssayeni, Muayad~S Croock, Aymen~D Salman, Hassan~Falah Al-khafaji,
  Zakaria~A Yahya, and Behnaz Ghoraani.
\newblock Computed tomography images for intracranial hemorrhage detection and
  segmentation (version 1.3.1).

\bibitem{hssayeni2020intracranial}
Murtadha~D Hssayeni, Muayad~S Croock, Aymen~D Salman, Hassan~Falah Al-khafaji,
  Zakaria~A Yahya, and Behnaz Ghoraani.
\newblock Intracranial hemorrhage segmentation using a deep convolutional
  model.
\newblock {\em Data}, 5(1):14, 2020.

\bibitem{hullermeier2021aleatoric}
Eyke H{\"u}llermeier and Willem Waegeman.
\newblock Aleatoric and epistemic uncertainty in machine learning: An
  introduction to concepts and methods.
\newblock {\em Machine Learning}, 110(3):457--506, 2021.

\bibitem{huynh2015estimating}
Tri Huynh, Yaozong Gao, Jiayin Kang, Li~Wang, Pei Zhang, Jun Lian, and Dinggang
  Shen.
\newblock Estimating ct image from mri data using structured random forest and
  auto-context model.
\newblock {\em IEEE transactions on medical imaging}, 35(1):174--183, 2015.

\bibitem{isola2017image}
Phillip Isola, Jun-Yan Zhu, Tinghui Zhou, and Alexei~A Efros.
\newblock Image-to-image translation with conditional adversarial networks.
\newblock In {\em Proceedings of the IEEE conference on computer vision and
  pattern recognition}, pages 1125--1134, 2017.

\bibitem{kabir2018neural}
HM~Dipu Kabir, Abbas Khosravi, Mohammad~Anwar Hosen, and Saeid Nahavandi.
\newblock Neural network-based uncertainty quantification: A survey of
  methodologies and applications.
\newblock {\em IEEE access}, 6:36218--36234, 2018.

\bibitem{kaji2019overview}
Shizuo Kaji and Satoshi Kida.
\newblock Overview of image-to-image translation by use of deep neural
  networks: denoising, super-resolution, modality conversion, and
  reconstruction in medical imaging, 2019.

\bibitem{kamnitsas2017efficient}
Konstantinos Kamnitsas, Christian Ledig, Virginia~FJ Newcombe, Joanna~P
  Simpson, Andrew~D Kane, David~K Menon, Daniel Rueckert, and Ben Glocker.
\newblock Efficient multi-scale 3d cnn with fully connected crf for accurate
  brain lesion segmentation.
\newblock {\em Medical image analysis}, 36:61--78, 2017.

\bibitem{kendall2017uncertainties}
Alex Kendall and Yarin Gal.
\newblock What uncertainties do we need in bayesian deep learning for computer
  vision?
\newblock {\em Advances in neural information processing systems}, 30, 2017.

\bibitem{kustner2017mr}
Thomas K{\"u}stner, Martin Schwartz, Petros Martirosian, Sergios Gatidis,
  Ferdinand Seith, Christopher Gilliam, Thierry Blu, Hadi Fayad, Dimitris
  Visvikis, Fritz Schick, et~al.
\newblock Mr-based respiratory and cardiac motion correction for pet imaging.
\newblock {\em Medical image analysis}, 42:129--144, 2017.

\bibitem{lakshminarayanan2017simple}
Balaji Lakshminarayanan, Alexander Pritzel, and Charles Blundell.
\newblock Simple and scalable predictive uncertainty estimation using deep
  ensembles.
\newblock {\em Advances in neural information processing systems}, 30, 2017.

\bibitem{litjens2017survey}
Geert Litjens, Thijs Kooi, Babak~Ehteshami Bejnordi, Arnaud Arindra~Adiyoso
  Setio, Francesco Ciompi, Mohsen Ghafoorian, Jeroen~Awm Van Der~Laak, Bram
  Van~Ginneken, and Clara~I S{\'a}nchez.
\newblock A survey on deep learning in medical image analysis.
\newblock {\em Medical image analysis}, 42:60--88, 2017.

\bibitem{liu2017unsupervised}
Ming-Yu Liu, Thomas Breuel, and Jan Kautz.
\newblock Unsupervised image-to-image translation networks.
\newblock {\em Advances in neural information processing systems}, 30, 2017.

\bibitem{mobiny2021dropconnect}
Aryan Mobiny, Pengyu Yuan, Supratik~K Moulik, Naveen Garg, Carol~C Wu, and Hien
  Van~Nguyen.
\newblock Dropconnect is effective in modeling uncertainty of bayesian deep
  networks.
\newblock {\em Scientific reports}, 11(1):1--14, 2021.

\bibitem{nair2020exploring}
Tanya Nair, Doina Precup, Douglas~L Arnold, and Tal Arbel.
\newblock Exploring uncertainty measures in deep networks for multiple
  sclerosis lesion detection and segmentation.
\newblock {\em Medical image analysis}, 59:101557, 2020.

\bibitem{nie2018medical}
Dong Nie, Roger Trullo, Jun Lian, Li~Wang, Caroline Petitjean, Su~Ruan, Qian
  Wang, and Dinggang Shen.
\newblock Medical image synthesis with deep convolutional adversarial networks.
\newblock {\em IEEE Transactions on Biomedical Engineering}, 65(12):2720--2730,
  2018.

\bibitem{park2020contrastive}
Taesung Park, Alexei~A Efros, Richard Zhang, and Jun-Yan Zhu.
\newblock Contrastive learning for unpaired image-to-image translation.
\newblock In {\em European conference on computer vision}, pages 319--345.
  Springer, 2020.

\bibitem{rueda2013single}
Andrea Rueda, Norberto Malpica, and Eduardo Romero.
\newblock Single-image super-resolution of brain mr images using overcomplete
  dictionaries.
\newblock {\em Medical image analysis}, 17(1):113--132, 2013.

\bibitem{tanno2017bayesian}
Ryutaro Tanno, Daniel~E Worrall, Aurobrata Ghosh, Enrico Kaden, Stamatios~N
  Sotiropoulos, Antonio Criminisi, and Daniel~C Alexander.
\newblock Bayesian image quality transfer with cnns: exploring uncertainty in
  dmri super-resolution.
\newblock In {\em International Conference on Medical Image Computing and
  Computer-Assisted Intervention}, pages 611--619. Springer, 2017.

\bibitem{tran2018dist}
Ngoc-Trung Tran, Tuan-Anh Bui, and Ngai-Man Cheung.
\newblock Dist-gan: An improved gan using distance constraints.
\newblock In {\em Proceedings of the European conference on computer vision
  (ECCV)}, pages 370--385, 2018.

\bibitem{upadhyay2019mixed}
Uddeshya Upadhyay and Suyash~P Awate.
\newblock A mixed-supervision multilevel gan framework for image quality
  enhancement.
\newblock In {\em International Conference on Medical Image Computing and
  Computer-Assisted Intervention}, pages 556--564. Springer, 2019.

\bibitem{upadhyay2019robust}
Uddeshya Upadhyay and Suyash~P Awate.
\newblock Robust super-resolution gan, with manifold-based and perception loss.
\newblock In {\em 2019 IEEE 16th International Symposium on Biomedical Imaging
  (ISBI 2019)}, pages 1372--1376. IEEE, 2019.

\bibitem{upadhyay2021uncertainty}
Uddeshya Upadhyay, Yanbei Chen, and Zeynep Akata.
\newblock Uncertainty-aware generalized adaptive cyclegan.
\newblock {\em arXiv preprint arXiv:2102.11747}, 2021.

\bibitem{upadhyay2021uncertainty2}
Uddeshya Upadhyay, Yanbei Chen, Tobias Hepp, Sergios Gatidis, and Zeynep Akata.
\newblock Uncertainty-guided progressive gans for medical image translation.
\newblock In {\em International Conference on Medical Image Computing and
  Computer-Assisted Intervention}, pages 614--624. Springer, 2021.

\bibitem{wang2019aleatoric}
Guotai Wang, Wenqi Li, Michael Aertsen, Jan Deprest, S{\'e}bastien Ourselin,
  and Tom Vercauteren.
\newblock Aleatoric uncertainty estimation with test-time augmentation for
  medical image segmentation with convolutional neural networks.
\newblock {\em Neurocomputing}, 338:34--45, 2019.

\bibitem{wang2018automatic}
Guotai Wang, Wenqi Li, S{\'e}bastien Ourselin, and Tom Vercauteren.
\newblock Automatic brain tumor segmentation using convolutional neural
  networks with test-time augmentation.
\newblock In {\em International MICCAI Brainlesion Workshop}, pages 61--72.
  Springer, 2018.

\bibitem{wang2019automatic}
Guotai Wang, Wenqi Li, S{\'e}bastien Ourselin, and Tom Vercauteren.
\newblock Automatic brain tumor segmentation based on cascaded convolutional
  neural networks with uncertainty estimation.
\newblock {\em Frontiers in computational neuroscience}, 13:56, 2019.

\bibitem{welander2018generative}
Per Welander, Simon Karlsson, and Anders Eklund.
\newblock Generative adversarial networks for image-to-image translation on
  multi-contrast mr images-a comparison of cyclegan and unit.
\newblock {\em arXiv preprint arXiv:1806.07777}, 2018.

\bibitem{wen2018flipout}
Yeming Wen, Paul Vicol, Jimmy Ba, Dustin Tran, and Roger Grosse.
\newblock Flipout: Efficient pseudo-independent weight perturbations on
  mini-batches.
\newblock {\em arXiv preprint arXiv:1803.04386}, 2018.

\bibitem{wolterink2017deep}
Jelmer~M Wolterink, Anna~M Dinkla, Mark~HF Savenije, Peter~R Seevinck,
  Cornelis~AT van~den Berg, and Ivana I{\v{s}}gum.
\newblock Deep mr to ct synthesis using unpaired data.
\newblock In {\em International workshop on simulation and synthesis in medical
  imaging}, pages 14--23. Springer, 2017.

\bibitem{yang2020mri}
Qianye Yang, Nannan Li, Zixu Zhao, Xingyu Fan, Eric~I Chang, Yan Xu, et~al.
\newblock Mri cross-modality image-to-image translation.
\newblock {\em Scientific reports}, 10(1):1--18, 2020.

\bibitem{yang2018low}
Qingsong Yang, Pingkun Yan, Yanbo Zhang, Hengyong Yu, Yongyi Shi, Xuanqin Mou,
  Mannudeep~K Kalra, Yi~Zhang, Ling Sun, and Ge~Wang.
\newblock Low-dose ct image denoising using a generative adversarial network
  with wasserstein distance and perceptual loss.
\newblock {\em IEEE transactions on medical imaging}, 37(6):1348--1357, 2018.

\bibitem{zhang2019image}
He~Zhang, Vishwanath Sindagi, and Vishal~M Patel.
\newblock Image de-raining using a conditional generative adversarial network.
\newblock {\em IEEE transactions on circuits and systems for video technology},
  30(11):3943--3956, 2019.

\bibitem{zhifei2017cvpr}
Zhifei Zhang, Yang Song, and Hairong Qi.
\newblock Age progression or regression by conditional adversarial autoencoder.
\newblock In {\em IEEE Conference on Computer Vision and Pattern Recognition
  (CVPR)}. IEEE, 2017.

\bibitem{zhu2017unpaired}
Jun-Yan Zhu, Taesung Park, Phillip Isola, and Alexei~A Efros.
\newblock Unpaired image-to-image translation using cycle-consistent
  adversarial networks.
\newblock In {\em Proceedings of the IEEE international conference on computer
  vision}, pages 2223--2232, 2017.

\end{thebibliography}

\FloatBarrier
\appendix
\section{Uncertainty Quantification Methods}
\label{unc_methods}
We employ multiple uncertainty quanitification methods to achieve a more holistic view of their ability to disentangle predictive uncertainty into epistemic and aleatoric uncertainty.

\textbf{MC-Dropout} This method acts as a regularizer which is applied at both training and testing time. It zeros out random activations in a layer. Since MC-Dropout\cite{gal2016dropout} is active during test time as well, the model becomes stochastic. Hence, each forward pass produces a unique sample from the Bayesian posterior distribution. For our experiments we take the drop probability as $p=0.5$

\textbf{Flipout} This is a variational inference method that models weights as an approximate Gaussian distribution \cite{blundell2015weight}. Flipout \cite{wen2018flipout} is employed to reduce training process variance and to improve learning stability and performance. The resulting model that uses Flipout is stochastic in nature. For the purpose of our experiments, we use Flipout in mulitple layers with a disabled prior.

\textbf{Ensembles} This method\cite{lakshminarayanan2017simple} involves mulitple training runs of the same model with different random weight initializations each time. The final output is predicted by considering the individual outputs from each of the member models of the ensemble. We make use of an ensemble of 5 member models for our experiments.

\textbf{MC-DropConnect} This method\cite{mobiny2021dropconnect} sets random weights to zero. It is in a sense similar to MC-Dropout\cite{gal2016dropout} and has a similar regularization effect. This method allows weights to be zeroed out at inference time which produces samples from the Bayesian posterior distribution as in the case of MS-Dropout\cite{gal2016dropout}. We use DropConnect layers with a drop probability of $p = 0.05$

\section{Potential Negative Societal Impact}
While the sole purpose of this work is to improve the quality of predictions made by medical generative models and make them more trustworthy, we can foresee some negative societal impacts as well.

A model is only as good as the data it is trained on hence unknown correlations within the data can reflect on the model. For uncertainty in particular, all existing uncertainty quantification methods are approximations and thus do not provide the true predictive posterior distribution which in turn brings into question the trustworthiness of predicted uncertainty values. Such methods could at times prove to be misleading and thus lead to wrong medical diagnoses.

In addition to this, it ultimately comes down to the medical professionals to consider the uncertainty predictions from the model while making their diagnoses. If medical professionals decide not to consider model uncertainty, it could lead to wrong diagnoses and further complications in treatment. Another important point of concern is privacy ad anonymity. In order to make useful predictions, the model needs to train on patient data and learn from it and this can put personal medical records in a vulnerable position.

\FloatBarrier
\section{Qualitative Result Comparison}
This section presents visual comparison of predicted image translation results, including disentangled aleatoric and epistemic uncertainty. We present outputs produced using three different inputs, one is in-distribution, and two are out of distribution.
\begin{table}[ht]
	\caption{Qualitative results for IXI Brain Scans: Since the these results are for ID data we note the output is more defined compared to the other results. Moreover, the epistemic and aleatoric uncertainty images also have a definite structure to them which is not the case for the OOD data results which follow.}
	\label{ixi_table}
	\centering
	\begin{tabular}{cc}
		\toprule
		Method     & Results   \\
		\midrule
		\rotatebox{90}{MC-Dropout} & \includegraphics[width=0.8\textwidth]{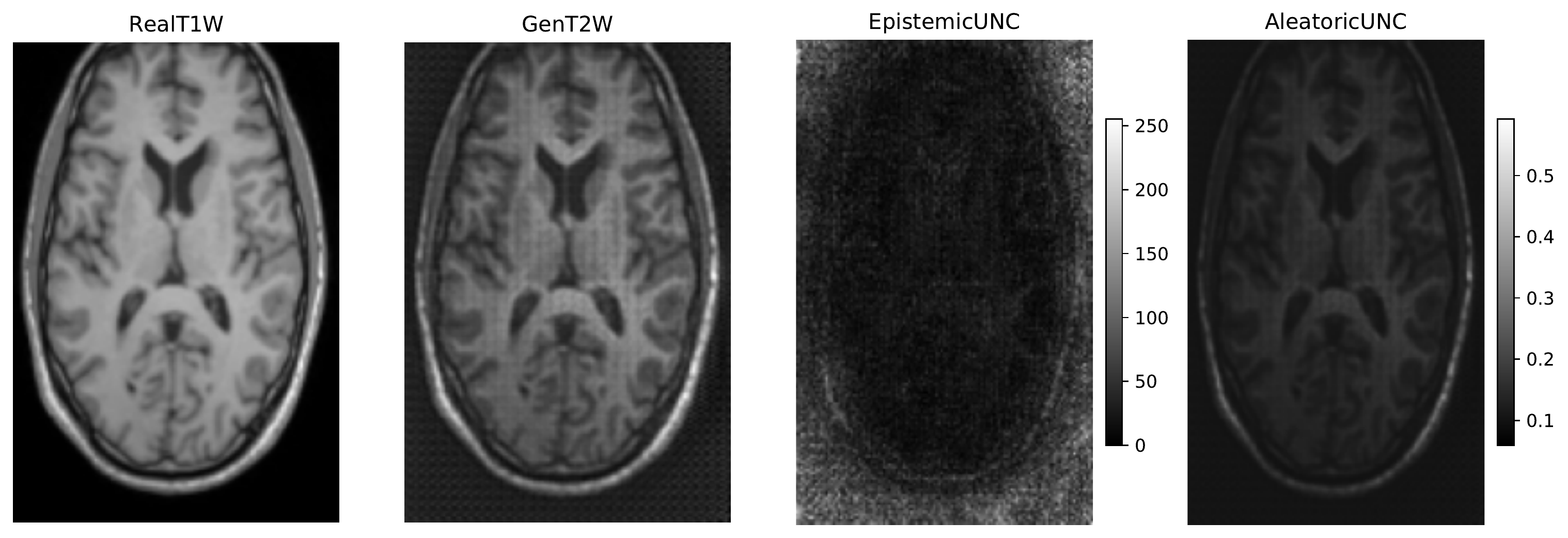} \\
		\rotatebox{90}{Flipout}     & \includegraphics[width=0.8\textwidth]{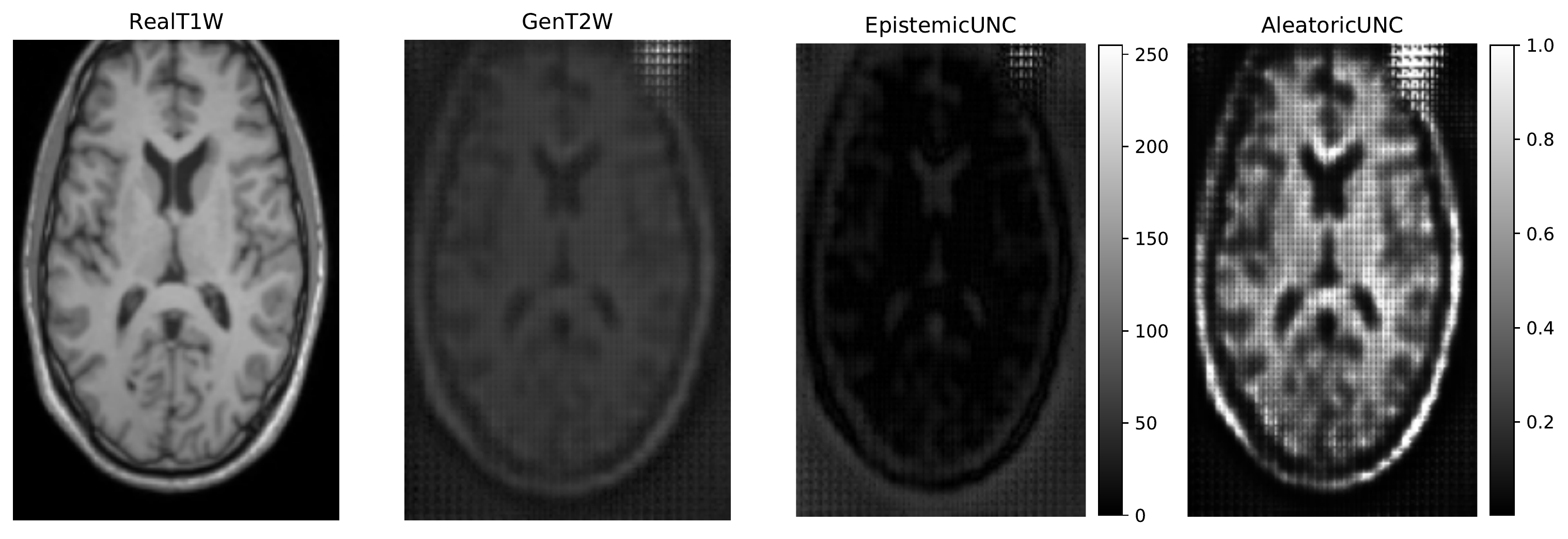} \\
		\rotatebox{90}{Ensemble}     &  \includegraphics[width=0.8\textwidth]{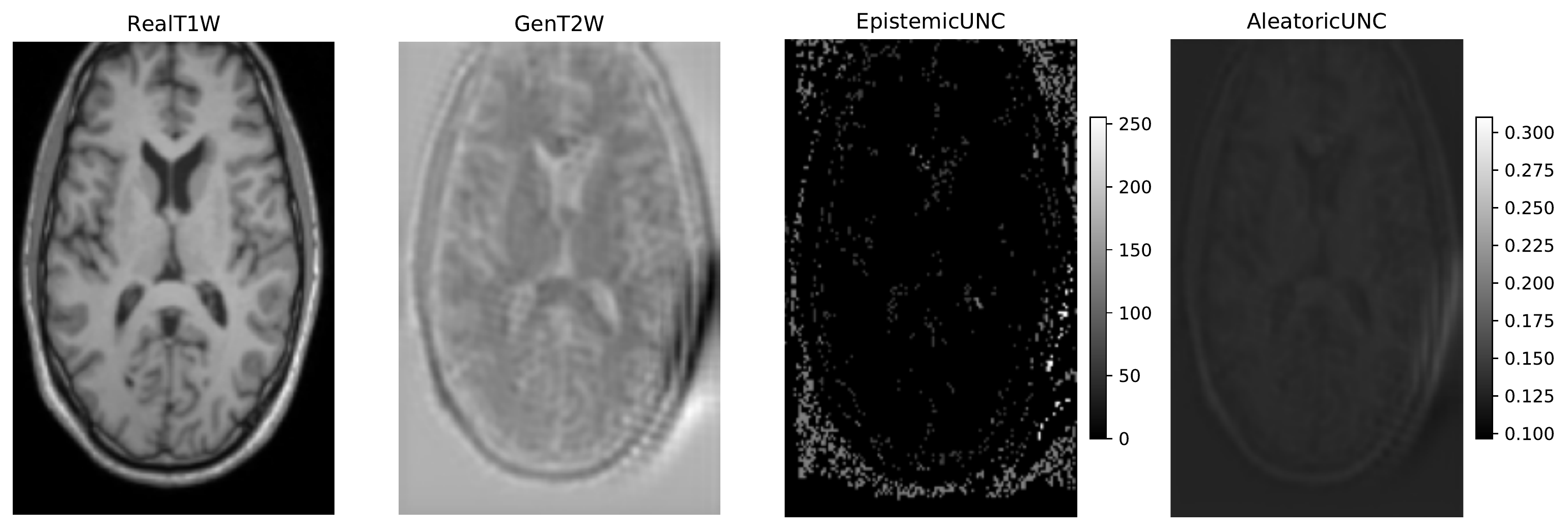}      \\
		\rotatebox{90}{MC-DropConnect}	  & \includegraphics[width=0.8\textwidth]{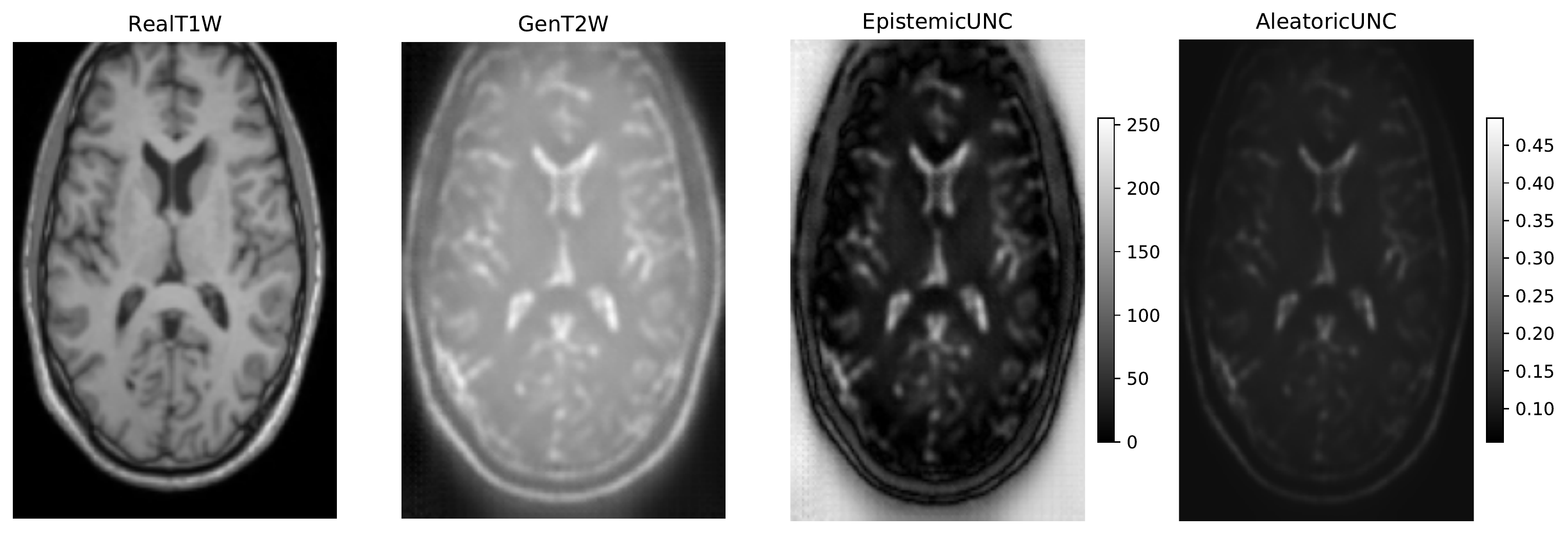} \\
		\bottomrule
	\end{tabular}
\end{table}

\begin{table}[ht]
    \caption{Qualitative results for Brain CT Scans: The model predictions make sense to some extent however since the CT data is OOD we can notice strange patterns in the images for epistemic and aleatoric uncertainties.}
    \label{ct_table}
    \centering
    \begin{tabular}{cc}
        \toprule
        Method     & Results   \\
        \midrule
        \rotatebox{90}{MC-Dropout} & \includegraphics[width=0.8\textwidth]{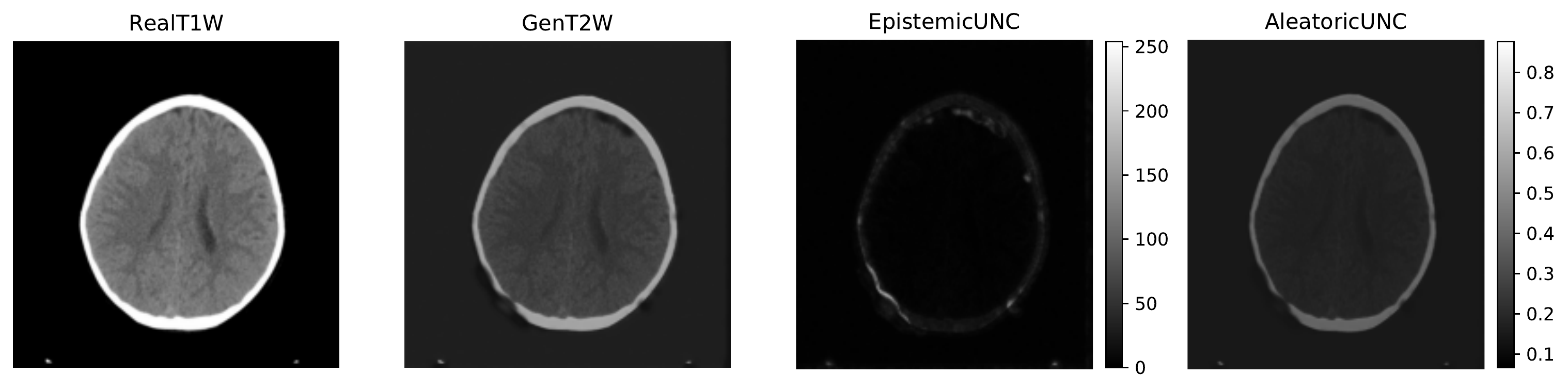} \\
        \rotatebox{90}{Flipout}     & \includegraphics[width=0.8\textwidth]{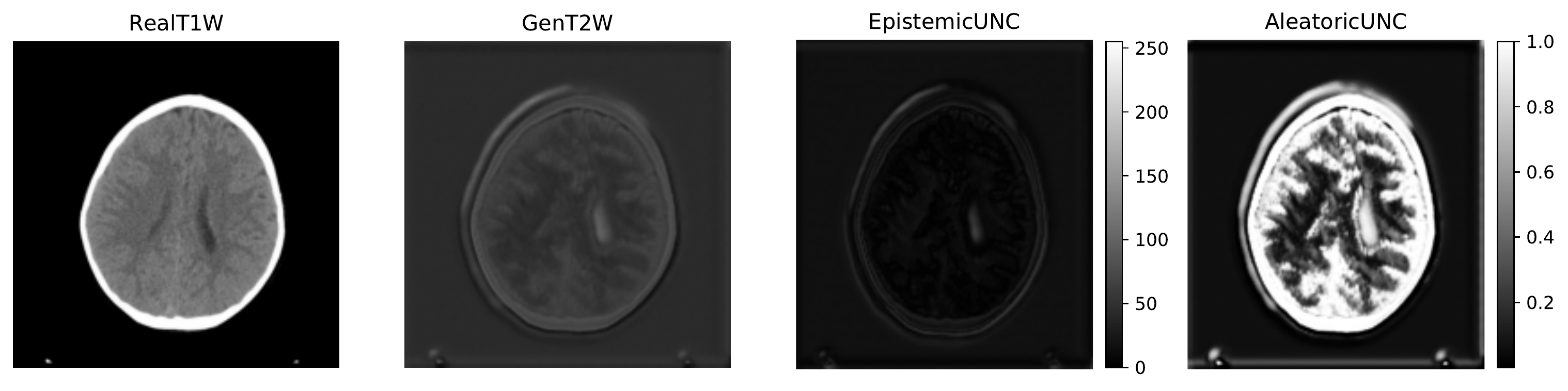} \\
        \rotatebox{90}{Ensemble}     &  \includegraphics[width=0.8\textwidth]{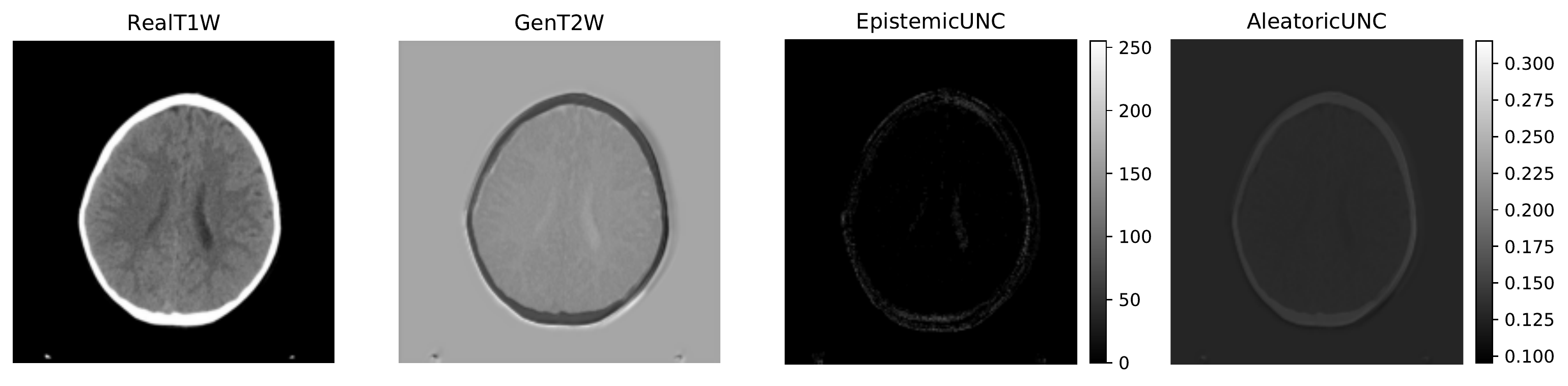}      \\
        \rotatebox{90}{MC-DropConnect}	  & \includegraphics[width=0.8\textwidth]{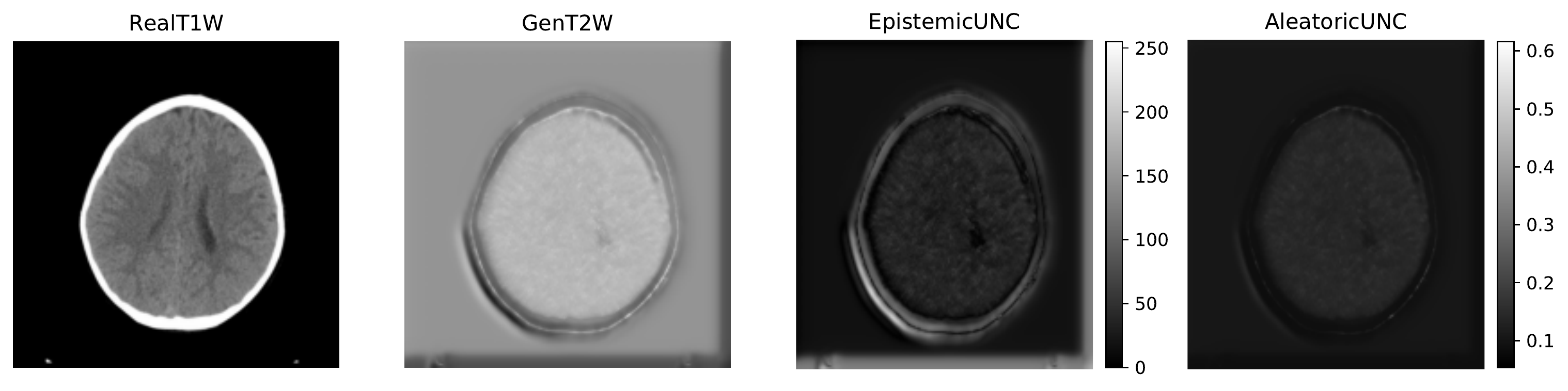} \\
        \bottomrule
    \end{tabular}
\end{table}

\begin{table}
    \caption{Qualitative results for UTKFace Dataset: The output images here do not make a lot sense and their quality is not as good as the ID resultant images. In addition to this, the epistemic and aleatoric uncertainty images in particular have strange patterns which are key to detecting out of distribution data.}
    \label{faces_table}
    \centering
    \begin{tabular}{cc}
        \toprule
        Method     & Results   \\
        \midrule
        \rotatebox{90}{MC-Dropout} & \includegraphics[width=0.8\textwidth]{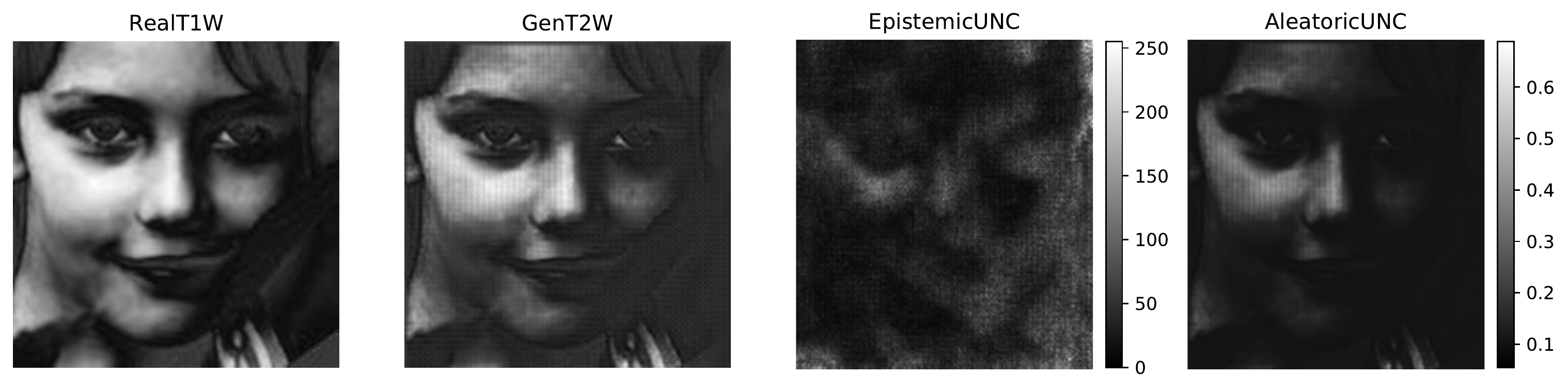} \\
        \rotatebox{90}{Flipout}     & \includegraphics[width=0.8\textwidth]{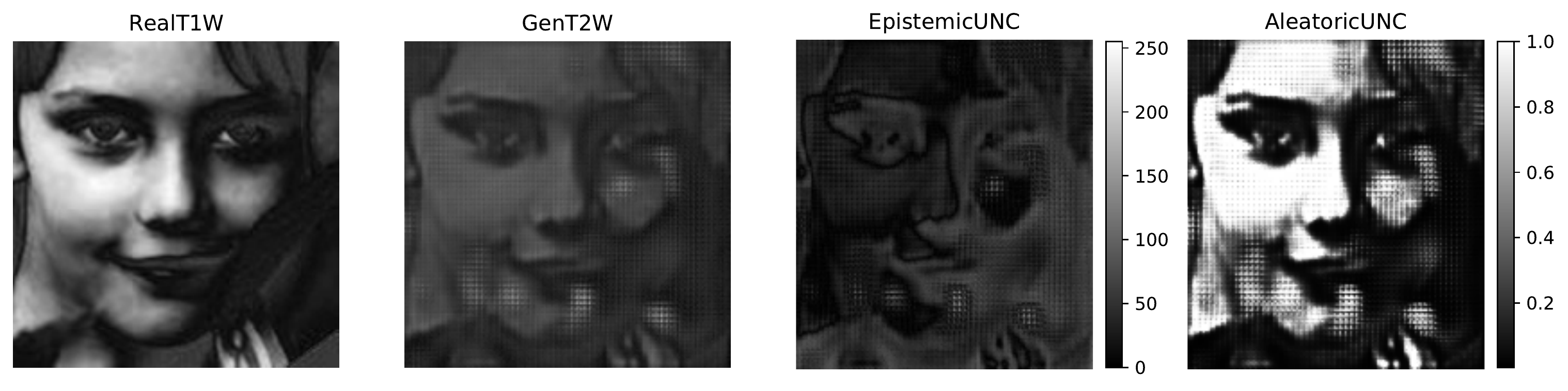} \\
        \rotatebox{90}{Ensemble}     &  \includegraphics[width=0.8\textwidth]{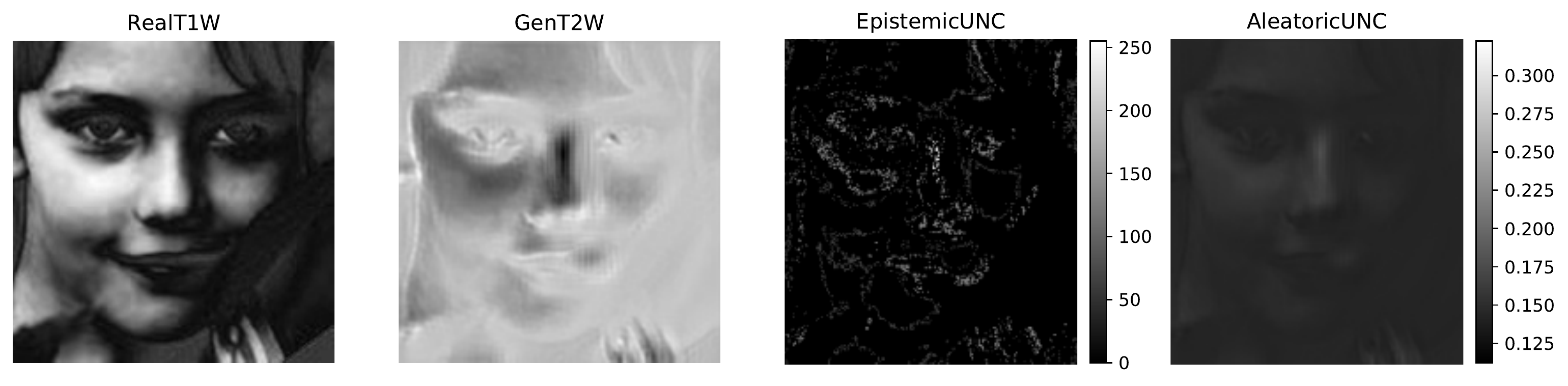}      \\
        \rotatebox{90}{MC-DropConnect}	  & \includegraphics[width=0.8\textwidth]{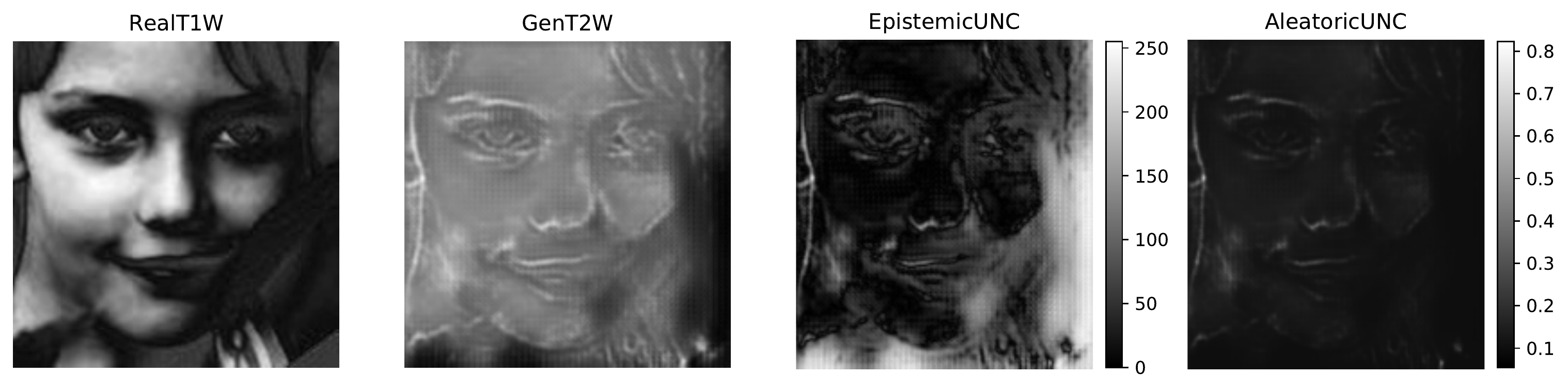} \\
        \bottomrule
    \end{tabular}
\end{table}

\FloatBarrier
\section{Out of Distribution Detection Results}

This section presents all out of distribution detection results, including numerical AUC scores in Table \ref{auc_ct}, and ROC curves in Table \ref{roc_table}. Each ID-OOD dataset combination is shown in the header of each table.

\begin{table}[ht]
    \caption{ROC curves for IXI-CT (OOD) and IXI-UTKFace (OOD) dataset combinations}
    \label{roc_table}
    \centering
    \begin{tabular}{ccccc}
        \toprule
        \rotatebox{45}{Dataset}	&	\multicolumn{2}{c}{IXI-CT}	&	\multicolumn{2}{c}{IXI-UTKFace}	\\
        \midrule
        Method     & Aleatoric Uncert & Epistemic Uncert & Aleatoric Uncert & Epistemic Uncert  \\
        \midrule
        \rotatebox{90}{MC-Dropout} & \includegraphics[width=0.2\textwidth]{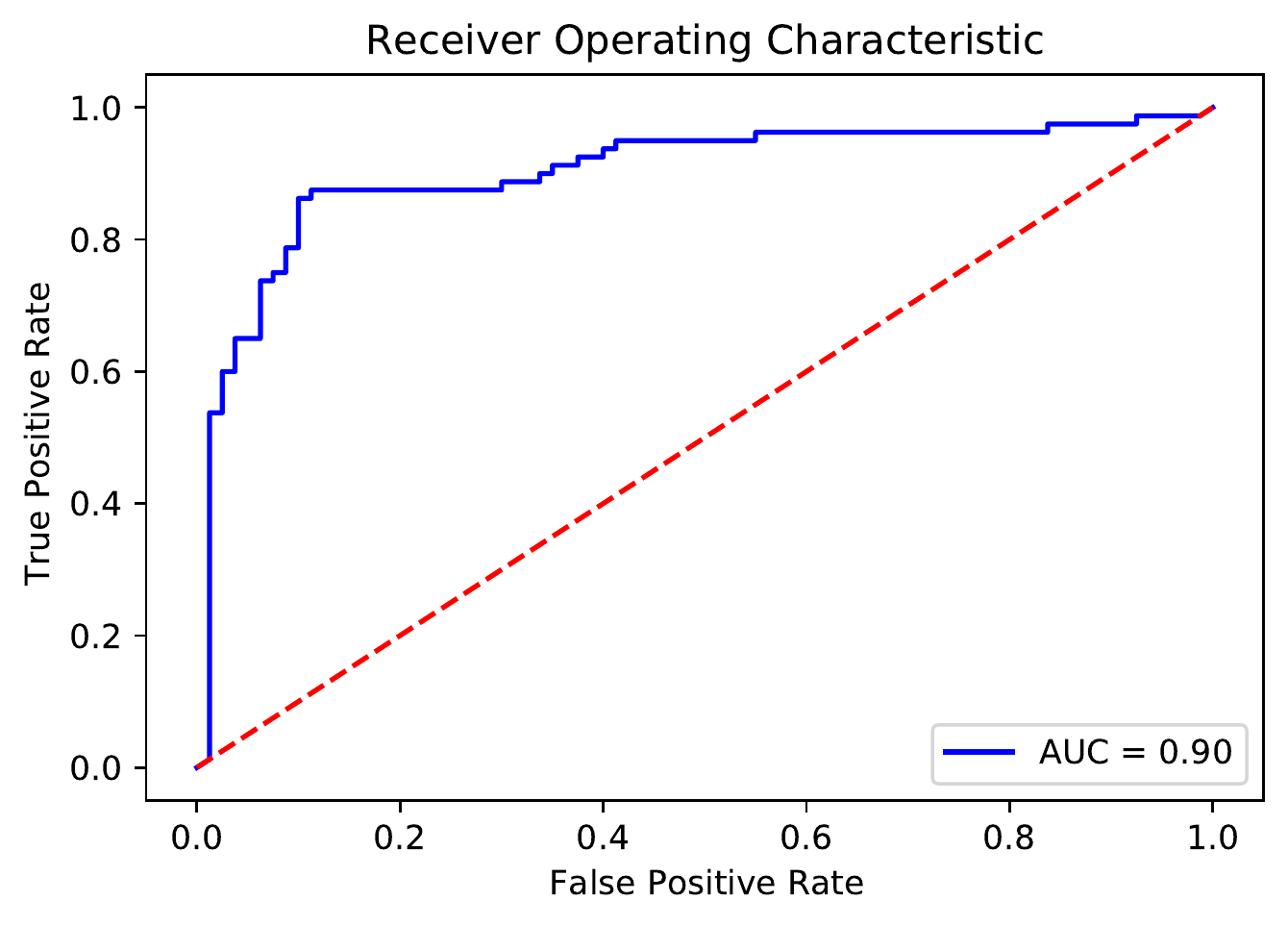} & \includegraphics[width=0.2\textwidth]{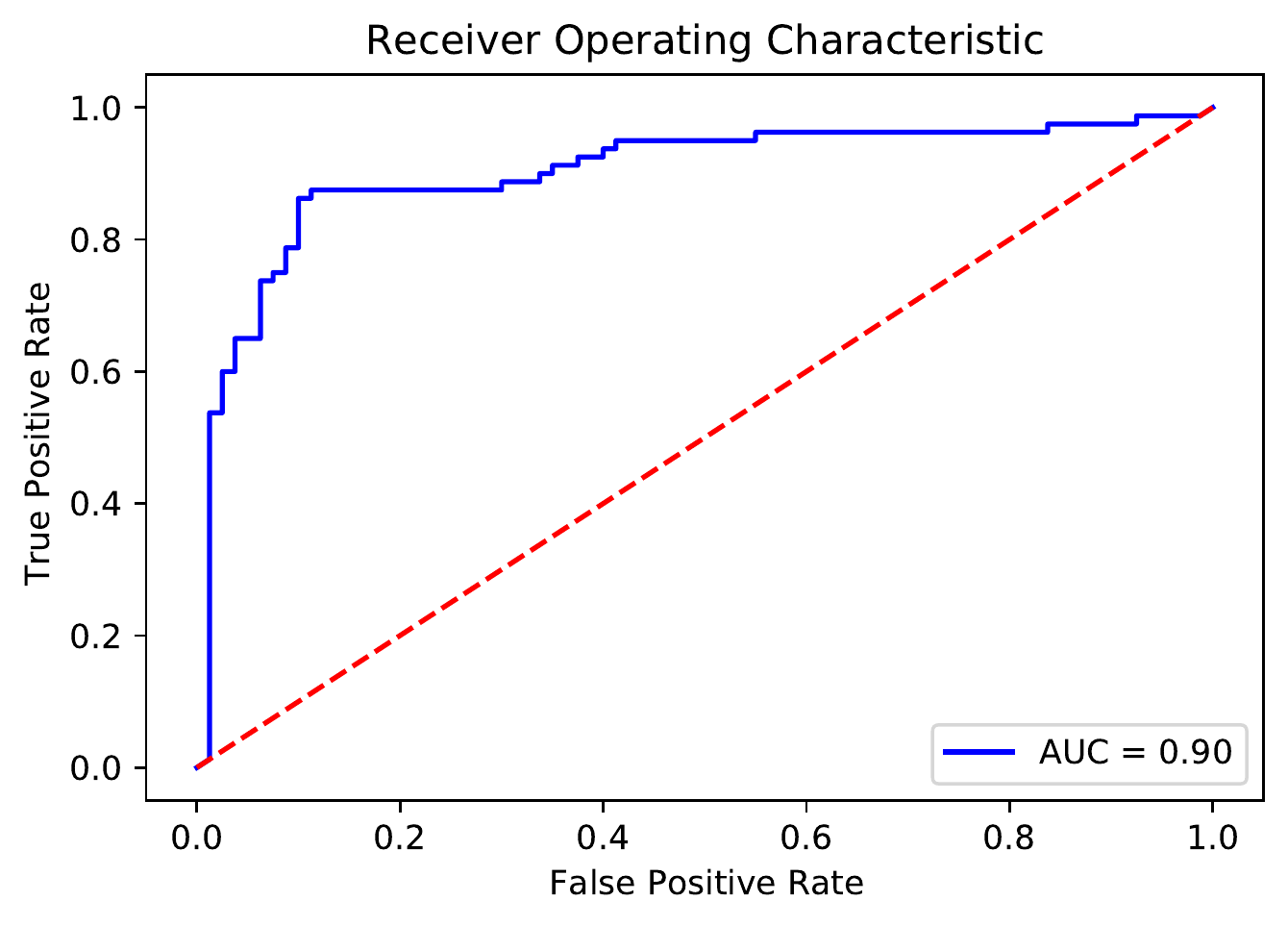} & \includegraphics[width=0.2\textwidth]{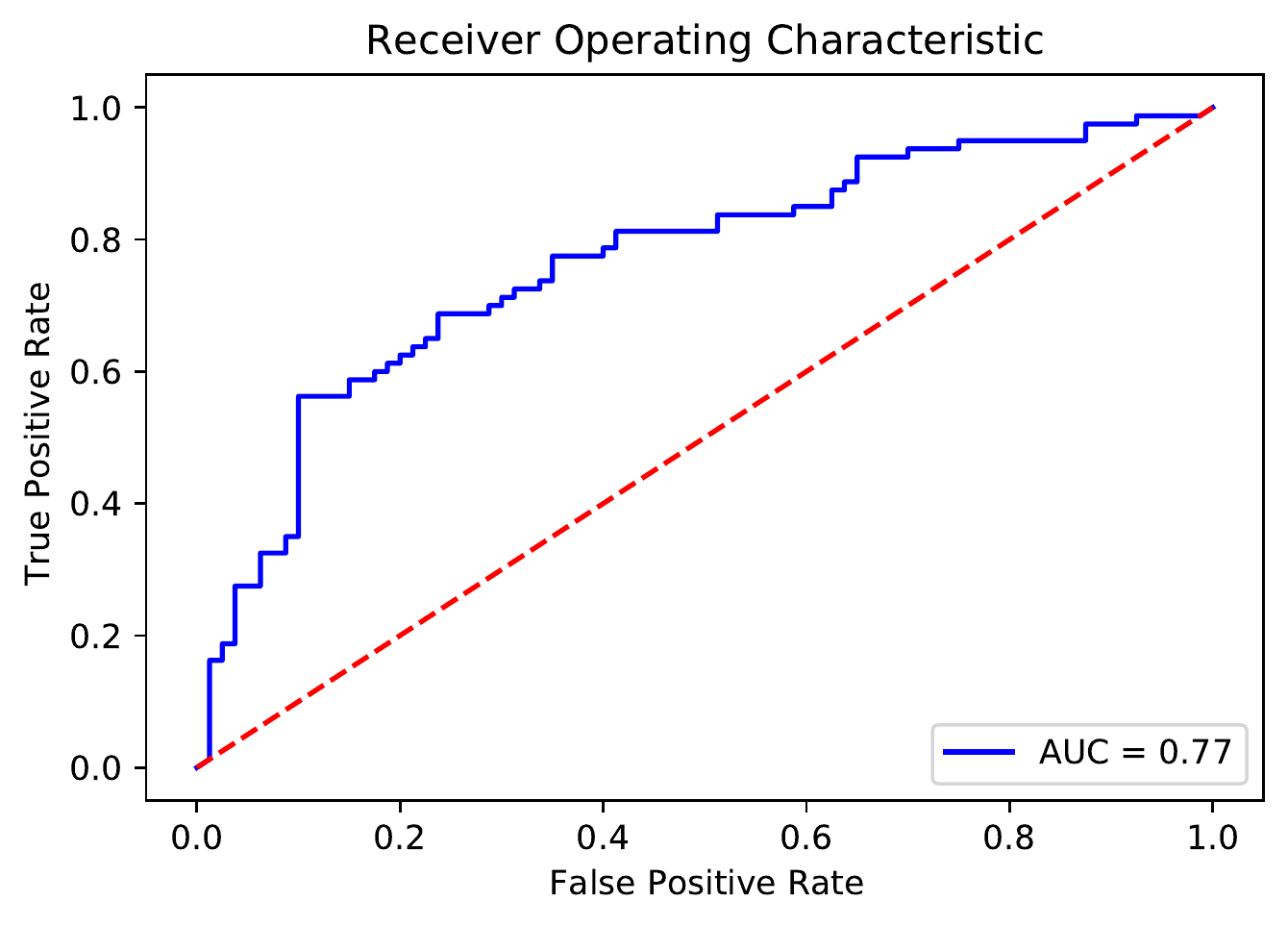} & \includegraphics[width=0.2\textwidth]{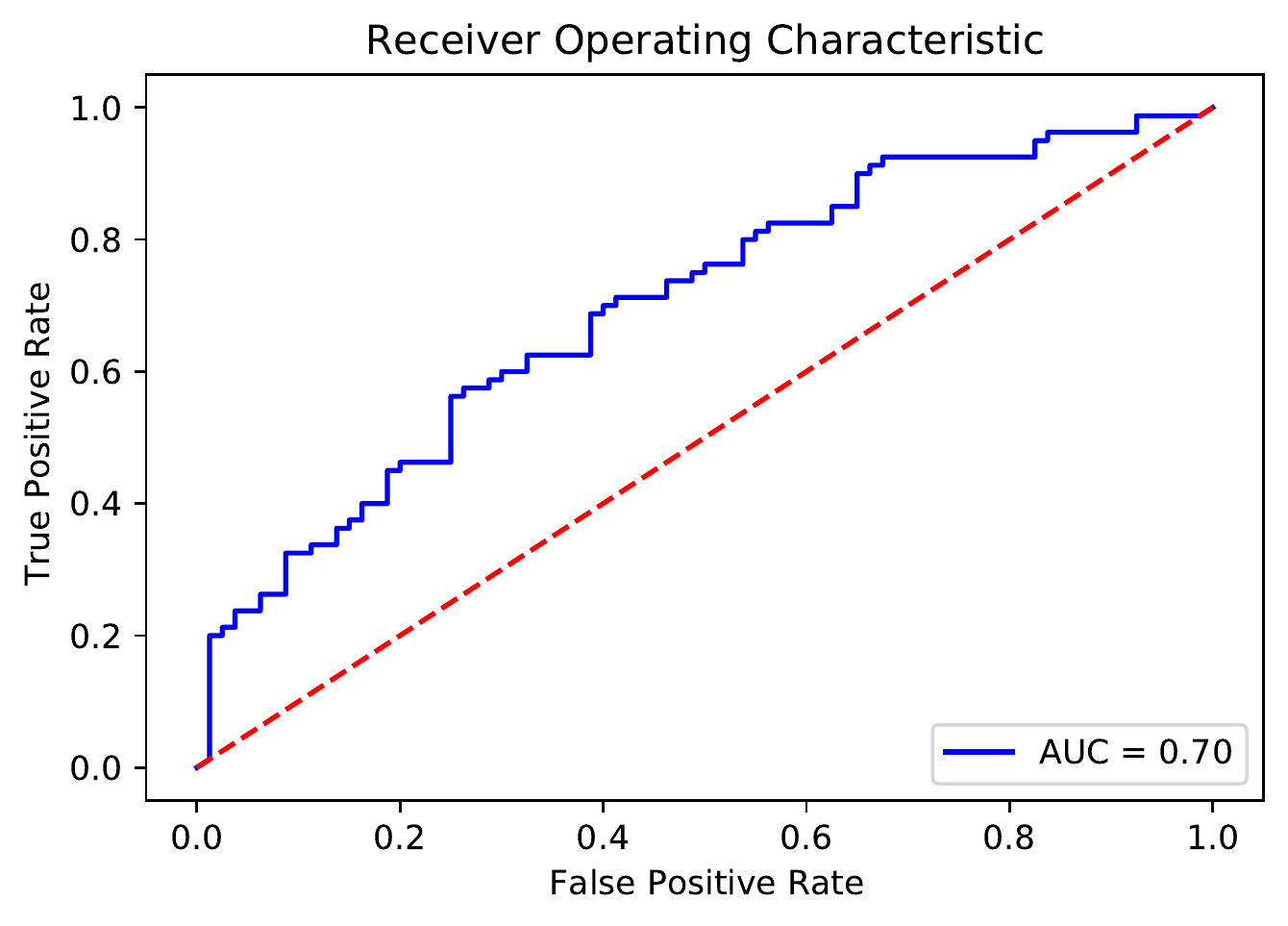} \\
        \rotatebox{90}{Flipout}   & \includegraphics[width=0.2\textwidth]{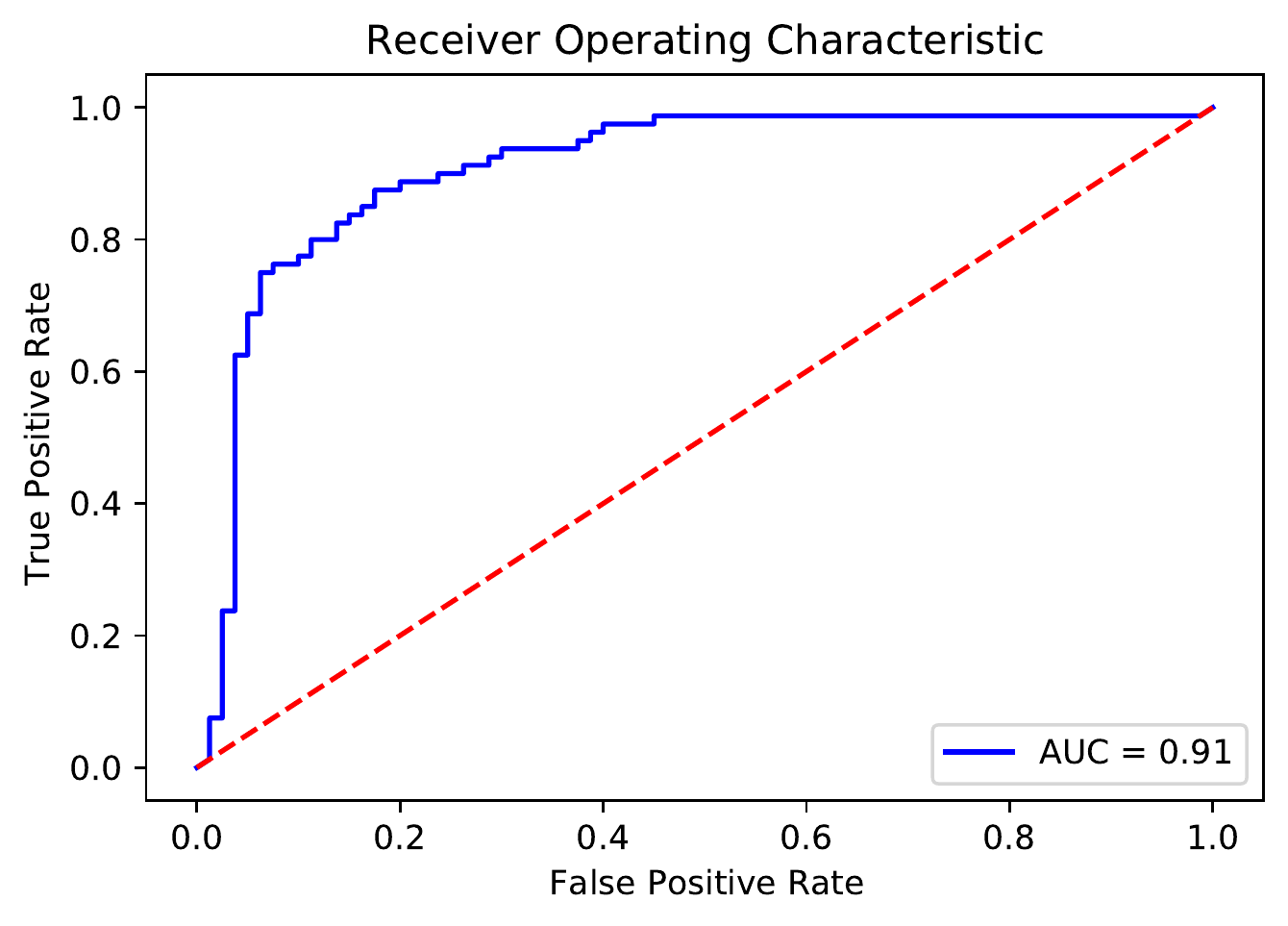} & \includegraphics[width=0.2\textwidth]{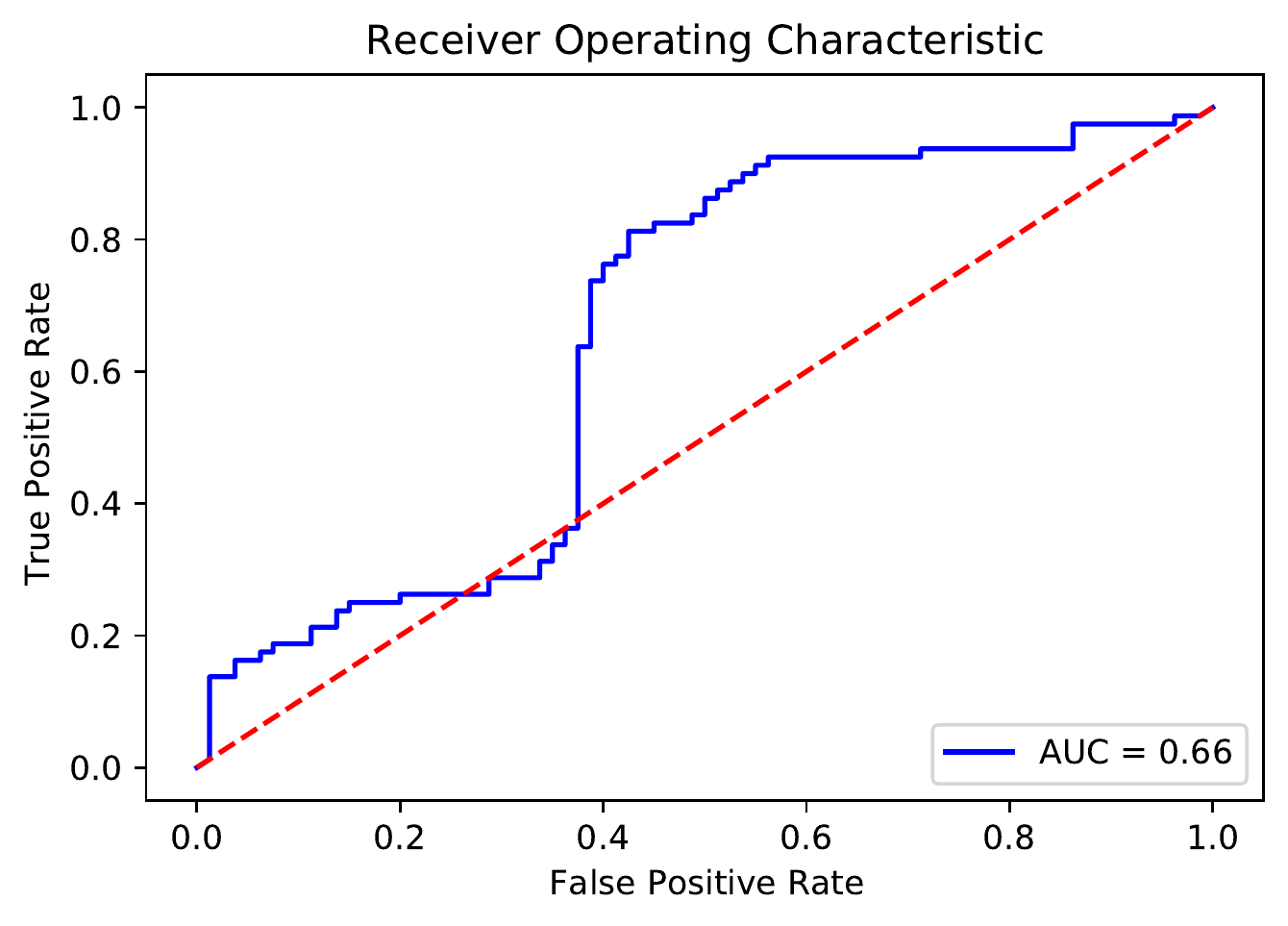}   & \includegraphics[width=0.2\textwidth]{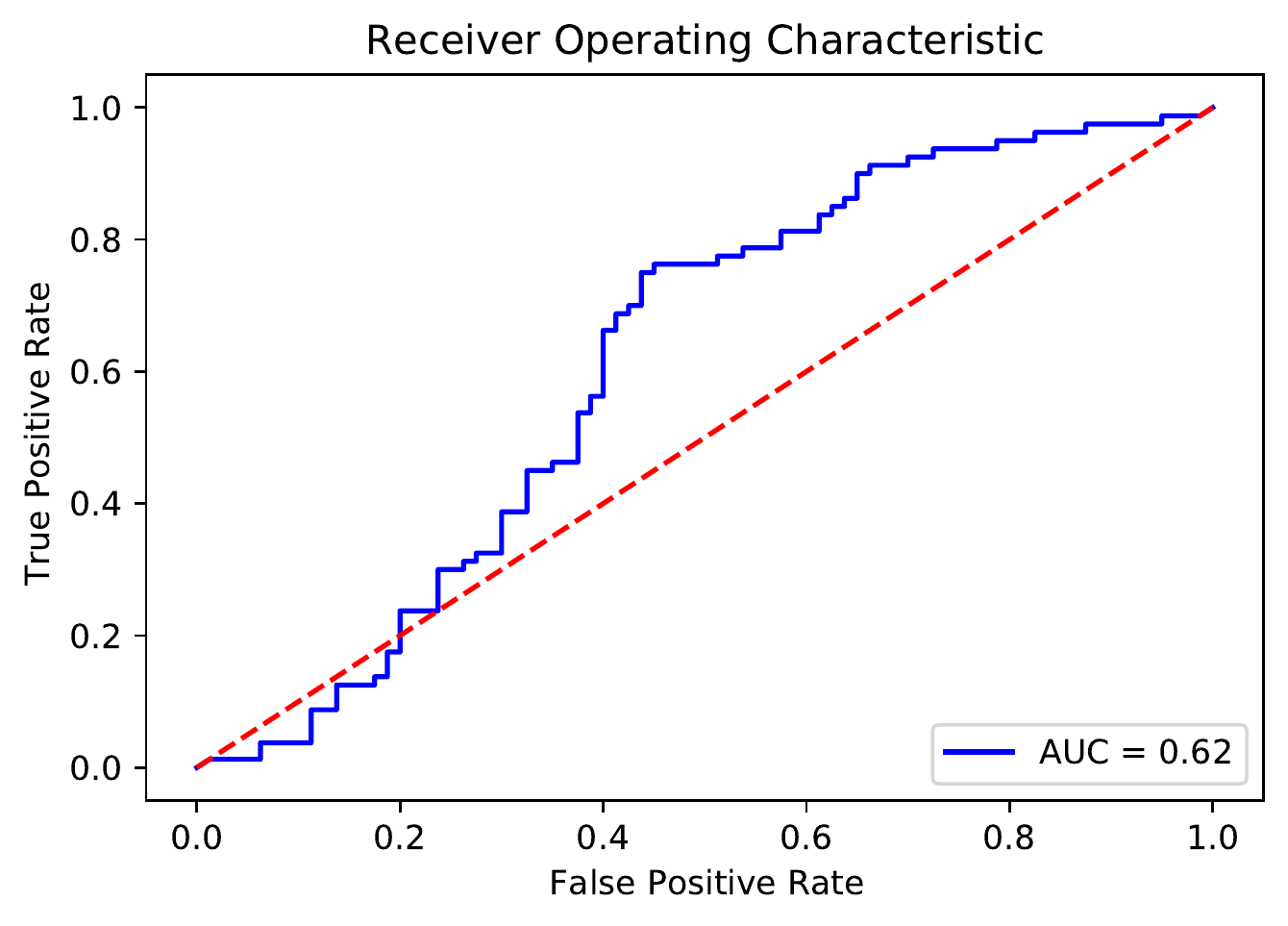} & \includegraphics[width=0.2\textwidth]{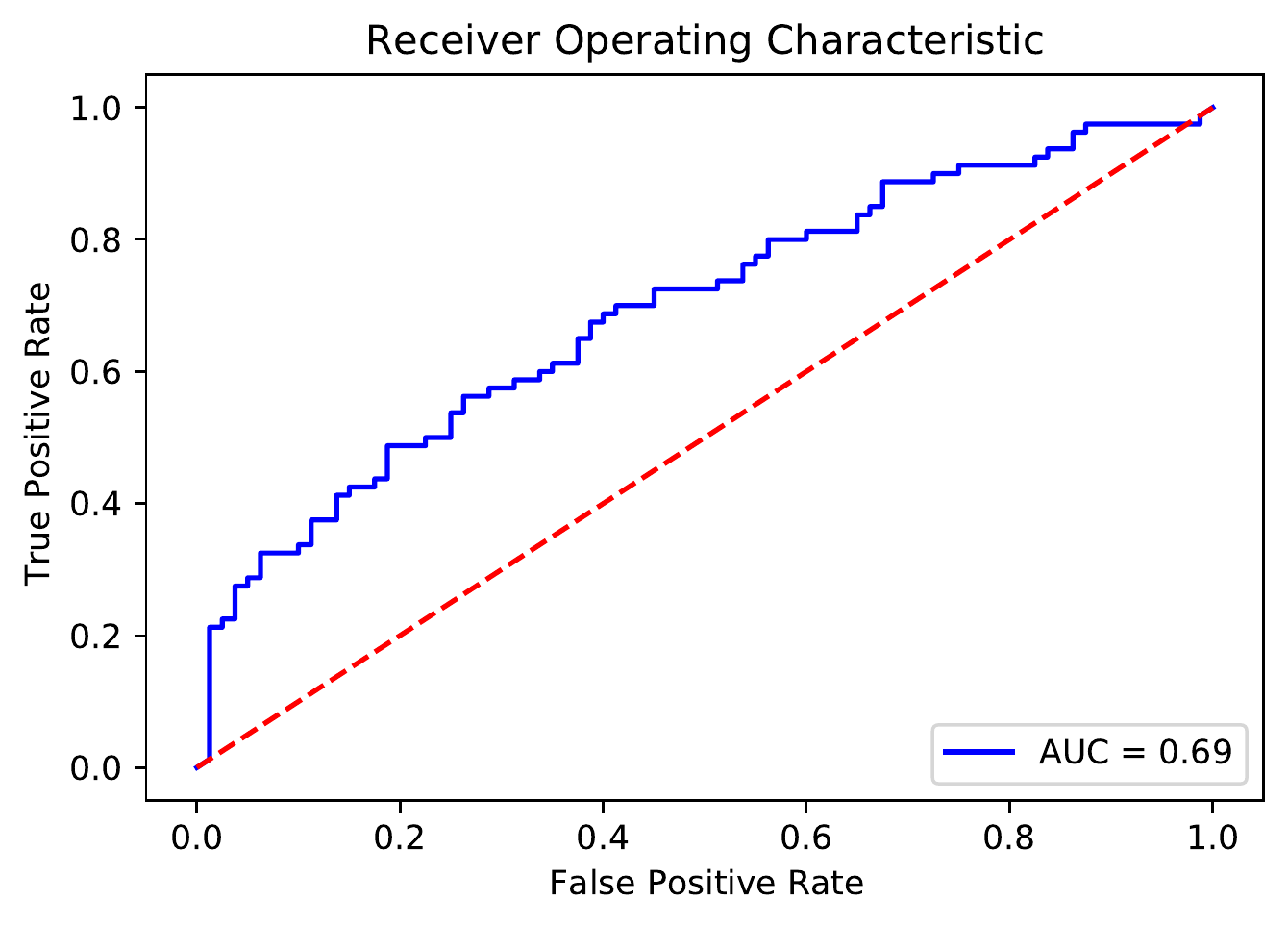} \\
        \rotatebox{90}{Ensemble}   &  \includegraphics[width=0.2\textwidth]{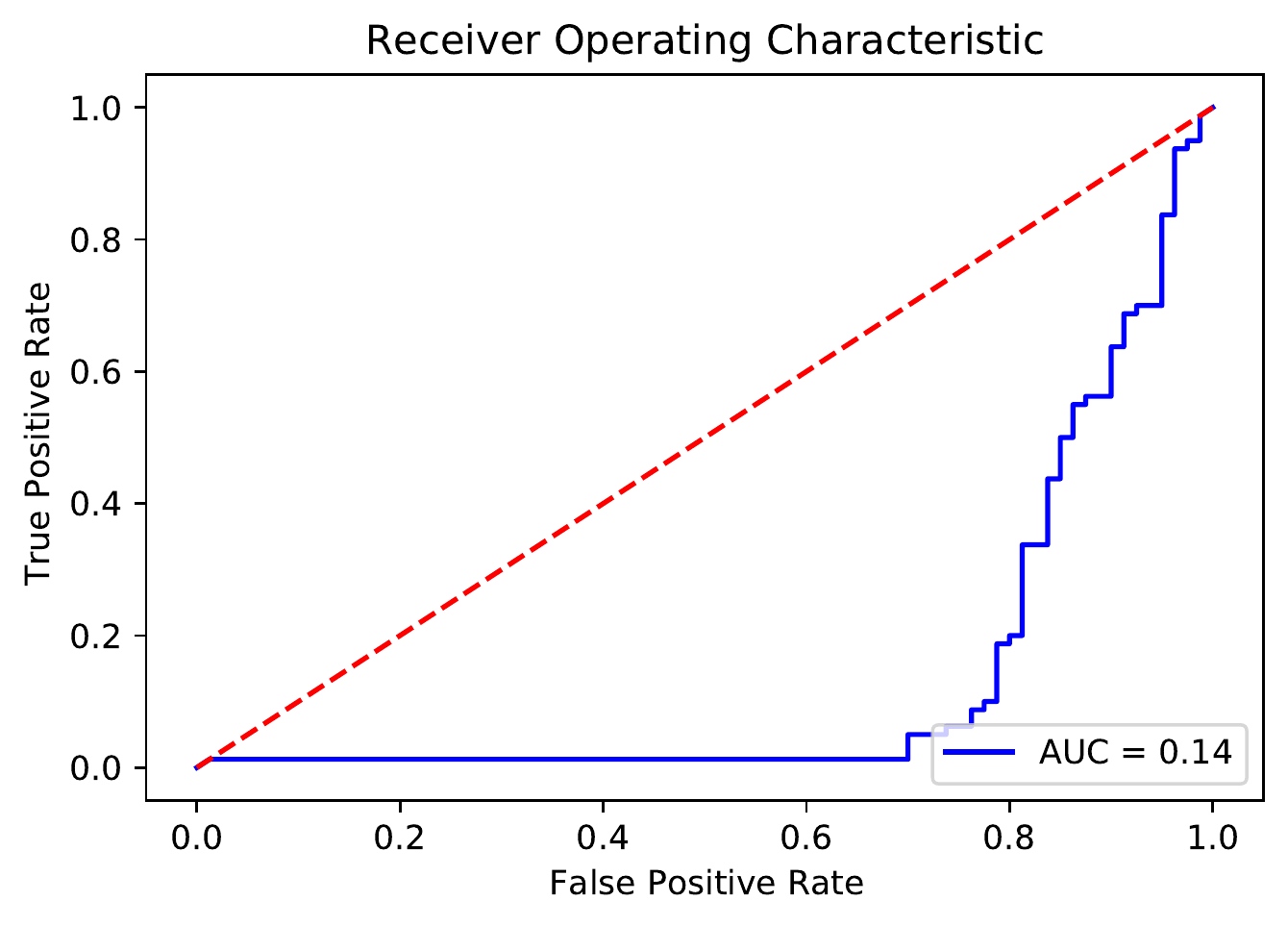} & \includegraphics[width=0.2\textwidth]{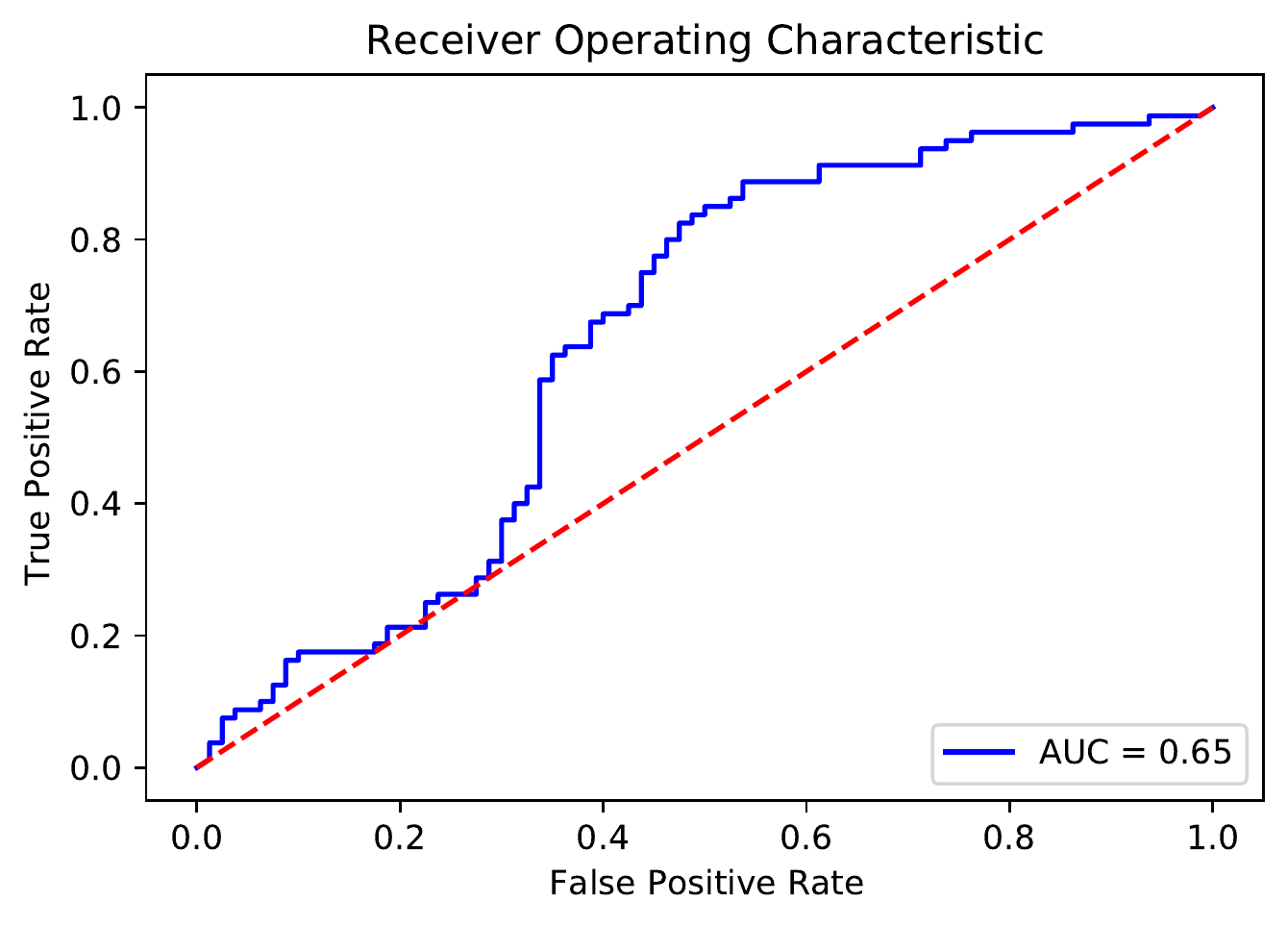}  &  \includegraphics[width=0.2\textwidth]{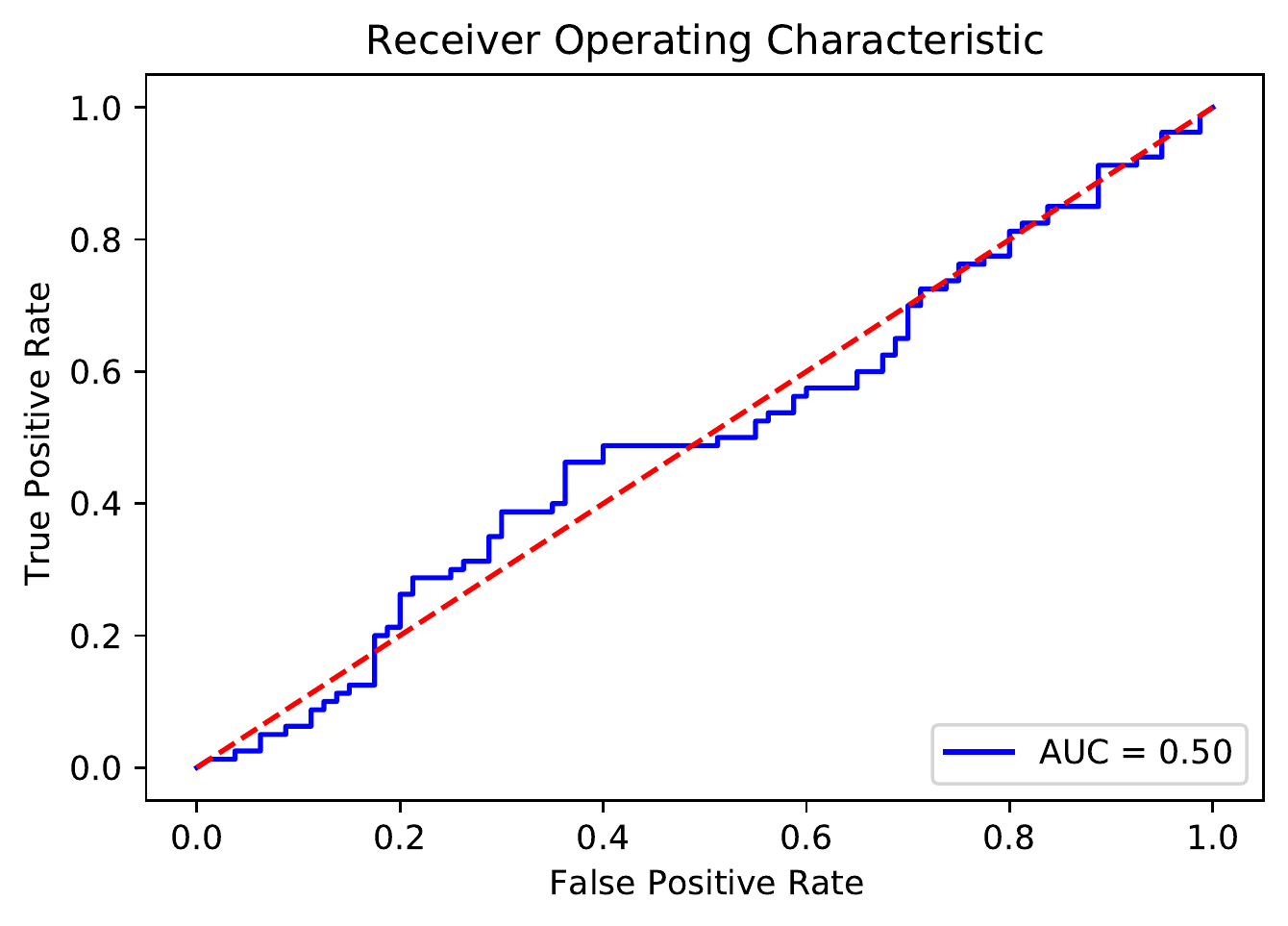} & \includegraphics[width=0.2\textwidth]{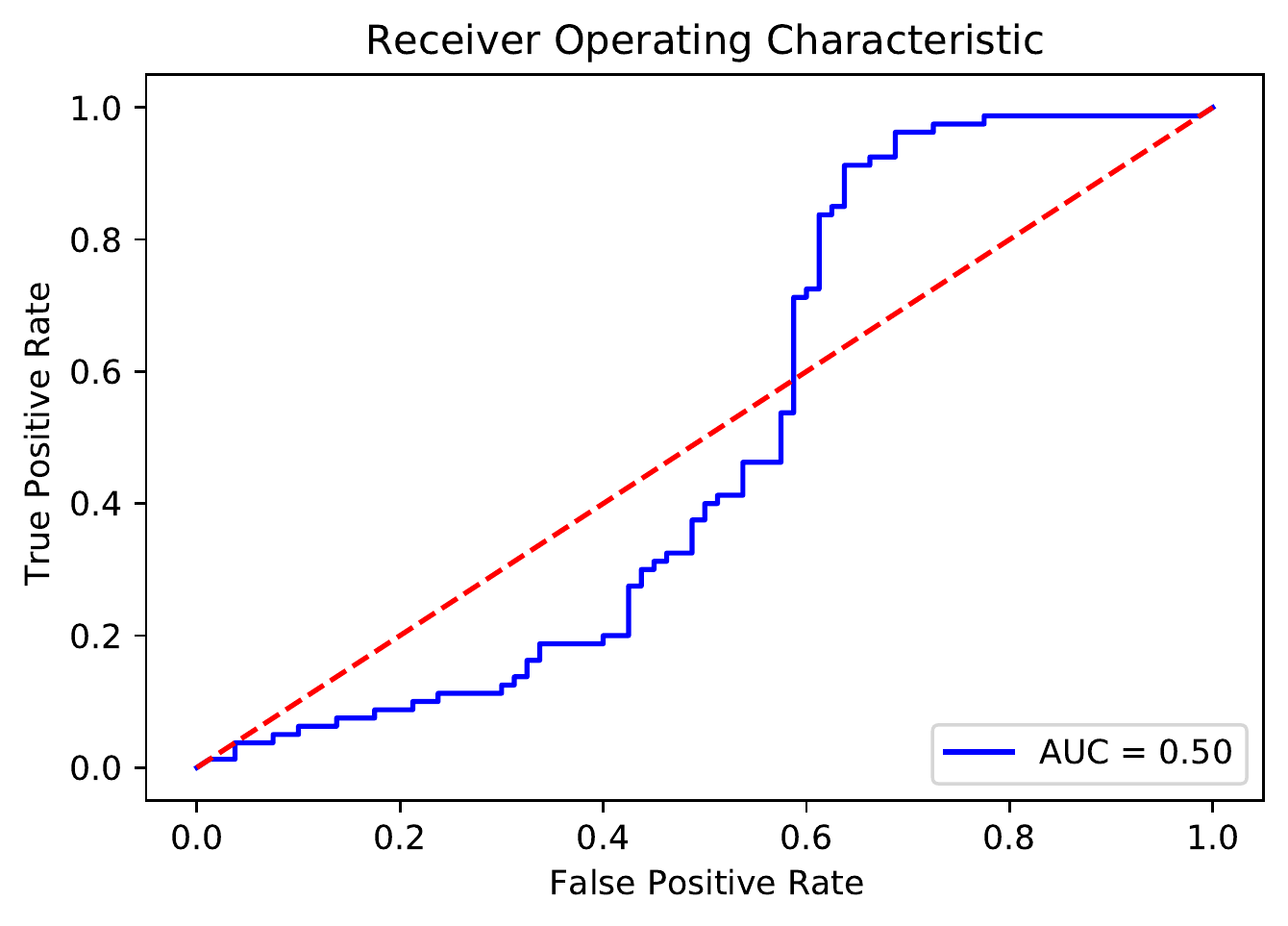}      \\
        \rotatebox{90}{MC-DropConnect}	 & \includegraphics[width=0.2\textwidth]{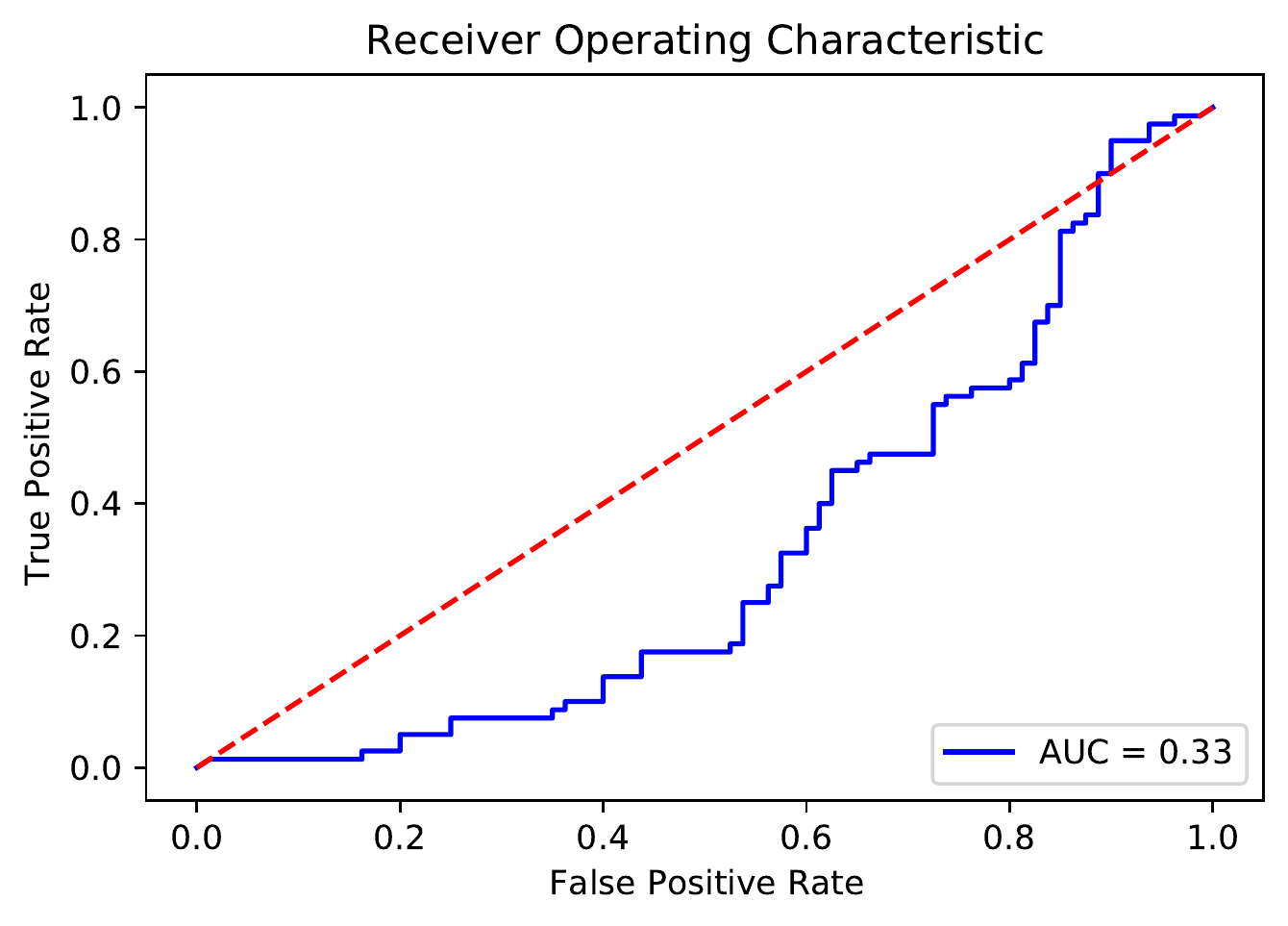} & \includegraphics[width=0.2\textwidth]{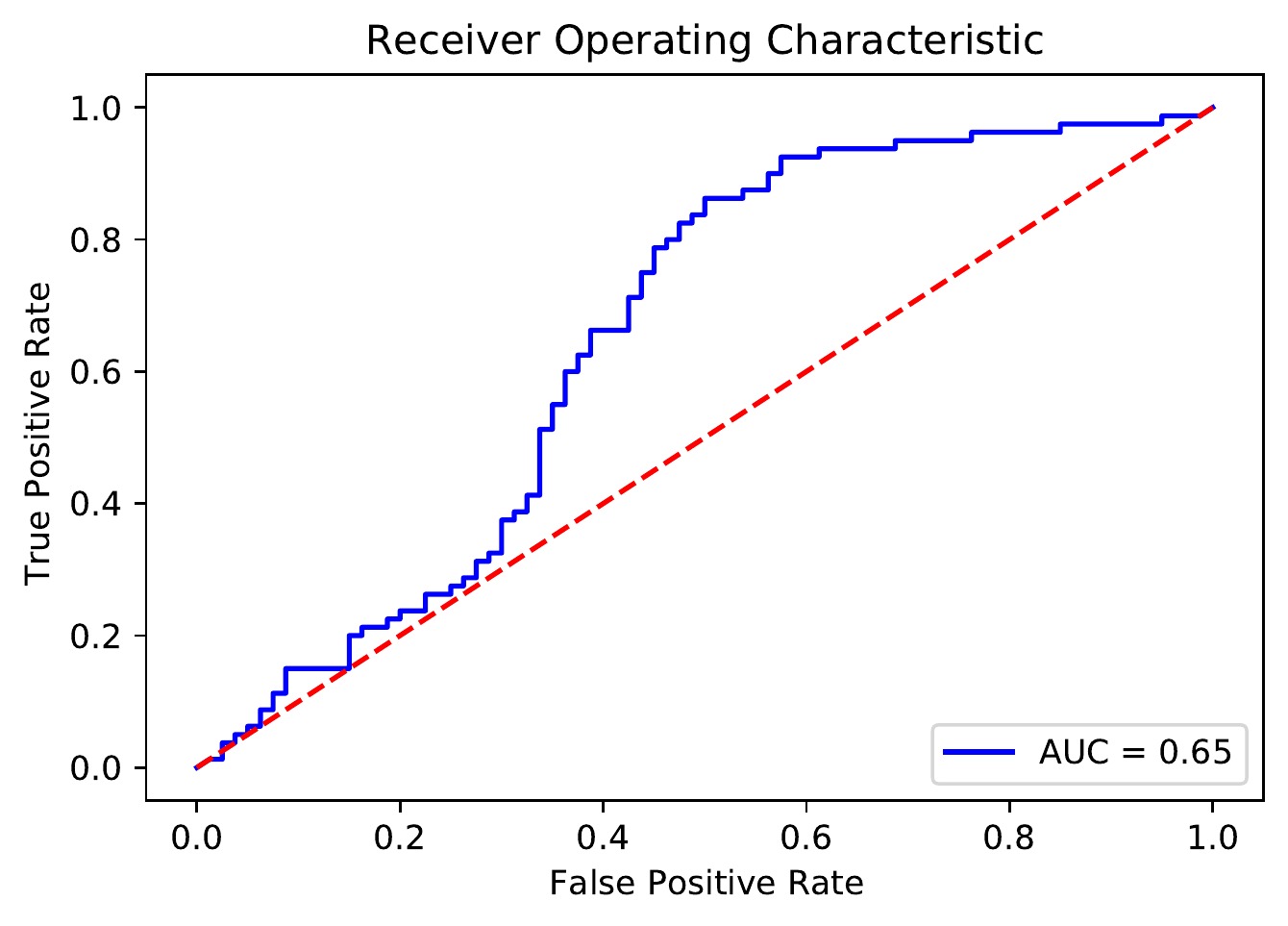} & \includegraphics[width=0.2\textwidth]{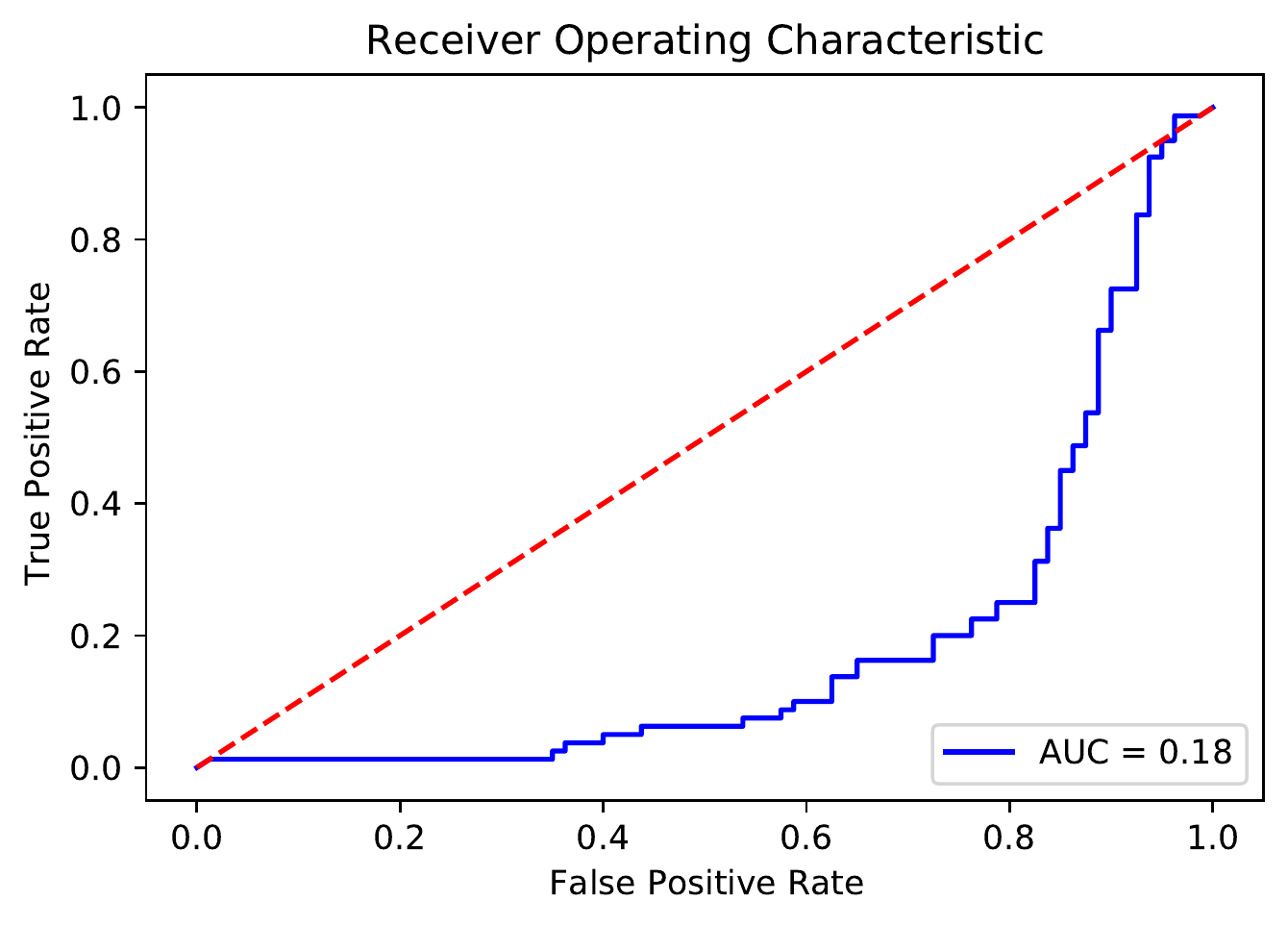} & \includegraphics[width=0.2\textwidth]{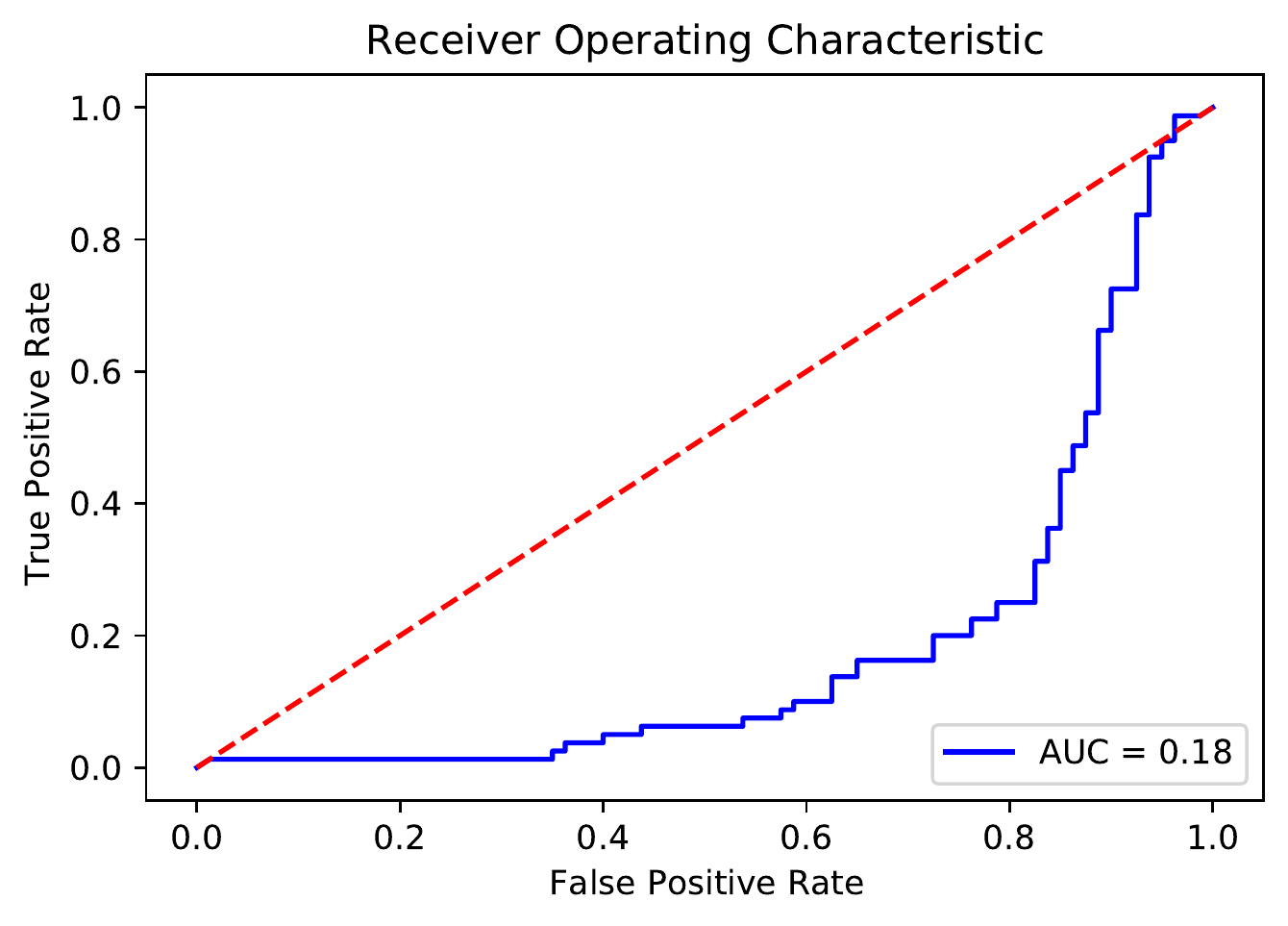} \\
        \bottomrule
    \end{tabular}
\end{table}

\begin{table}
    \caption{AUC values for IXI-CT (OOD) and IXI-UTKFace (OOD) dataset combinations}
    \label{auc_ct}
    \centering
    \begin{tabular}{lllll}
        \toprule
        \rotatebox{45}{Dataset}	&	\multicolumn{2}{c}{IXI-CT}	&	\multicolumn{2}{c}{IXI-UTKFace}	\\
        \midrule
        Method     & Aleatoric Uncert & Epistemic Uncert & Aleatoric Uncert & Epistemic Uncert  \\
        \midrule
        MC-Dropout & 0.90 & 0.90  & 0.77 & 0.70\\
        Flipout     & 0.91 & 0.66 & 0.62 & 0.69\\
        Ensemble     &  0.14 & 0.65	&  0.50 & 0.50      \\
        DropConnect	  & 0.33 & 0.65 & 0.18 & 0.18\\
        \bottomrule
    \end{tabular}
\end{table}

\FloatBarrier
\section{Histograms for Aleatoric and Epistemic Uncertainty}

\end{document}